\documentclass[11pt]{article}
\usepackage{graphicx}
\usepackage{amsmath}
\usepackage{amsfonts}
\usepackage{amssymb}
\usepackage{epsfig}
\usepackage{times}
\usepackage{cite}
\usepackage{calc}
\usepackage{version}
\usepackage[english]{babel}
\begin{document}

%%%%%%%%%%%%%%%%%%%%%%%%%%%%%%%%%%%%%%%%%%%%%%%%%%%%%%%%%%%%%%%%%%%%%%%%%%%%%%
%%%%%%%%%%%%%%%%%%%%%%%%%%%%%%%%%%%%%%%%%%%%%%%%%%%%%%%%%%%%%%%%%%%%%%%%%%%%%%
\begin{titlepage}

\null

\vskip 1.5cm
\begin{flushright}
Report: IFIC/11-18, FTUV-11-0419\\
\end{flushright}

\vskip 1.cm

{\bf\large\baselineskip 20pt
\begin{center}
\begin{Large}
Three-particle correlations in QCD jets and beyond
\end{Large}
\end{center}
}
\vskip 1cm

\begin{center}
Redamy P\'erez-Ramos
\footnote{e-mail: redamy.perez@uv.es}, 
Vincent Mathieu
\footnote{e-mail: vincent.mathieu@ific.uv.es} and Miguel-Angel Sanchis-Lozano
\footnote{e-mail: miguel.angel.sanchis@ific.uv.es}\\
\medskip
Departament de F\'{\i}sica Te\`orica and IFIC, 
Universitat de Val\`encia - CSIC\\
Dr. Moliner 50, E-46100 Burjassot, Spain
\end{center}

\baselineskip=15pt

\vskip 3.5cm

{\bf Abstract}: In this paper, we present a more detailed version of our 
previous work for three-particle correlations in quark and gluon jets \cite{Ramos:2011tw}.
We give theoretical results for this observable in the double 
logarithmic approximation and the modified leading logarithmic 
approximation. In both resummation schemes, we use the formalism
of the generating functional and solve the evolution
equations analytically from the steepest descent evaluation of
the one-particle distribution. 
In addition, in this paper we include predictions beyond the limiting
spectrum approximation and study this observable near
the hump of the single inclusive distribution. 
We thus provide a further test of 
the local parton hadron duality (LPHD)  and make predictions for the
LHC. The computation of higher rank correlators is presented in the double 
logarithmic approximation and shown to be rather cumbersome.
\end{titlepage}

\section{Introduction}The observation of quark and gluon jets has played a crucial role
in establishing Quantum Chromodynamics (QCD) as the theory of strong interaction within
the Standard Model of particle physics \cite{Fritzsch:1972jv,Gross:1973id}. 
Jets, narrowly collimated bundles of hadrons, reflect configurations of quarks 
and gluons at short distances \cite{Hanson:1975fe,Berger:1978rr}.

The evolution of gluon and quark initiated jets is dominated by soft gluon 
bremsstrahlung. Powerful schemes, like the Double Logarithmic Approximation (DLA)
and the Modified Leading Logarithmic Approximation (MLLA), which allow for the 
perturbative resummation of soft-collinear and hard-collinear gluons before the hadronization occurs, 
have been developed over the past thirty years (for a review see \cite{Dokshitzer:1991wu}). 
In the frame of high energy jets, one of 
the strikest predictions of perturbative QCD (pQCD),
which follows as a consequence of Angular Ordering (AO) within the MLLA and
the Local Parton Hadron Duality (LPHD) hypothesis \cite{Azimov:1984np}, is the existence of the hump-backed 
shape \cite{Azimov:1985by} of the inclusive energy distribution of hadrons,  later confirmed 
by experiments at colliders like the LEP \cite{Akrawy:1990ha,Abbiendi:1999ki} and the Tevatron \cite{:2008ec}. 
Within the same formalism,
the transverse momentum distribution, or $k_\perp$-spectra of hadrons produced in $p\bar p$ 
collisions at center of mass energy $\sqrt{s}=1.96\text{ TeV}$ at the Tevatron \cite{Aaltonen:2008yn},
was well described by MLLA \cite{PerezRamos:2005nh} and next-to-MLLA (NMLLA) \cite{PerezRamos:2007cr,Arleo:2007wn} 
predictions inside the validity ranges provided by such schemes, both supported by the LPHD.
Thus, the study and tests of enough inclusive observables like the inclusive energy distribution and the 
inclusive transverse momentum $k_\perp$ spectra of hadrons have shown that the perturbative stage of the process, which
evolves from the hard scale or leading parton virtuality $Q\sim E$ to the hadronization
scale $Q_0$, is dominant. In particular, these issues suggest that the hadronization stage 
of the QCD cascade do not affect pQCD predictions and therefore, 
that the LPHD hypothesis is successful while treating 
one-particle inclusive observables.

The study of particle correlations in intrajet cascades, which are less inclusive observables, 
provide a refined test of the partonic dynamics and the LPHD. In \cite{Dokshitzer:1982ia}, the 
two-particle correlations inside quark and gluon jets were first computed at DLA. In \cite{Fong:1990nt,Fong:1990ph},
this observable was computed for the first time at MLLA for such particles, whose energy or $x$
(energy fraction of the jet carried away by one parton) 
stays close to the maximum of the one-particle distribution. In \cite{Ramos:2006dx}, the
previous solutions were extended, at MLLA, to all possible values of $x$ by exactly solving
the QCD evolution equations. This observable was measured by the OPAL collaboration in the 
$e^+e^-$ annihilation at the $Z^0$ peak, that is for $\sqrt{s}=91.2\text{ GeV}$ at LEP \cite{Acton:1992gd}. 
Though the agreement with predictions presented in 
\cite{Ramos:2006dx} turned out to be rather good for the description
of the data \cite{Acton:1992gd}, a discrepancy still subsists pointing out a possible failure of 
the LPHD for less inclusive observables. However, these measurements were redone by the CDF collaboration in $p\bar p$ collisions
at the Tevatron for mixed samples of quark and gluon jets \cite{:2008ec}. The agreement with predictions presented 
in \cite{Fong:1990nt,Fong:1990ph} turned out to be rather good, in particular 
for very soft particles ($x\ll0.1$) having very close energy fractions ($x_1\approx x_2$). A discrepancy between the
OPAL and CDF analysis showed up and still stays unclear. That is why, the measurement of  
two-particle correlations at the LHC becomes crucial. 

By going one step beyond, in this paper we give predictions for three-particle 
correlations inside quark and gluon jets. This observable together with two-particle correlations 
can be measured in equal footing at the LHC. Such tests will provide further verifications of the LPHD
for less inclusive observables and shed more light on the role of confinement in jet evolution. 
Further issues on the importance of correlations versus single-particle distributions studies have
been presented in \cite{De Wolf:1995pc,Kittel:2005fu}.

The paper is organized as follows.
\begin{itemize}
\item in section \ref{sec:formalism} we recall the formalism of jet generating functionals and 
their evolution equations; 
\item the kinematics and the process for the inclusive production of three particles
inside the jet are specified in subsection \ref{subsec:kinemat} and \ref{sub:bounds} respectively;
\item in subsection \ref{subsec:threeparteveq}, we obtain the MLLA exact system of integro-differential
evolution equations for the three-particle correlations and in subsection \ref{eq:approxeveq}, 
the single logarithms (SLs) contributions are obtained from the exact
evolution equations;
\item in subsection \ref{subsec:dlasol}, we obtain the DLA solution of the evolution equations and study
the shape and overall normalization of this observable;
\item in subsection \ref{subsec:exactsol} the evolution equations are solved iteratively and the solution
are expressed in terms of the logarithmic derivatives of the one-particle distribution and the two-particle
correlations;
\item in subsection \ref{subsec:MLLAsteepdesc}, we finally give the analytical predictions which will be 
displayed in order to provide predictions for the Tevatron and the LHC;
\item in subsection \ref{subsec:humpapp}, the hump approximation is applied to this observable;
\item in subsection \ref{subsec:smallx}, the region in $x$ where the emission of three correlated particles
becomes dominant is discussed;
\item in subsection \ref{subsec:beyond}, we give the analytical solution of the DLA four-particle correlator 
and show that including higher order corrections for differential higher rank correlators would become a 
cumbersome task;
\item in subsection \ref{subsec:pheno}, the predictions are displayed and the phenomenology is applied to
the Tevatron and the LHC;
\item a conclusion summarizes this work; the appendices are written as complements of 
the main core of the paper.
\end{itemize}

\section{Formalism of the generating functional}
\label{sec:formalism}

A generating functional $Z(E,\Theta;
\{u\})$ can be constructed \cite{Konishi:1979cb} that describes the azimuth averaged
 parton content of a jet of energy $E$ with a given opening half-angle $\Theta$;
by virtue of the exact angular ordering (MLLA),
it satisfies the following integro-differential evolution equation 
\cite{Dokshitzer:1991wu}

\vbox{
\begin{eqnarray}
\frac{d}{d\ln\Theta}Z_A\left(p,\Theta;\{u\}\right)
\!\!&\!\!=\!\!&\!\!\frac{1}{2}\sum_{B,C}\int_{0}^{1}dz\>
\Phi_A^{B[C]}(z)\ \frac{\alpha_s\left(k^2_{\perp}\right)}{\pi} \cr
&&
\Big(Z_B\big(zp,\Theta;\{u\}\big)\ Z_C\big((1-z)p,\Theta;\{u\}\big)
-Z_A\big(p,\Theta;\{u\}\big)\Big); 
\label{eq:red1}
\end{eqnarray}
}
in (\ref{eq:red1}), $z$ and $(1-z)$ are the energy-momentum fractions
carried away by the two
offspring in the $A\to B C$ parton decay described by the standard
one loop splitting functions \cite{Dokshitzer:1978hw}
\begin{eqnarray}
\label{eq:split}
&& \Phi_q^{q[g]}(z) = C_F\, \frac{1+z^2}{1-z} , \quad
\Phi_q^{g[q]}(z)=C_F\, \frac{1+(1-z)^2}{z} , \\
\label{eq:cst}
&& \Phi_g^{q[\bar{q}]}(z) = T_R\left( z^2+(1-z)^2\right) , \quad
\Phi_g^{g[g]}(z)= 2C_A\left(\frac{1-z}{z}+ \frac{z}{1-z} +
z(1-z)\right),\\
\label{eq:cst2}
&& C_A=N_c,\quad C_F=(N_c^2-1)/2N_c,\quad T_R=1/2,
\end{eqnarray}
where $N_c$ is the number of colors;
$Z_A$ in the integral in the r.h.s. of (\ref{eq:red1})
 accounts for 1-loop virtual corrections, which exponentiate into
Sudakov form factors.
$\alpha_s(q^2)$ is the running coupling constant of QCD
\begin{equation}
\alpha_s(q^2)= \frac{4\pi}{4N_c\beta_0
\ln\displaystyle\frac{q^2}{\Lambda^2_{QCD}}},
\label{eq:alphas}
\end{equation}
where $\Lambda_{QCD} \approx$ a few hundred MeV's is the intrinsic scale
of QCD, and
\begin{equation}
\beta_0 = \frac{1}{4N_c}\Big(\frac{11}{3}N_c - \frac{4}{3} n_f T_R\Big)
\label{eq:beta}
\end{equation}
is the first term in the perturbative expansion of the $\beta$ function,
$n_f$ the number of light quark flavors.

If the radiated parton with 4-momentum $k=(k_0,\vec k)$
 is emitted with an angle $\Theta$ with respect to
the direction of the jet, one has ($k_\perp$ is the modulus of the
transverse trivector $\vec k_\perp$ orthogonal to the direction of the jet)
$k_\perp\simeq  |\vec k| \Theta \approx k_0 \Theta \approx 
        z E\Theta$ when $z \ll 1$  or $k_\perp\approx(1-z)E\Theta$ when $z\to 1$,
and a collinear cutoff $k_\perp\geq Q_0$ is imposed. 

In (\ref{eq:red1}) the symbol $\{u\}$ denotes a set of {\em probing
functions} $u_a(k)$ with $k$ the 4-momentum of a secondary parton of
type $a$.
The jet functional is normalized to the total jet production cross
section such that 
\begin{equation}
Z_A(p,\Theta;u\equiv1)=1;
\label{eq:norm}
\end{equation}
for vanishingly small opening
angle it reduces to the probing function of the single initial parton
\begin{equation}
Z_A(p,\Theta\to 0;\{u\})= u_A(k\equiv p).
\end{equation}

To obtain {\em exclusive} $n$-particle distributions one takes $n$
variational derivatives of $Z_A$ over $u(k_i)$ with appropriate
particle momenta, $i=1 \ldots n$, and sets $u\equiv 0$ afterwards;
{\em inclusive} distributions are generated by taking variational
derivatives around $u\equiv 1$. We introduce the n-particle differential
inclusive distribution, also known as parton densities, as \cite{Dokshitzer:1991wu}
\begin{equation}\label{eq:zfunction}
x_1\ldots x_nD_A^{(n)}(x_1,\ldots, x_n,Y)=
E_1\ldots E_n\frac{\delta^n}{\delta u(k_1)\ldots \delta u(k_n)}
Z_A(k_1,\ldots,k_n,\Theta;\left\{u(k)\right\})\bigg|_{u=1}.
\end{equation}
Accordingly, we introduce the following notations for gluon and quark jets $A=G,Q,\bar Q$
\begin{equation}\label{eq:Anotation}
A_{1\ldots n}^{(n)}(z)\equiv\frac{x_1}{z}\ldots\frac{x_n}{z}D_A^{(n)}(\frac{x_1}{z},\ldots,\frac{x_n}{z},Y+\ln z),\;
A_{1\ldots n}^{(n)}\equiv x_1\ldots x_nD_A^{(n)}(x_1,\ldots, x_n,Y),
\end{equation}
which we will use hereafter; $x_i$ corresponds to the Feynman energy fraction of the jet taken 
away by one particle ``$i$".  In the case of three-particle correlations $n=3$, 
the observable to be measured experimentally reads 
$$
{\cal C}^{(3)}_{A_{123}}=\frac{A^{(3)}_{123}}{A_1A_2A_3}.
$$

\subsection{Kinematics and variables}
\label{subsec:kinemat}
The probability of soft gluon radiation off a color charge
(moving in the $z$ direction) has the polar angle dependence 
\begin{equation*}
\frac{\sin\Theta\,d\Theta}{2(1-\cos\Theta)}
= \frac{d\sin(\Theta/2)}{\sin(\Theta/2)}
 \simeq\frac{d\Theta}{\Theta};
\end{equation*}
therefore, we choose the angular evolution parameter to be
\begin{equation}
Y = \ln \frac{2E\sin(\Theta/2)}{Q_0}
\Rightarrow dY = \frac{d\sin(\Theta/2)}{\sin(\Theta/2)};
\label{eq:Ydef}
\end{equation}
note that this choice accounts for finite angles ${\cal O}(1)$ up to
the full opening half-angle $\Theta=\pi$, at which
\begin{equation*}
 Y_{\Theta=\pi}=\ln\frac{2E}{Q_0},
\end{equation*}
where $2E$ is the center-of-mass annihilation energy of the process
$e^+e^- \to q\bar{q}$. 
For small angles (\ref{eq:Ydef}) reduces to
\begin{equation}
Y\simeq \ln\frac{Q}{Q_0}, \quad \Theta\ll 1,
\quad \frac{d}{dY} = \frac{d}{d\ln\Theta},
\label{eq:Ydef2}
\end{equation}
where $Q=E\Theta$, defined as the virtuality of the jet,
is the maximal transverse momentum of a parton inside the jet 
with opening half-angle $\Theta$. Moreover, we make use of 
variables known from previous works \cite{Ramos:2006dx,Ramos:2006mk},
\begin{eqnarray}
&&\ell=\ln\frac{z}{x_i},\quad y=\ln\frac{x_jE\Theta_1}{Q_0},\quad\lambda=\frac{Q_0}{\Lambda_{QCD}},\\
&&\ell_i=\ln\frac{1}{x_i},\quad y_j=\ln\frac{x_jE\Theta_0}{Q_0},\quad 
\eta_{ij}=\ln\frac{x_i}{x_j},\quad Y=\ell_i+y_j+\eta_{ij}. 
\end{eqnarray}
Since $\frac{d}{dy}=\frac{d}{d\ln\Theta_1}$, $y$ could also be used as the evolution-time variable
in forthcoming quark and gluon jet evolution equations.
Accordingly, the anomalous dimension, related to the coupling constant (\ref{eq:alphas}), 
can be parametrized as follows
\begin{equation}\label{eq:parandimen}
\gamma_0^2(q^2)=2N_c\frac{\alpha_s(q^2)}{\pi}\Rightarrow\gamma_0^2(\ell+y)=\frac1{\beta_0(\ell+y+\eta_{ij}+\lambda)},
\end{equation}
such that, 
\begin{itemize}
\item for one particle \cite{Dokshitzer:1991wu}, the denominator in (\ref{eq:parandimen}) is simply
$\ell+y+\lambda$, with \cite{Dokshitzer:1988bq}
$\ell=\ln\frac{z}{x}$, $y=\ln\frac{xE\Theta}{Q_0}$, $\eta=0$;
\smallskip
\item for two-particle correlation \cite{Ramos:2006dx,Ramos:2006mk}, $\ell+y+\eta_{12}$, with $\ell=\ln\frac{z}{x_1}$, 
$y=\ln\frac{x_2E\Theta_1}{Q_0}$, $\eta_{12}=\ln\frac{x_1}{x_2}$;
\smallskip
\item for three-particle correlation, $\ell+y+\eta_{13}$, with $\ell=\ln\frac{z}{x_1}$, 
$y=\ln\frac{x_3E\Theta_1}{Q_0}$, $\eta_{13}=\eta_{12}+\eta_{23}=\ln\frac{x_1}{x_3}$.
\end{itemize}
\subsubsection{Integration bounds for three-particle evolution equations}
\label{sub:bounds}
The production of three hadrons is displayed in Fig.\ref{fig:three-part} after a 
quark or a gluon (A) jet of energy $E$, half opening angle $\Theta_0$ and virtuality $Q=E\Theta_0$
has been produced in a high energy collision. The kinematical variable 
characterizing the process is given by the transverse momentum $k_\perp=zE\Theta_1\geq Q_0$ 
(or $(1-z)E\Theta_1\geq Q_0$) of the first splitting $A\to BC$. The parton C fragments into three offspring such that
three hadrons of energy fractions $x_1$, $x_2$ and $x_3$ can be triggered from the same 
cascade following the condition:
\begin{figure}
\begin{center}
\epsfig{file=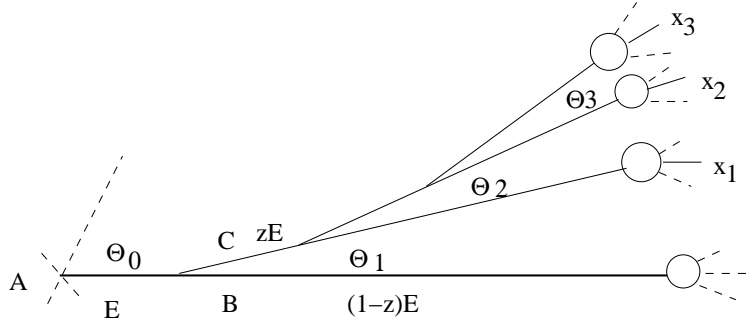,width=10truecm}
\caption{Three-particle yield and angular ordering inside a high energy jet.
\label{fig:three-part}
}
\end{center}
\end{figure}
\begin{equation}\label{eq:3partao}
\Theta_0\geq\Theta_1\geq\Theta_2\geq\Theta_3,
\end{equation}
which arises from the exact AO in MLLA \cite{Dokshitzer:1991wu}. In particular, the condition
$\Theta_0\geq\Theta_1$ is kinematical rather than supported by the AO; it states that every collinear gluon
is emitted inside the jet of half opening angle $\Theta_0$. The two variables entering the 
evolution equations are $z$ and $\Theta_1$, such that 
\begin{equation}
x_1\leq z\leq1\quad\Rightarrow\quad 0\leq\ell\leq\ell_1.
\end{equation}
From (\ref{eq:3partao}) and the initial condition at threshold 
$x_3E\Theta_0\geq x_3E\Theta_1\geq x_3E\Theta_3\geq Q_0$, 
one has
\begin{equation}
\frac{Q_0}{x_3E}\leq\Theta_1\leq\Theta_0\quad\Rightarrow\quad0\leq y\leq y_3.
\end{equation}
\subsection{From single inclusive distribution and two-particle correlation
to three-particle correlation}
\label{subsec:threeparteveq}
The evolution equations satisfied by (\ref{eq:zfunction}) are derived from the MLLA master equation 
for the generating functional $Z_A(u(k_i))$
(\ref{eq:red1}). In this case, one takes the first $\frac{\delta Z_A}{\delta u(k_1)}$, second 
$\frac{\delta^2 Z_A}{\delta u(k_1)\delta u(k_2)}$, and finally third 
$\frac{\delta^3 Z_A}{\delta u(k_1)\ldots\delta u(k_3)}$ 
functional derivatives of $Z_A(u(k_i))$ over the probing functions 
$u(k_i)$ so as to obtain the system of evolution equations for 3-particle correlations.
Following from (\ref{eq:red1}), after applying the operator 
$\frac{\delta^3}{\delta u(k_1)\ldots\delta u(k_3)}$ to both members of the equation,
according to (\ref{eq:zfunction}) and (\ref{eq:Anotation}) together with the initial 
condition (\ref{eq:norm}), it is straightforward to get the coupled system of evolution equations
\begin{subequations}
\begin{eqnarray}
Q^{(3)}_y\!\!&\!\!=\!\!&\!\!\int_{x_1}^1dz\frac{\alpha_s}{\pi}\Phi_q^g(z)
\left[G^{(3)}(z)+\left(Q^{(3)}(1-z)-Q^{(3)}\right)+G_{12}^{(2)}(z)Q_3(1-z)
+G_3(z)Q_{12}^{(2)}\right.\cr
\!\!&\!\!+\!\!&\!\!\left.G_{13}^{(2)}(z)Q_2(1-z)
+G_2(z)Q_{13}^{(2)}+G_{23}^{(2)}(z)Q_1(1-z)+G_1(z)Q_{23}^{(2)}\right]\label{eq:3correlq},\\
G^{(3)}_y\!\!&\!\!=\!\!&\!\!\int_{x_1}^1dz\frac{\alpha_s}{\pi}\Phi_g^g(z)
\left[G^{(3)}(z)-zG^{(3)}+G_{12}^{(2)}(z)G_3(1-z)+G_{13}^{(2)}(z)G_2(1-z)\right.\cr
\!\!&\!\!+\!\!&\!\!\left.G_{23}^{(2)}(z)G_1(1-z)\right]+
\int_{x_1}^1dz\frac{\alpha_s}{\pi}n_f\Phi_g^q(z)\left[\left(2Q^{(3)}(z)-G^{(3)}\right)+
2Q_{12}^{(2)}(z)Q_3(1-z)\right.\cr
\!\!&\!\!+\!\!&\!\!\left.2Q_{13}^{(2)}(z)Q_2(1-z)+2Q_{23}^{(2)}(z)Q_1(1-z)\right]\label{eq:3correlg}.
\end{eqnarray}
\end{subequations}
The l.h.s. of the equations (\ref{eq:3correlq}) and (\ref{eq:3correlg}) can be written in the 
convenient form
\begin{equation}\label{eq:hatAb}
\hat{A}^{(3)}=A^{(3)}-A_1A_2A_3-(A^{(2)}_{12}-A_1A_2)A_3-(A^{(2)}_{13}-A_1A_3)A_2-(A^{(2)}_{23}-A_2A_3)A_1,
\end{equation}
where$A=G,Q,\bar Q$ is the leading parton of the jet. Moreover, we have introduced the notations
$A^{(n)}_{1\ldots n}=A^{(n)}_{1\ldots n}(1)$, where
$$
A^{(n)}_{1\ldots n}\equiv A^{(n)}_{1\ldots n}(1)=x_1\ldots x_nD^{(n)}(x_1,\ldots,x_n,Y),
$$
for the sake of simplicity. The evolution equations for the single inclusive distribution and the two-particle 
correlation are written in \cite{Ramos:2006dx} in the form
\begin{subequations}
\begin{eqnarray}
Q_{y}\!\!&\!\!=\!\!&\!\!\int_{x_1}^1 dz\>  \frac{\alpha_s}{\pi} \>\Phi_q^g(z)\>
\bigg[ \Big(Q(1-z)-Q\Big) + G(z) \bigg],
\label{eq:qpr}\\
G_{y}\!\!&\!\!=\!\!&\!\!\int_{x_1}^1 dz\> \frac{\alpha_s}{\pi} \>
\bigg[\Phi_g^g(z) \Big(G(z)-zG\Big)
+n_f\; \Phi_g^q(z)\, \Big(2Q(z)-G\Big) \bigg],
\label{eq:gpr}
\end{eqnarray}
\end{subequations}
and
\begin{subequations}
\begin{eqnarray}
\label{eq:Q2prsub}
(Q^{(2)}-Q_1Q_2)_{y}  \!\!&\!\!=\!\!& \!\! \int_{x_1}^1 \!dz\>\frac{\alpha_s}{\pi}\> 
 \Phi_q^g(z)\>\bigg[ G^{(2)}(z)+ \Big(Q^{(2)}(1-z)-Q^{(2)}\Big) \cr
&& \hskip -1cm + \Big(G_1(z)-Q_1\Big)\Big(Q_2(1-z) - Q_2\Big) +  \Big(G_2(z)-
Q_2\Big)\Big(Q_1(1-z)- Q_1\Big)\bigg],\\
\label{eq:G2prsub}
(G^{(2)}-G_1G_2)_{y} \!\!&\!\!=\!\!&\!\!
\int_{x_1}^1 dz\> \frac{\alpha_s}{\pi}
\Phi_g^g(z)\>\bigg[ \Big(G^{(2)}(z)-zG^{(2)}\Big) 
  + \Big(G_1(z)-G_1\Big)\Big(G_2(1-z)-G_2\Big) \bigg] \cr
\!\!&\!\!+\!\!&\!\!  \int_{x_1}^1 dz\>\frac{\alpha_s}{\pi}\, n_f\Phi_g^q(z)\>
\bigg[ 2\Big(Q^{(2)}(z)-Q_1(z)Q_2(z)\Big) - \Big(G^{(2)}-G_1G_2\Big)\cr
\!\!&\!\! +\!\!&\!\! \Big(2Q_1(z)-G_1\Big)\Big(2Q_2(1-z)-G_2\Big) \bigg],
\end{eqnarray}
\end{subequations}
respectively. Making use of the equations (\ref{eq:qpr},\ref{eq:gpr}) 
and (\ref{eq:Q2prsub},\ref{eq:G2prsub}),
one can then construct the total derivatives $\left[A_1A_2A_3\right]_y$, 
$\left[(A^{(2)}_{12}-A_1A_2)A_3\right]_y$, $\left[(A^{(2)}_{13}-A_1A_3)A_2\right]_y$,
$\left[(A^{(2)}_{23}-A_2A_3)A_1\right]_y$ as they appear in (\ref{eq:hatAb}), 
which are to be subtracted, term by term from the system of 
equations (\ref{eq:3correlq},\ref{eq:3correlg}). Therefore, we get the equivalent system for the
three-particle correlations inside quark and gluon jets:
\begin{subequations}
\begin{eqnarray}\label{eq:correl3q}
\hat{Q}_y^{(3)}\!\!&\!\!=\!\!&\!\!\int_{x_1}^1dz\frac{\alpha_s}{\pi}\Phi_q^g(z)\left[G^{(3)}(z)+
\left(Q^{(3)}(1-z)-Q^{(3)}\right)\right.\\
\!\!&\!\!+\!\!&\!\!\left.\left(Q^{(2)}_{12}(1-z)-Q^{(2)}_{12}\right)\left(G_3(z)-Q_3\right)
+\left(G^{(2)}_{12}(z)-Q^{(2)}_{12}\right)\left(Q_3(1-z)-Q_3\right)\right.\cr
\!\!&\!\!+\!\!&\!\!\left.\left(Q^{(2)}_{13}(1-z)-Q^{(2)}_{13}\right)\left(G_2(z)-Q_2\right)+
\left(G^{(2)}_{13}(z)-Q^{(2)}_{13}\right)\left(Q_2(1-z)-Q_2\right)\right.\cr
\!\!&\!\!+\!\!&\!\!\left.\left(Q^{(2)}_{23}(1-z)-Q^{(2)}_{23}\right)\left(G_1(z)-Q_1\right)
+\left(G^{(2)}_{23}(z)-Q^{(2)}_{23}\right)\left(Q_1(1-z)-Q_1\right)\right.\cr
\!\!&\!\!+\!\!&\!\!\left.\left(\left(Q_1-G_1(z)\right)\left(Q_2(1-z)-Q_2\right)+
\left(Q_2-G_2(z)\right)\left(Q_1(1-z)-Q_1\right)\right)Q_3\right.\cr
\!\!&\!\!+\!\!&\!\!\left.\left(\left(Q_1-G_1(z)\right)\left(Q_3(1-z)-Q_3\right)+\left(Q_3-G_3(z)\right)
\left(Q_1(1-z)-Q_1\right)\right)Q_2\right.\cr
\!\!&\!\!+\!\!&\!\!\left.\left(\left(Q_2-G_2(z)\right)\left(Q_3(1-z)-Q_3\right)+
\left(Q_3-G_3(z)\right)\left(Q_2(1-z)-Q_2\right)\right)Q_1\right],\notag\\\notag\\
\hat{G}_y^{(3)}\!\!&\!\!=\!\!&\!\!\int_{x_1}^1dz\frac{\alpha_s}{\pi}\Phi_g^g(z)\left[\left(G^{(3)}(z)-zG^{(3)}\right)+
\left(G^{(2)}_{12}(z)-G^{(2)}_{12}\right)\left(G_3(1-z)-G_3\right)\right.\label{eq:correl3g}\\
\!\!&\!\!+\!\!&\!\!\left.\left(G^{(2)}_{13}(z)-G^{(2)}_{13}\right)\left(G_2(1-z)-G_2\right)+
\left(G^{(2)}_{23}(z)-G^{(2)}_{23}\right)\left(G_1(1-z)-G_1\right)\right.\cr
\!\!&\!\!+\!\!&\!\!\left.(G_1-G_1(z))(G_2(1-z)-G_2)G_3+(G_1-G_1(z))(G_3(1-z)-G_3)G_2\right.\cr
\!\!&\!\!+\!\!&\!\!\left.(G_2-G_2(z))(G_3(1-z)-G_3)G_1\right]
+\int_{x_1}^1dz\frac{\alpha_s}{\pi}n_f\Phi_g^q(z)\left[\left(2Q^{(3)}(z)-G^{(3)}\right)\right.\cr
\!\!&\!\!+\!\!&\!\!\left.2\left(Q^{(2)}_{12}(z)-G^{(2)}_{12}\right)(Q_3(1-z)-G_3)+
(2Q_1(z)Q_2(z)-G_1G_2)G_3\right.\cr
\!\!&\!\!+\!\!&\!\!\left.2\left(Q^{(2)}_{13}(z)-G^{(2)}_{13}\right)(Q_2(1-z)-G_2)+
(2Q_1(z)Q_3(z)-G_1G_3)G_2\right.\cr
\!\!&\!\!+\!\!&\!\!\left.2\left(Q^{(2)}_{23}(z)-G^{(2)}_{23}\right)(Q_1(1-z)-G_1)+
(2Q_2(z)Q_3(z)-G_2G_3)G_1\right.\cr
\!\!&\!\!+\!\!&\!\!\left.(G_1-2Q_1(z))(2Q_2(1-z)-G_2)G_3+(G_1-2Q_1(z))(2Q_3(1-z)-G_3)G_2\right.\cr
\!\!&\!\!+\!\!&\!\!\left.(G_2-2Q_2(z))(2Q_3(1-z)-G_3)G_1\right].\notag
\end{eqnarray}
\end{subequations}
The system of evolution equations (\ref{eq:correl3q},\ref{eq:correl3g}), which 
appears as a consequence of the exact AO in intra-jet cascades, provides
the complete theoretical picture of the three-particle correlations as a function of $x_i$
and the characteristic hardness of the jet $Q$; this is the first new result of this paper. However, since these
equations could only be solved numerically, we will extract the SLs contributions 
${\cal O}(\sqrt{\alpha_s})$ in order to provide an approximated analytical solution 
in the following.
\subsection{Approximate evolution equations}
\label{eq:approxeveq}
Let us start with equation (\ref{eq:correl3q}). We proceed to cast all SLs 
contributions corresponding to hard-collinear parton splittings in the shower.
In the hard parton fragmentation region one has $z\sim(1-z)\sim1$, such that the 
second contribution in (\ref{eq:correl3q}) can be approximated through a Taylor 
series for $\ln z\sim\ln(1-z)\ll\ell_1$, written in the appendix \ref{appendix:taylorseries}.
Therefore, one obtains the simplified system of evolution equations
\begin{eqnarray}\label{eq:qy3}
\hat{Q}_y^{(3)}\!\!&\!\!=\!\!&\!\!\int_{x_1}^1\!\!dz\frac{\alpha_s}{\pi}\Phi_q^g(z)G^{(3)}(z),\\
\hat{G}_y^{(3)}\!\!&\!\!=\!\!&\!\!\int_{x_1}^1\!\!dz\frac{\alpha_s}{\pi}(1-z)\Phi_g^g(z)G^{(3)}(z)
+\int_{x_1}^1dz\frac{\alpha_s}{\pi}n_f\Phi_g^q(z)\left[\left(2Q^{(3)}-G^{(3)}\right)+
2\left(Q^{(2)}_{12}-G^{(2)}_{12}\right)\right.\label{eq:gy3}\\
\!\!&\!\!\times\!\!&\!\!\left.(Q_3-G_3)+(2Q_1Q_2-G_1G_2)G_3+2\left(Q^{(2)}_{13}-G^{(2)}_{13}\right)(Q_2-G_2)+
(2Q_1Q_3-G_1G_3)G_2\right.\cr
\!\!&\!\!+\!\!&\!\!\left.2\left(Q^{(2)}_{23}-G^{(2)}_{23}\right)(Q_1-G_1)+
(2Q_2Q_3-G_2G_3)G_1+(G_1-2Q_1)(2Q_2-G_2)G_3\right.\cr
\!\!&\!\!+\!\!&\!\!\left.
(G_1-2Q_1)(2Q_3-G_3)G_2+(G_2-2Q_2)(2Q_3-G_3)G_1\right],\notag
\end{eqnarray}
where we have kept all terms of order ${\cal O}(\sqrt{\alpha_s})$, which contribute 
to MLLA. In addition, from the DLA relation $Z_A=Z_G^{C_A/N_c}$ \cite{Dokshitzer:1982xr}, 
and Eqs.(\ref{eq:zfunction}-\ref{eq:Anotation}), one has
the useful expressions for the single inclusive distribution, two- and three-particle correlations:
\begin{eqnarray}\label{eq:12dist}
Q_i\!\!&\!\!=\!\!&\!\!\frac{C_F}{N_c}G_i,\quad Q^{(2)}_{ij}=\frac{C_F}{N_c}G^{(2)}_{ij}+
\frac{C_F}{N_c}\left(\frac{C_F}{N_c}-1\right)G_iG_j,\quad i\ne j,\\
Q^{(3)}\!\!&\!\!=\!\!&\!\!\frac{C_F}{N_c}G^{(3)}+\frac{C_F}{N_c}\!\!\left(\frac{C_F}{N_c}
-1\right)\!\!\left(G^{(2)}_{12}G_3+G^{(2)}_{13}G_2+G^{(2)}_{23}G_1\right)+\frac{C_F}{N_c}
\!\!\left(\frac{C_F}{N_c}-1\right)\!\!\left(\frac{C_F}{N_c}-2\right)\cr
\!\!&\!\!\times\!\!&\!\!G_1G_2G_3,\label{eq:3dist}
\end{eqnarray}
which in turn can be replaced in (\ref{eq:gy3}). The two expressions
written in (\ref{eq:12dist}) are known from previous works at DLA \cite{Dokshitzer:1982ia,Dokshitzer:1982xr}, 
while (\ref{eq:3dist}) will be used for the first time in this context. After integrating over the regular 
part of the splitting functions (\ref{eq:split}), (\ref{eq:cst}) and (\ref{eq:cst2}), one obtains the
integro-differential system of equations ($\eta_{13}=\eta_{12}+\eta_{23}$),
\begin{eqnarray}\label{eq:mlla3correlq}
\hat{Q}_y^{(3)}\!\!&\!\!=\!\!&\!\!\frac{C_F}{N_c}\int_{0}^{\ell_1}\!\!d\ell\gamma_0^2(\ell+y_3)G^{(3)}(\ell,y_3;\eta_{13})
-\frac34\frac{C_F}{N_c}\gamma_0^2(\ell_1+y_3)G^{(3)}(\ell_1,y_3;\eta_{13}),\\
\hat{G}_y^{(3)}\!\!&\!\!=\!\!&\!\!\int_{0}^{\ell_1}\!\!d\ell\gamma_0^2(\ell+y_3)G^{(3)}(\ell,y_3;\eta_{13})
-a\gamma_0^2(\ell_1+y_3)G^{(3)}(\ell_1,y_3;\eta_{13})+(a-b)\gamma_0^2(\ell_1+y_3)\label{eq:mlla3correlg}\\
\!\!&\!\!\times\!\!&\!\!\left[\left(G^{(2)}_{12}(\ell_1,y_3+\eta_{23};\eta_{12})-G_1(\ell_1,y_3+\eta_{13})
G_2(\ell_1+\eta_{12},y_3+\eta_{23})\right)G_3(\ell_1+\eta_{13},y_3)\right.\cr
\!\!&\!\!+\!\!&\!\!\left.\left(G^{(2)}_{13}(\ell_1,y_3;\eta_{13})
-G_1(\ell_1,y_3+\eta_{13})G_3(\ell_1+\eta_{13},y_3)\right)G_2(\ell_1+\eta_{12},y_3+\eta_{23})\right.\cr
\!\!&\!\!+\!\!&\!\!\left.\left(G^{(2)}_{23}(\ell_1+\eta_{12},y_3;\eta_{23})-G_2(\ell_1+\eta_{12},y_3+\eta_{23})
G_3(\ell_1+\eta_{13},y_3)\right)G_1(\ell_1,y_3+\eta_{13})\right]\cr
\!\!&\!\!+\!\!&\!\!(a-c)\gamma_0^2(\ell_1+y_3)G_1(\ell_1,y_3+\eta_{13})G_2(\ell_1+\eta_{12},y_3+\eta_{23})
G_3(\ell_1+\eta_{13},y_3),\notag
\end{eqnarray}
%or written in the following form, convenient for numerical analysis
%\begin{eqnarray}
%\hat{Q}_y^{(3)}\!\!&\!\!=\!\!&\!\!\frac{C_F}{N_c}\int_{0}^{\ell_1}d\ell\gamma_0^2(\ell+y_3)G^{(3)}(\ell,y_3;\eta_{13})
%-\frac34\frac{C_F}{N_c}\gamma_0^2(\ell_1+y_3)G^{(3)}(\ell_1,y_3;\eta_{13})\\
%\hat{G}_y^{(3)}\!\!&\!\!=\!\!&\!\!\int_{0}^{\ell_1}d\ell\gamma_0^2(\ell+y_3)G^{(3)}(\ell,y_3;\eta_{13})
%-a\gamma_0^2(\ell_1+y_3)G^{(3)}(\ell_1,y_3;\eta_{13})+(a-b)\gamma_0^2(\ell_1+y_3)\\
%\!\!&\!\!\times\!\!&\!\!\left[G^{(2)}_{12}(\ell_1,y_3+\eta_{23};\eta_{12})G_3(\ell_1+\eta_{13},y_3))+
%G^{(2)}_{13}(\ell_1,y_3;\eta_{13})G_2(\ell_1+\eta_{12},y_3+\eta_{23})\right.\cr
%\!\!&\!\!+\!\!&\!\!\left.G^{(2)}_{23}(\ell_1+\eta_{12},y_3;\eta_{23})G_1(\ell_1,y_3+\eta_{13})\right]
%-(2a-3b+c)\gamma_0^2(\ell_1+y_3)G_1(\ell_1,y_3+\eta_{13})\cr
%\!\!&\!\!\times\!\!&\!\!G_2(\ell_1+\eta_{12},y_3+\eta_{23})
%G_3(\ell_1+\eta_{13},y_3),\notag
%\end{eqnarray}
%where
with the following hard constants,
\begin{eqnarray}
a(n_f)\!\!&\!\!=\!\!&\!\!\frac{1}{4N_c}\left[\frac{11}3N_c+\frac43n_fT_R\left(1-2\frac{C_F}{N_c}\right)\right]
\stackrel{n_f=3}{=}0.935,\\
b(n_f)\!\!&\!\!=\!\!&\!\!\frac{1}{4N_c}\left[\frac{11}3N_c-\frac43n_fT_R\left(1-2\frac{C_F}{N_c}\right)^2\right]
\stackrel{n_f=3}{=}0.915,\\
c(n_f)\!\!&\!\!=\!\!&\!\!\frac{1}{4N_c}\left[\frac{11}3N_c+\frac43n_fT_R\left(1-2\frac{C_F}{N_c}\right)^3\right]
\stackrel{n_f=3}{=}0.917,
\end{eqnarray}
where $n_f=3$ corresponds to the number of light active flavors of quarks $u,d,s$.
As an example of such procedure, one could write the example,
$$
a(n_f)=\int_0^1dz\left[(1-z)\Big(2-z(1-z)\Big)+\frac{n_fT_R}{2C_A}\Big(z^2+(1-z)^2\Big)\left(1-2\frac{C_F}{N_c}\right)\right].
$$
The first integral terms of the equations in 
(\ref{eq:mlla3correlq}) and (\ref{eq:mlla3correlg}) are of classical origin and therefore, universal.
Corrections $\propto-\frac34$, $a$, $(a-b)$ and $(a-c)$, which are ${\cal O}(\sqrt{\alpha_s})$ suppressed, 
better account for energy conservation at each vertex of the splitting process, as compared
with the DLA. Notice that the form of the quark initiated jet 
equation (\ref{eq:mlla3correlq}) is universal at MLLA (see (\ref{eq:solq}) and (\ref{eq:eveeqq} 
in the appendix \ref{subsec:oneandtwopartdist} for the single inclusive distribution and 
two-particle correlation respectively), that is, invariant with respect to the number of 
particles considered in the cascade. In the equation for the gluon 
initiated jet (\ref{eq:mlla3correlg}), the first and second constants $a(n_f)$ and $b(n_f)$ 
were obtained in the frame of the single inclusive 
distribution and two-particle correlations respectively \cite{Fong:1990ph,Fong:1990nt}. 
The third constant $c(n_f)$ appears in this paper for the 
first time for the three-particle correlation. In particular, notice that a certain recurrency shows up 
in the coefficients combining the colour factors $(-1)^{n-1}\left(1-2\frac{C_F}{N_c}\right)^n$, as a function of 
the number $n$ of particles considered in the shower.

\subsection{DLA solution of the evolution equations}
\label{subsec:dlasol}
\begin{figure}
\begin{center}
\epsfig{file=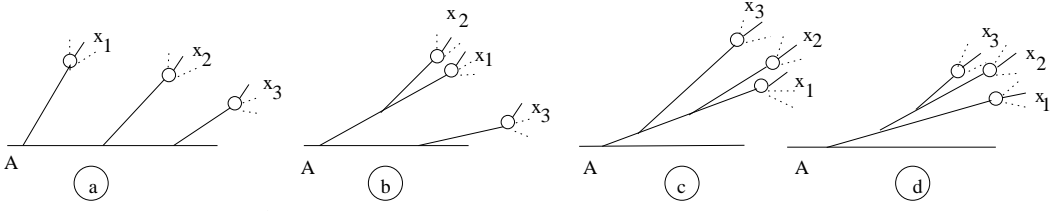,width=14truecm}
\caption{Three particles emitted inside the shower with color factors for the square of the amplitudes: 
$C_A^3$, $C_A^2N_c$, $C_AN_c^2$ and $C_AN_c^2$ for a, b, c and d respectively.
\label{fig:physpic3}
}
\end{center}
\end{figure}

In this subsection we compute the leading order DLA contributions 
in order to provide general features concerning the
the shape and overall normalization of three-particle correlations. This
procedure is equivalent to cast the leading order (LO) solution of the equations 
(\ref{eq:mlla3correlq},\ref{eq:mlla3correlg}).
We differentiate (\ref{eq:mlla3correlq}) 
and (\ref{eq:mlla3correlg}) with respect to ``$_\ell$", such that after setting  hard corrections 
$\propto3/4, a, b, c=0$, the MLLA evolution equations are reduced to the new
DLA compact differential equation
\begin{equation}\label{eq:A3DLA}
\left[\tilde A^{(3)}\right]_{\ell y}=\frac{C_A}{N_c}\gamma_0^2G^{(3)},
\end{equation}
with
\begin{equation}\label{eq:hatA3}
\left[\hat A^{(3)}\right]_{\ell y}=\left\{\left[\left({\cal C}_{A_{123}}^{(3)}-1\right)-
\left({\cal C}_{A_{12}}^{(2)}-1\right)-\left({\cal C}_{A_{13}}^{(2)}-1\right)
-\left({\cal C}_{A_{23}}^{(2)}-1\right)
\right]A_1A_2A_3\right\}_{\ell y},
\end{equation}
after having set $A^{(3)}={\cal C}_{A_{123}}^{(3)}A_1A_2A_3$ for the three-particle correlator and 
$A_{ij}^{(2)}={\cal C}_{A_{ij}}^{(2)}A_iA_j$ for the two-particle correlator. 
We fix the anomalous dimension to the characteristic hardness of the jet $Q\approx E\Theta_0$
($\gamma_0^2(E\Theta_0)=const$) and solve this equation iteratively by derivating the r.h.s. 
of (\ref{eq:hatA3}) with respect to $\ell$ and $y$, such that the solution of (\ref{eq:A3DLA}) reads
\begin{eqnarray}\label{eq:dla3partsol}
\left(\dot{{\cal C}}_{A_{123}}^{(3)}-1\right)\!\!&\!\!-\!\!&\!\!\left(\dot{{\cal C}}_{A_{12}}^{(2)}-1\right)
-\left(\dot{{\cal C}}_{A_{13}}^{(2)}-1\right)-\left(\dot{{\cal C}}_{A_{23}}^{(2)}-1\right)\\
\!\!&\!\!=\!\!&\!\!\frac{N_c}{C_A}\frac{\left(\dot{{\cal C}}_{A_{12}}^{(2)}-1\right)+\left(\dot{{\cal C}}_{A_{13}}^{(2)}-1\right)
+\left(\dot{{\cal C}}_{A_{23}}^{(2)}-1\right)}{2+\tilde\Delta_{12}+\tilde\Delta_{13}+\tilde\Delta_{23}}
+\frac{N_c^2}{C_A^2}\frac1{2+\tilde\Delta_{12}
+\tilde\Delta_{13}+\tilde\Delta_{23}},\notag
\end{eqnarray}
which have been written in terms of the logarithmic derivatives of the one-particle spectrum,
\begin{equation}\label{eq:logders}
\tilde\Delta_{ij}=\gamma_0^{-2}\left(\psi_{A_i,\ell}\psi_{A_j,y}+\psi_{A_i,y}\psi_{A_j,\ell}\right),\quad 
\psi_{A_i,\ell}=\frac{1}{A_i}\frac{\partial A_i}{\partial\ell},\;\psi_{A_i,y}
=\frac{1}{A_i}\frac{\partial A_i}{\partial y}
\end{equation}
and the DLA two-particle correlator \cite{Dokshitzer:1991wu,Dokshitzer:1982ia} (for a review see also 
\cite{Khoze:1996dn})
\begin{equation}\label{eq:2partcorrdla}
\dot{{\cal C}}_{A_{ij}}^{(2)}-1=\frac{N_c}{C_A}\frac{1}{1+\Delta_{ij}}.
\end{equation}
The dot over ${\cal C}^{(n)}$ differentiates the DLA correlators from the MLLA correlators obtained below.
In DLA however, since the single inclusive distribution 
satisfies $Q=\frac{C_F}{N_c}G$ \cite{Dokshitzer:1982xr}, one has
$$
\psi_{Q_i,\ell}=\psi_{G_i,\ell}\equiv\psi_{i,\ell},\quad \psi_{Q_i,y}=\psi_{G_i,y}\equiv\psi_{i,y}.
$$
That is why, we will use the much simplest notation 
$\psi_{G_i,\ell}=\psi_{i,\ell}$, $\psi_{G_i,y}=\psi_{i,y}$. It is worth
giving the order of magnitude of some quantities that will be considered in forthcoming calculations. 
In DLA, the one-particle inclusive distribution can be written as $A_i(\ell,y)\propto\exp\left(2\gamma_0\sqrt{\ell y}\right)$
asymptotically for fixed running coupling $\gamma_0=const$ \cite{Dokshitzer:1982xr}. Though the solution with fixed coupling constant
provides general features of the single inclusive distribution, it is not enough for the description of a more
realistic picture at colliders. However, from its simplicity, it can be used to give the order of magnitude of terms
involved in the solution of the DLA and MLLA evolution equations. 
Therefore, making use of (\ref{eq:logders}), one has
\begin{eqnarray}\label{eq:ordermag1}
\psi_{A_i,\ell}\!\!&\!\!=\!\!&\!\!{\cal O}(\gamma_0),\;
\psi_{A_i,y}={\cal O}(\gamma_0),\; \psi_{A_i,\ell\ell}={\cal O}(\gamma_0^2),\;  
\psi_{A_i,\ell y}={\cal O}(\gamma_0^2),\;\psi_{A_i,yy}={\cal O}(\gamma_0^2),\\
\tilde\Delta_{ij}\!\!&\!\!=\!\!&\!\!{\cal O}(1),\;\tilde\Delta_{ij,\ell}={\cal O}(\gamma_0^2),\;
\tilde\Delta_{ij,y}={\cal O}(\gamma_0^2),\label{eq:ordermag2}
\end{eqnarray}
where $\psi_{A_i,\ell\ell}$, $\psi_{A_i,\ell y}$ and $\psi_{A_i,yy}$ are double derivatives of 
$\psi_{A_i}=\ln A_i(\ell,y)$. The DLA solution (\ref{eq:dla3partsol}) describes the
following picture:  the first term $(=-1)$ in the l.h.s.  translates the independent or decorrelated 
emission of three hadrons in the shower like depicted by Fig.\ref{fig:physpic3}a. After inserting
the two-particle correlator (\ref{eq:2partcorrdla}) in the l.h.s. of (\ref{eq:dla3partsol}),
terms $\propto\frac{N_c}{C_A}$ correpond to the case where two partons 
are correlated inside the same subjet, while the other one is emitted independently from the rest
like in Fig.\ref{fig:physpic3}b. Next, replacing (\ref{eq:2partcorrdla}) in the r.h.s. of
(\ref{eq:dla3partsol}) one obtains a contribution $\propto\frac{N_c^2}{C_A^2}$ described by 
Fig.\ref{fig:physpic3}c, where two partons are emitted independently inside the same subjet.
The last term $\propto\frac{N_c^2}{C_A^2}$ depicted by Fig.\ref{fig:physpic3}d, 
involves three particles strongly correlated inside the same partonic 
shower and corresponds to the cumulant of genuine correlations. 
Actually, this interpretation has been given after computing the color factors of such Feynman diagrams 
describing the process, normalized by $C_A^3$ in the end. Notice that diagrams displayed in
Fig.\ref{fig:physpic3}c and Fig.\ref{fig:physpic3}d present the same color factors but different 
Lorentz structure. In both cases, the DLA strong AO $\Theta\gg\Theta'\gg\Theta''$ and strong energy ordering
$x_1\gg x_2\gg x_3$ are necessary conditions satisfied by (\ref{eq:A3DLA}) \cite{Fadin:1983dl}.

Performing the steepest descent evaluation of the DLA single inclusive distribution 
from an integral representation, which was written in Mellin space in the form 
\cite{Dokshitzer:1982xr,Dokshitzer:1982ia},
\begin{equation}\label{eq:dlaintegrep}
G(\ell,y)=(\ell+y+\lambda)\!\!\iint\frac{d\omega d\nu}{(2\pi i)^2}e^{\omega\ell+\nu y}\int_0^{\infty}
\frac{ds}{\nu+s}\left(\frac{\omega(\nu+s)}{(\omega+s)\nu}\right)^{1/\beta_0(\omega-\nu)}e^{-\lambda s},
\;Q=\frac{C_F}{N_c}G.
\end{equation}
and which accounts for the running of the coupling $\alpha_s$, the energy of most particles inside the jet was proved to be 
close to the maximum of the distribution, which shapes like a Gaussian in this 
region \cite{Dokshitzer:1982xr},
\begin{equation}
A_i(\ell_i, Y)\simeq\exp\left[-\frac{3}{\sqrt{\beta_0}}\frac{(\ell_i-\ell_{max})^2}{Y^{3/2}}\right],\quad
\ell_{max}\approx \frac{Y}2.
\end{equation}
From this method \cite{Dokshitzer:1982ia}, the expressions 
of the logarithmic derivative of the one particle distribution were written as,
\begin{equation}
\label{eq:dlalogder}
\psi_{i,\ell}(\mu_i,\nu_i)=\gamma_0e^{\mu_i},\quad\psi_{i,y}(\mu,\nu)=\gamma_0e^{-\mu_i}.
\end{equation}
such that $\Delta_{ij}$ and the correlator were 
given in the form \cite{Dokshitzer:1982ia}, 
\begin{equation}\label{eq:simpledla}
\Delta_{ij}=2\cosh(\mu_i-\nu_j),\qquad  
\dot{{\cal C}}_{A_{ij}}^{(2)}=1+\frac{N_c}{C_A}\frac1{1+2\cosh(\mu_i-\mu_j)}
\end{equation}
respectively, where $(\mu_i,\nu_i)$ were related to $(\ell_i,y_i)$ through the 
2x2 non-linear system of equations \cite{Dokshitzer:1982ia},
\begin{equation}\label{eq:systemunubis}
\frac{y_i-\ell_i}{y_i+\ell_i}=\frac{(\sinh2\mu_i-2\mu_i)-(\sinh2\nu_i-2\nu_i)}{2(\sinh^2\mu_i-\sinh^2\nu_i)},\quad
\frac{\sinh\nu_i}{\sqrt{\lambda}}=\frac{\sinh\mu_i}{\sqrt{\ell_i+y_i+\lambda}}.
\end{equation} 
Therefore, the DLA three-particle correlator reads in this approximation
\begin{eqnarray}\label{eq:dlasolmunuA}
\dot{{\cal C}}_{A_{123}}^{(3)}\!\!&\!\!=\!\!&\!\!1+\left(\dot{{\cal C}}_{A_{12}}^{(2)}-1\right)
+\left(\dot{{\cal C}}_{A_{13}}^{(2)}-1\right)+\left(\dot{{\cal C}}_{A_{23}}^{(2)}-1\right)\\
\!\!&\!\!+\!\!&\!\!\frac{N_c}{2C_A}\frac{\left(\dot{{\cal C}}_{A_{12}}^{(2)}-1\right)
+\left(\dot{{\cal C}}_{A_{13}}^{(2)}-1\right)
+\left(\dot{{\cal C}}_{A_{23}}^{(2)}-1\right)}{1+\cosh(\mu_1-\mu_2)+\cosh(\mu_1-\mu_3)+\cosh(\mu_2-\mu_3)}\cr
\!\!&\!\!+\!\!&\!\!\frac{N_c^2}{2C_A^2}\frac1{1+\cosh(\mu_1-\mu_2)+\cosh(\mu_1-\mu_3)+\cosh(\mu_2-\mu_3)}.
\notag
\end{eqnarray}
with $\dot{{\cal C}}_{A_{ij}}^{(2)}$ extracted from (\ref{eq:simpledla}). Taking 
$\mid\ell_i-\ell_{max}\mid\ll\sigma\propto Y^{3/2}$ for $i=1,2,3$, one has in this approximation 
(see appendix \ref{subappendix:humpapprox})
\begin{equation}
\Delta_{ij}\approx 2+9\left(\frac{\ell_i-\ell_j}{Y}\right)^2=2+9\left[\frac{\ln(x_j/x_i)}{\ln(Q/Q_0)}\right]^2,
\end{equation}
so that,
\begin{eqnarray}
\Delta_{12}+\Delta_{13}+\Delta_{23}\!\!&\!\!\approx\!\!&\!\! 6+9\left(\frac{\ell_1-\ell_2}{Y}\right)^2
+9\left(\frac{\ell_1-\ell_3}{Y}\right)^2+9\left(\frac{\ell_2-\ell_3}{Y}\right)^2\cr
\!\!&\!\!=\!\!&\!\!6+9\left[\frac{\ln(x_2/x_1)}{\ln(Q/Q_0)}\right]^2+9\left[\frac{\ln(x_3/x_1)}{\ln(Q/Q_0)}\right]^2
+9\left[\frac{\ln(x_3/x_2)}{\ln(Q/Q_0)}\right]^2.
\end{eqnarray}
Therefore, the shape of the three-particle correlator can be expected to be quadratic as a function of the difference
$(\ell_i-\ell_j)$, as for the two-particle correlator. Thus, the correlator is strongest when particles 
have the same energy $x_1=x_2=x_3$. 

Moreover, the decreasing behavior of the correlator as one parton gets much harder than the 
others $x_i\gg x_j$ shows that QCD coherence effects dominate this region of the phase space as 
interferences between such gluons occur. New kinds of contributions like the one in the first term of the 
r.h.s. of (\ref{eq:dla3partsol}) appear in this context. 
%Such is the case of terms of the type 
%$(\ell_i-\ell_j)^4$, which being suppressed as compared to $(\ell_i-\ell_j)^2$ in the above mentioned 
%region $\mid\ell_i-\ell_{max}\mid\ll\sigma\propto Y^{3/2}$, will in fact provide a more flattened aspect of 
%the three-particle correlator for $x_i\gg x_j$, as compared to the two-particle correlator.

The overall normalization of the $n$-particle correlator is fixed by that of the same rank multiplicity-correlator 
determining the multiplicity fluctuations inside the jet \cite{Dokshitzer:1982ia},
$$
{{\cal C}_{A}^{(k)}}(x_1,\ldots,x_k)\leq\frac{\left<n(n-1)\ldots(n-k+1)\right>}{\left<n\right>^k}.
$$
Then, one has
\begin{equation}
{{\cal C}_{A}^{(2)}}(x_1,x_2)-1\leq\frac{N_c}{3C_A},
\quad {{\cal C}_{A}^{(3)}}(x_1,x_2,x_3)-1\leq\frac{N_c}{C_A}+\frac{N_c^2}{4C_A^2}.
\end{equation}
These bounds can also be obtained by setting $\Delta(x_i,x_j)=2$ (for $x_i=x_j$) in 
(\ref{eq:2partcorrdla}) and (\ref{eq:dla3partsol}) respectively.  Since DLA neglects the 
energy balance, it is not realistic and does not provide the real physical picture of any
jet process in the frame of jet calculus. 

\subsection{Iterative solution of the evolution equations}
\label{subsec:exactsol}
As we can see, the computation of three-particle correlations requires a mastering knowledge of the one-particle
inclusive energy distribution and two-particle correlations. 
The behavior of the two-particle 
correlators as shown by these solutions was proved to be quadratic as a function of 
$(\ell_i-\ell_j)$ and increasing as a function of $(\ell_i+\ell_j)$ like in the Fong-Webber
approximation \cite{Fong:1990ph,Fong:1990nt}. However, the solutions (\ref{eq:CGfull},\ref{eq:CQfull})
(see appendix \ref{subsec:oneandtwopartdist}) were 
shown to better account for soft gluon coherence effects, by describing the flatting of the slopes 
as ($\ell_i+\ell_j$) increases. In \cite{Ramos:2006mk}, the solution was obtained by 
the steepest descent evaluation of the spectrum $G_i(\ell,y)$, while in \cite{Ramos:2006dx}, the 
evaluation was performed by taking the expression of $G_i(\ell,y)$ given by (\ref{eq:ifD})
in the appendix \ref{subsec:oneandtwopartdist}.
In \cite{Ramos:2006dx}, the solution of the
evolution equations for two-particle correlation were obtained from the differential version of the 
equations (\ref{eq:diffcorrel2q},\ref{eq:diffcorrel2g}) 
over $\ell$ and $y$ written in the appendix \ref{subsec:oneandtwopartdist}. 
Therefore, in this subsection, we will make some transformations in order to simplify this 
cumbersome task without adding further information. In the appendix \ref{subsec:oneandtwopartdist}, 
we briefly summarize what should be known in order
to complete the solution of the evolution equations for the three-particle correlations.

Differentiating (\ref{eq:mlla3correlq}) and (\ref{eq:mlla3correlg}) with respect to $``\ell"$, 
one has the differential system of evolution equations for three-particle correlations,
\begin{eqnarray}\label{eq:mlla3correlqdiff}
\hat{Q}_{\ell y}^{(3)}\!\!&\!\!=\!\!&\!\!\frac{C_F}{N_c}\gamma_0^2G^{(3)}
-\frac34\frac{C_F}{N_c}\gamma_0^2\left(G^{(3)}_\ell-\beta_0\gamma_0^2G^{(3)}\right),\\
\hat{G}_{\ell y}^{(3)}\!\!&\!\!=\!\!&\!\!\gamma_0^2G^{(3)}
\!-\!a\gamma_0^2\left(G^{(3)}_\ell\!-\!\beta_0\gamma_0^2G^{(3)}\right)\!+\!(a-b)\gamma_0^2\left\{
\left[\left(G^{(2)}_{12}-G_1G_2\right)G_3\right.\right.\label{eq:mlla3correlgdiff}\\
\!\!&\!\!+\!\!&\!\!\left.\left.\left(G^{(2)}_{13}-G_1G_3\right)G_2+\left(G^{(2)}_{23}-G_2G_3\right)G_1\right]_\ell
\!-\!\beta_0\gamma_0^2\left[\left(G^{(2)}_{12}-G_1G_2\right)G_3\right.\right.\cr
\!\!&\!\!+\!\!&\!\!\left.\left.\left(G^{(2)}_{13}-G_1G_3\right)G_2+\left(G^{(2)}_{23}-G_2G_3\right)G_1\right]\right\}
\!+\!(a\!-\!c)\gamma_0^2\left[(G_1G_2G_3)_\ell\!-\!\beta_0\gamma_0^2G_1G_2G_3\right],\notag
\end{eqnarray}
which is written in this paper for the first time.
The equation (\ref{eq:mlla3correlgdiff}) is self-contained and can be solved iteratively like (\ref{eq:A3DLA}).
For this purpose, one sets $G^{(3)}=C_{G_{123}}^{(3)}G_1G_2G_3$ and $G_{ij}^{(2)}=C_{G_{ij}}^{(2)}G_iG_j$ in the left and right 
hand sides of (\ref{eq:mlla3correlgdiff}), such that the solution obtained in the appendix \ref{appendix:iteratsolgq}
can be written in the compact form
\begin{eqnarray}\label{eq:sol3partg}
{\cal C}_{G_{123}}^{(3)}-1=
\left({\cal C}_{G_{12}}^{(2)}-1\right)\!\!F_{12}^{(2)}+\left({\cal C}_{G_{13}}^{(2)}-1\right)\!\!F_{13}^{(2)}
+\left({\cal C}_{G_{23}}^{(2)}-1\right)\!\!F_{23}^{(2)}+F_{123}^{(3)},
\end{eqnarray}
where,
\begin{equation}\label{eq:sol3partgbis}
F_{ij}^{(2)}=1+\frac{N_{G_{ij}}^{(2)}}{D_{G}^{(2)}},\quad
F_{123}^{(3)}=\frac{N_{G}^{(3)}}{D_{G}^{(3)}},
\end{equation}
with
\begin{subequations}
\begin{eqnarray}
\label{eq:NG2}
N_{G_{ij}}^{(2)}\!\!&\!\!=\!\!&\!\!1
-b\left(\psi_{1,\ell}+\psi_{2,\ell}+\psi_{3,\ell}-\beta_0\gamma_0^2\right)-a\zeta_\ell+(a-b)\chi_\ell^{ij}+\xi_1^{ij}+\delta_2^{ij}-\epsilon_1-\epsilon_2,\\
D_{G}^{(2)}\!\!&\!\!=\!\!&\!\!2+\Delta_{12}+\Delta_{13}+\Delta_{23}+a\zeta_\ell+2a\beta_0\gamma_0^2+\epsilon_1+\epsilon_2,\\
N_{G}^{(3)}\!\!&\!\!=\!\!&\!\!1
-c\left(\psi_{1,\ell}+\psi_{2,\ell}+\psi_{3,\ell}-\beta_0\gamma_0^2\right)-a\zeta_\ell+(a-b)(\chi_\ell^{12}+\chi_\ell^{13}+\chi_\ell^{23})+(\xi_1^{12}+\delta_2^{12})\\
\!\!&\!\!+\!\!&\!\!(\xi_1^{13}+\delta_2^{13})+(\xi_1^{23}+\delta_2^{23})-\epsilon_1-\epsilon_2,\cr
D_{G}^{(3)}\!\!&\!\!=\!\!&\!\!D_{G}^{(2)}=2+\Delta_{12}+\Delta_{13}+\Delta_{23}+a\zeta_\ell+2a\beta_0\gamma_0^2+\epsilon_1+\epsilon_2.
\label{eq:DG3}
\end{eqnarray}
\end{subequations}
The solution (\ref{eq:sol3partg}) can be checked to recover the DLA result (\ref{eq:dla3partsol}) 
inside a gluon jet, that is for $C_A=N_c$. Since DLA neglects recoil effects at each splitting inside
the cascade, one should expect the DLA three-particle correlation to be much larger than MLLA predictions
and therefore to overestimate the data. 
We introduce the following notations and give the order of magnitude of each contribution following 
from (\ref{eq:ordermag1}) and (\ref{eq:ordermag2}),
\begin{subequations}
\begin{eqnarray}
\label{eq:zetaell}
\zeta\!\!&\!\!=\!\!&\!\!\ln \dot{{\cal C}}_{G_{123}}^{(3)},\quad 
\zeta_\ell=\frac{\dot{{\cal C}}_{G_{123},\ell}^{(3)}}{\dot{{\cal C}}_{G_{123}}^{(3)}}={\cal O}(\gamma_0^2),\quad
\zeta_y=\frac{\dot{{\cal C}}_{G_{123},y}^{(3)}}{\dot{{\cal C}}_{G_{123}}^{(3)}}={\cal O}(\gamma_0^2),\;\\ 
\chi_\ell^{ij}\!\!&\!\!=\!\!&\!\!\frac{\dot{{\cal C}}_{G_{ij},\ell}^{(2)}}{\dot{{\cal C}}_{G_{ij}}^{(2)}}
={\cal O}(\gamma_0^2),\quad
\chi_y^{ij}=\frac{\dot{{\cal C}}_{G_{ij},y}^{(2)}}{\dot{{\cal C}}_{G_{ij}}^{(2)}}={\cal O}(\gamma_0^2),\\
\xi_1^{ij}\!\!&\!\!=\!\!&\!\!\frac{1}{\gamma_0^2}\left[\chi_\ell^{ij}(\psi_{1,y}+\psi_{2,y}+\psi_{3,y})+
\chi_y^{ij}(\psi_{1,\ell}+\psi_{2,\ell}+\psi_{3,\ell})\right]={\cal O}(\gamma_0),\label{eq:xiell}\\
\delta_2^{ij}\!\!&\!\!=\!\!&\!\!\frac{1}{\gamma_0^2}\left(\chi_\ell^{ij}\chi_y^{ij}
+\chi_{\ell,y}^{ij}\right)={\cal O}(\gamma_0^2),\\
\epsilon_1\!\!&\!\!=\!\!&\!\!\frac{1}{\gamma_0^2}\left[\zeta_\ell(\psi_{1,y}+\psi_{2,y}+\psi_{3,y})+
\zeta_y(\psi_{1,\ell}+\psi_{2,\ell}+\psi_{3,\ell})\right]={\cal O}(\gamma_0),\label{eq:epsilon1}\\
\epsilon_2\!\!&\!\!=\!\!&\!\!\frac{1}{\gamma_0^2}\left(\zeta_\ell\zeta_y+\zeta_{\ell,y}\right)={\cal O}(\gamma_0^2).
\label{eq:epsilon2}
\end{eqnarray}
\end{subequations}
The solution of the gluon evolution equation for the correlator can be either obtained numerically by solving 
(\ref{eq:mlla3correlgdiff}) or by performing the evaluation from the previous solution (\ref{eq:sol3partg}). However,
in this paper, we will directly compute the solution (\ref{eq:sol3partg}) from the steepest descent method introduced
in \cite{Ramos:2006mk} and make some approximations in subsection \ref{subsec:MLLAsteepdesc}.
Accordingly, the solution of (\ref{eq:mlla3correlqdiff}) is also obtained in the appendix 
\ref{appendix:iteratsolgq} by setting $Q^{(3)}=C_{Q_{123}}^{(3)}Q_1Q_2Q_3$ and $Q_{ij}^{(2)}=C_{Q_{ij}}^{(2)}Q_iQ_j$
in the l.h.s. of (\ref{eq:mlla3correlqdiff}) and $G^{(3)}=C_{G_{123}}^{(3)}G_1G_2G_3$ in the r.h.s. of 
the same equation, such that,
\begin{eqnarray}\label{eq:sol3partq}
{\cal C}_{Q_{123}}^{(3)}-1=\left({\cal C}_{Q_{12}}^{(2)}-1\right)\tilde F_{12}^{(2)}+
\left({\cal C}_{Q_{13}}^{(2)}-1\right)\tilde F_{13}^{(2)}
+\left({\cal C}_{Q_{23}}^{(2)}-1\right)\tilde F_{23}^{(2)}+\tilde F_{123}^{(3)},
\end{eqnarray}
where,
\begin{equation}\label{eq:sol3partqbis}
\tilde F_{ij}^{(2)}=1+\frac{N_{Q_{ij}}^{(2)}}{D_{Q}^{(2)}},\quad
\tilde F_{123}^{(3)}=\frac{N_{Q}^{(3)}}{D_{Q}^{(3)}},
\end{equation}
with
\begin{subequations}
\begin{eqnarray}
\label{eq:NQ2}
N_{Q_{ij}}^{(2)}\!\!&\!\!=\!\!&\!\!\tilde\xi_1^{ij}+\tilde\delta_2^{ij}-\tilde\epsilon_1-\tilde\epsilon_2,\\
\label{eq:DQ2}
D_{Q}^{(2)}\!\!&\!\!=\!\!&\!\!\tilde\Delta_{12}+\tilde\Delta_{13}+\tilde\Delta_{23}+
\sum_i\frac{Q_{i\ell y}}{\gamma_0^2Q_i}+\tilde\epsilon_1+\tilde\epsilon_2,\\
\label{eq:NQ3}
N_{Q}^{(3)}\!\!&\!\!=\!\!&\!\!\frac{C_F}{N_c}{\cal C}_{G_{123}}^{(3)}\!\!\left[1-\frac34\left(\psi_{1,\ell}+\psi_{2,\ell}+\psi_{3,\ell}+\zeta_\ell-\beta_0\gamma_0^2\right)\right]\frac{G_1G_2G_3}{Q_1Q_2Q_3}+(\tilde\xi_1^{12}+\tilde\delta_2^{12})\\
\!\!&\!\!+\!\!&\!\!(\tilde\xi_1^{13}+\tilde\delta_2^{13})+(\tilde\xi_1^{23}+\tilde\delta_2^{23})-\tilde\epsilon_1-\tilde\epsilon_2,\cr
D_{Q}^{(3)}\!\!&\!\!=\!\!&\!\!D_{Q}^{(2)}=\tilde\Delta_{12}+\tilde\Delta_{13}+\tilde\Delta_{23}+
\sum_i\frac{Q_{i\ell y}}{\gamma_0^2Q_i}+\tilde\epsilon_1+\tilde\epsilon_2,
\label{eq:DQ3}
\end{eqnarray}
\end{subequations}
where one find the list of corrections,
\begin{subequations}
\begin{eqnarray}
\label{eq:tildezetaell}
\tilde\zeta\!\!&\!\!=\!\!&\!\!\ln \dot{{\cal C}}_{Q_{123}}^{(3)},\quad 
\tilde\zeta_\ell=\frac{\dot{{\cal C}}_{Q_{123},\ell}^{(3)}}{\dot{{\cal C}}_{Q_{123}}^{(3)}}={\cal O}(\gamma_0^2),\quad
\tilde\zeta_y=\frac{\dot{{\cal C}}_{Q_{123},y}^{(3)}}{\dot{{\cal C}}_{Q_{123}}^{(3)}}={\cal O}(\gamma_0^2),\;\\ 
\tilde\chi_\ell^{ij}\!\!&\!\!=\!\!&\!\!\frac{\dot{{\cal C}}_{Q_{ij},\ell}^{(2)}}
{\dot{{\cal C}}_{Q_{ij}}^{(2)}}={\cal O}(\gamma_0^2),\quad
\tilde\chi_y^{ij}=\frac{\dot{{\cal C}}_{Q_{ij},y}^{(2)}}{\dot{{\cal C}}_{Q_{ij}}^{(2)}}={\cal O}(\gamma_0^2),\\
\tilde\xi_1^{ij}\!\!&\!\!=\!\!&\!\!\frac{1}{\gamma_0^2}\left[\tilde\chi_\ell^{ij}(\psi_{Q_1,y}+\psi_{Q_2,y}+\psi_{Q_3,y})+
\tilde\chi_y^{ij}(\psi_{Q_1,\ell}+\psi_{Q_2,\ell}+\psi_{Q_3,\ell})\right]={\cal O}(\gamma_0),
\label{eq:tildexiell}\\
\tilde\delta_2^{ij}\!\!&\!\!=\!\!&\!\!\frac{1}{\gamma_0^2}\left(\tilde\chi_\ell^{ij}\tilde\chi_y^{ij}
+\tilde\chi_{\ell,y}^{ij}\right)={\cal O}(\gamma_0^2),\\
\tilde\epsilon_1\!\!&\!\!=\!\!&\!\!\frac{1}{\gamma_0^2}\left[\tilde\zeta_\ell(\psi_{Q_1,y}+\psi_{Q_2,y}+\psi_{Q_3,y})+
\tilde\zeta_y(\psi_{Q_1,\ell}+\psi_{Q_2,\ell}+\psi_{Q_3,\ell})\right]={\cal O}(\gamma_0),\label{eq:tildeepsilon1}\\
\tilde\epsilon_2\!\!&\!\!=\!\!&\!\!\frac{1}{\gamma_0^2}\left(\tilde\zeta_\ell\tilde\zeta_y
+\tilde\zeta_{\ell,y}\right)={\cal O}(\gamma_0^2).
\label{eq:tildeepsilon2}
\end{eqnarray}
\end{subequations}
The order of magnitude of these terms follows from (\ref{eq:ordermag1}) and (\ref{eq:ordermag2}). Setting all
corrections to zero, one recovers the DLA solution (\ref{eq:dla3partsol}) for $C_A=C_F$. The solutions
(\ref{eq:sol3partg}) and (\ref{eq:sol3partq}) of the evolution equations entangle corrections of order 
${\cal O}(\gamma_0)$ and ${\cal O}(\gamma_0^2)$, which are MLLA and NMLLA respectively. Furthermore, every term in 
(\ref{eq:sol3partg}) and (\ref{eq:sol3partq}) can be associated to a Feynman diagram of Fig.\ref{fig:physpic3} as was
explained in subsection \ref{subsec:dlasol}. The functions $F_{123}^{(3)}$ and $\tilde F_{123}^{(3)}$ in 
(\ref{eq:sol3partg}) and (\ref{eq:sol3partq})
correspond respectively to the cumulant of genuine correlations associated to the process displayed in 
Fig.\ref{fig:three-part} and Fig.\ref{fig:physpic3}d. 
These contributions, (\ref{eq:zetaell}-\ref{eq:epsilon2}) and 
(\ref{eq:tildezetaell}-\ref{eq:tildeepsilon2}) are small corrections arising from the 
iterative solution of the evolution equations because one takes the derivatives over the functions 
$\zeta=\ln \dot{{\cal C}}^{(3)}_{G_{123}},\tilde\zeta=\ln \dot{{\cal C}}^{(3)}_{Q_{123}}$ and 
$\chi^{ij} = \ln \dot{{\cal C}}_{G_{ij}}^{(2)},\tilde\chi^{ij} = \ln\dot{{\cal C}}_{Q_{ij}}^{(2)}$ 
for both quark and gluon jets. For the evaluation of such corrections one needs to take the DLA 
expressions of $\dot{{\cal C}}^{(3)}_{A_{123}}$ and $\dot{{\cal C}}_{A_{ij}}^{(2)}$ written in 
(\ref{eq:dla3partsol}) and (\ref{eq:2partcorrdla}) respectively.
\subsection{MLLA approximation and evaluation by the steepest descent method}
\label{subsec:MLLAsteepdesc}
In \cite{Ramos:2006dx}, the exact solutions of the two-particle evolution 
equations were compared with the MLLA solutions from the steepest descent method
for the one particle distribution. The agreement between both approaches was successful 
and made possible the fast computation of the correlators from the steepest descent. 
That is the reason for in this paper, we limit ourselves to this method.
Making use of the ratio (\ref{eq:ratioGQ}), it is easy to demonstrate that,
\begin{eqnarray}
%\tilde\xi_1^{ij}\!\!&\!\!\stackrel{mlla}{=}\!\!&\!\!\xi_1^{ij}+{\cal O}(\gamma_0^2),\\
%\tilde\epsilon_1\!\!&\!\!\stackrel{mlla}{=}\!\!&\!\!\epsilon_1+{\cal O}(\gamma_0^2),\\
\psi_{Q,\ell}=\psi_{\ell}+{\cal O}(\gamma_0^2),\quad
\psi_{Q,y}=\psi_{y}+{\cal O}(\gamma_0^2),\quad
\tilde\Delta_{ij}\!\!&\!\!\stackrel{mlla}{=}\!\!&\!\!\Delta_{ij}+{\cal O}(\gamma_0^2).
\end{eqnarray}
Dropping corrections of order ${\cal O}(\gamma_0^2)$, which go beyond the MLLA approximation, 
we obtain for the gluon jet
\begin{eqnarray}\label{eq:fij}
F_{ij}^{(2)}\!\!&\!\!\stackrel{mlla}{=}\!\!&\!\!
1+\frac{1-b\left(\psi_{1,\ell}+\psi_{2,\ell}+\psi_{3,\ell}\right)+\xi_1^{ij}-\epsilon_1}
{2+\Delta_{12}+\Delta_{13}+\Delta_{23}+\epsilon_1},\\
\label{eq:tildef123g}
F_{123}^{(3)}\!\!&\!\!\stackrel{mlla}{=}\!\!&\!\!\frac{1-c\left(\psi_{1,\ell}+\psi_{2,\ell}+\psi_{3,\ell}\right)
+\xi_1^{12}+\xi_1^{13}+\xi_1^{23}-\epsilon_1}
{2+\Delta_{12}+\Delta_{13}+\Delta_{23}+\epsilon_1}
\end{eqnarray}
and for the quark jet
\begin{eqnarray}\label{eq:tildefij}
\tilde F_{ij}^{(2)}\!\!&\!\!\stackrel{mlla}{=}\!\!&\!\!1+\frac{\tilde\xi_1^{ij}
-\tilde\epsilon_1}{3+\Delta_{12}+\Delta_{13}+\Delta_{23}-a\left(\psi_{1,\ell}+\psi_{2,\ell}+\psi_{3,\ell}\right)+\tilde\epsilon_1},\\
\tilde F_{123}^{(3)}\!\!&\!\!\stackrel{mlla}{=}\!\!&\!\!\frac{N_c^2}{C_F^2}\frac{{\cal C}_{G_{123}}^{(3)}\left[1-a\left(\psi_{1,\ell}+\psi_{2,\ell}+\psi_{3,\ell}\right)\right]+\tilde\xi_1^{12}+\tilde\xi_1^{13}+\tilde\xi_1^{23}-\tilde\tilde\epsilon_1}
{3+\Delta_{12}+\Delta_{13}+\Delta_{23}-a\left(\psi_{1,\ell}+\psi_{2,\ell}+\psi_{3,\ell}\right)+\tilde\epsilon_1}.
\label{eq:tildef123}
\end{eqnarray}
The subtracted terms $\propto-a$ in the denominators of (\ref{eq:tildefij}) and (\ref{eq:tildef123}) 
appear after having replaced (\ref{eq:ratioGQ}) and (\ref{eq:QilyQ}) in (\ref{eq:DQ2}) and (\ref{eq:NQ3}) respectively.
Such simplified expressions are useful for the steepest descent evaluation that proved successful while 
describing the single inclusive distribution and two-particle correlations in \cite{Ramos:2006mk}. Except 
the MLLA corrections $\epsilon_1$ and $\xi_1^{ij}$, all the other corrections and functions appearing in the solutions of the 
evolution equations were obtained in \cite{Ramos:2006mk}, which will allow for the straightforward computation
of the three-particle correlators in quark and gluon jets. We write some of these formul{\ae} 
for the evaluation in the appendix \ref{appendix:steepdesceval}. Integrating the equation (\ref{eq:solg}) 
over ``$y$", the solution for the single inclusive distribution is given 
by the following integral representation in Mellin space \cite{Ramos:2006mk},
\begin{equation}\label{eq:integrep}
G(\ell,y)=(\ell+y+\lambda)\iint\frac{d\omega d\nu}{(2\pi i)^2}e^{\omega\ell+\nu y}\int_0^{\infty}
\frac{ds}{\nu+s}\left(\frac{\omega(\nu+s)}{(\omega+s)\nu}\right)^{1/\beta_0(\omega-\nu)}
\left(\frac{\nu}{\nu+s}\right)^{a/\beta_0}e^{-\lambda s}.
\end{equation}
The integral representation (\ref{eq:integrep}) was estimated by the steepest descent method at small $x\ll1$
and high energy scale $Q\gg1$; the approached solution was compared with the exact solution (\ref{eq:ifD}) 
(see the appendix \ref{subsec:oneandtwopartdist})
in the limiting spectrum ($\lambda=0$) and beyond ($\lambda\ne0$). 
In particular, (\ref{eq:integrep}) was also demonstrated to be equivalent to (\ref{eq:ifD}) 
for $\lambda=0$ \cite{Ramos:2006dx}. The agreement between the approached and exact solutions
turned out to be good, such that the following expressions of the approached logarithmic derivatives from the 
steepest descent method were suited for the evaluation of the two-particle correlators \cite{Ramos:2006mk},
\begin{eqnarray}\label{eq:psiell}
\psi_{i,\ell}(\mu_i,\nu_i)\!\!&\!\!=\!\!&\!\!\gamma_0e^{\mu_i}+\frac12a\gamma_0^2
\left[e^{\mu_i}\tilde Q(\mu_i,\nu_i)-\tanh\nu_i-\tanh\nu_i\coth\mu_i\Big(1+e^{\mu_i}\tilde Q(\mu_i,\nu_i)\Big)\right]\\
\!\!&\!\!-\!\!&\!\!\frac12\beta_0\gamma_0^2\left[1+\tanh\nu_i\Big(1+K(\mu_i,\nu_i)\Big)+C(\mu_i,\nu_i)
\Big(1+e^{\mu_i}\tilde Q(\mu_i,\nu_i)\Big)\right]+{\cal O}(\gamma_0^2),\cr
\psi_{i,y}(\mu,\nu)\!\!&\!\!=\!\!&\!\!\gamma_0e^{-\mu_i}-\frac12a\gamma_0^2
\left[2+e^{-\mu_i}\tilde Q(\mu_i,\nu_i)+\tanh\nu_i-\tanh\nu_i\coth\mu_i\Big(1+e^{-\mu_i}\tilde Q(\mu_i,\nu_i)\Big)\right]\cr
\!\!&\!\!-\!\!&\!\!\frac12\beta_0\gamma_0^2\left[1+\tanh\nu_i\Big(1+K(\mu_i,\nu_i)\Big)-C(\mu_i,\nu_i)
\Big(1+e^{-\mu_i}\tilde Q(\mu_i,\nu_i)\Big)\right]+{\cal O}(\gamma_0^2),
\label{eq:psiy}
\end{eqnarray}
where the functions $\tilde Q(\mu_i,\nu_i)$, $C(\mu_i,\nu_i)$ and $K(\mu_i,\nu_i)$ 
are defined in the appendix \ref{appendix:steepdesceval}.
The term $\propto a$ in (\ref{eq:psiell}) and (\ref{eq:psiy}) accounts for energy conservation while that 
$\propto\beta_0$ accounts for the running of the coupling $\alpha_s$. The variables $(\mu_i,\nu_i)$ are related to
$(\ell_i,y_i)$ through the same 2x2 non-linear system of equations (\ref{eq:systemunubis}).
After inverting (\ref{eq:systemunubis}) numerically, $\mu_i(\ell_i,y_i)$ and $\nu_i(\ell_i,y_i)$ can be 
plugged into (\ref{eq:psiell}) and (\ref{eq:psiy}) so as to get the logarithmic derivatives of the
single inclusive spectrum as a function of the original kinematical variables $\ell_i$ and $y_i$ as
it was done in \cite{Ramos:2006mk}. The MLLA two-particle correlators involved in  
(\ref{eq:sol3partg}) and (\ref{eq:sol3partq}) are (\ref{eq:sol2partcorrstdescg}) and 
(\ref{eq:sol2partcorrstdescq}) and are written in the appendix \ref{appendix:steepdesceval}.
These expressions have been taken from reference \cite{Ramos:2006mk}.

Corrections $\xi_1^{ij},\tilde\xi_1^{ij}$ 
and $\epsilon_1,\tilde\epsilon_1$ are new for three-particle correlations. 
Such expressions are explicitly written in the appendix
\ref{subappex:corrxis} from the steepest 
descent evaluation of the single inclusive distribution (\ref{eq:integrep}). They 
are small and decrease the three-particle correlator for $\ell_i\ne\ell_j$, 
that is when one parton is much harder than the other.
Notice that the steepest descent method constitutes the only way for the disentanglement between 
MLLA ${\cal O}(\sqrt{\alpha_s})$ and NMLLA ${\cal O}(\alpha_s)$ corrections appearing in the
solution of the evolution equations for the two and three-particle correlations. It makes also 
possible to distinguish between corrections following from the energy balance and the running 
effects of the coupling constant $\alpha_s$. Finally, this method also allows for the application
of the hump approximation or Fong-Webber expansion of the solutions 
with MLLA ${\cal O}(\sqrt{\alpha_s})$ accuracy \cite{Fong:1990ph,Fong:1990nt}.

In this frame, the role of MLLA corrections should be expected to be larger than for the 
two-particle correlations. Indeed, higher order corrections increase with the
rank of the correlator, which is known from the Koba-Nielsen-Olesen (KNO) 
problem for intra-jet multiplicity fluctuations 
\cite{Khoze:1996dn,Koba:1972ng,Dokshitzer:1993dc}. 
For the 2-particle for instance one has 
$\propto-b(\psi_{1,\ell}+\psi_{2,\ell})$ and for the three-particle correlator one 
gets the larger correction $\propto-c(\psi_{1,\ell}+\psi_{2,\ell}+\psi_{3,\ell})$.
\subsection{Hump approximation}
\label{subsec:humpapp}
From the steepest descent evaluation introduced in \cite{Ramos:2006mk}, near the hump of the single
inclusive distribution $\mid\ell-Y/2\mid\ll\sigma\propto Y^{3/2}$ for $i=1,2,3$, 
corrections $\xi_1^{ij},\tilde\xi_1^{ij}$ and $\epsilon_1,\tilde\epsilon_1$ 
could be written in the symbolic form (see appendix \ref{subappendix:humpapprox}),  
\begin{eqnarray}
\xi_1^{ij},\tilde\xi_1^{ij}\!\!&\!\!\simeq\!\!&\!\!\left(\frac{\ell_i-\ell_j}{Y}\right)^2\gamma_0
+{\cal O}(\gamma_0^2),\\
\epsilon_1,\tilde\epsilon_1\!\!&\!\!\simeq\!\!&\!\!\left(\frac{\ell_1-\ell_2}{Y}\right)^2\gamma_0+
\left(\frac{\ell_1-\ell_3}{Y}\right)^2\gamma_0
+\left(\frac{\ell_2-\ell_3}{Y}\right)^2\gamma_0+{\cal O}(\gamma_0^2),
\label{eq:humpepsilon1}
\end{eqnarray}
such that both can be neglected $\xi_1^{ij}\approx0$, $\epsilon_1\approx0$ in this approximation, like
$\delta_1^{ij}$ was also in \cite{Ramos:2006mk}. In the appendix \ref{subappendix:humpapprox}, following
from the steepest descent method, the expressions of (\ref{eq:NG2}-\ref{eq:DG3})  are given and 
(\ref{eq:NQ2}-\ref{eq:DQ3}) expanded in $\sqrt{\alpha_s}$.
In particular, the expressions (\ref{eq:humpCG2}) and (\ref{eq:humpCQ2}), after being expanded in $\gamma_0$, can be 
demonstrated to recover the Fong-Webber results for the two-particle correlations 
\cite{Fong:1990ph,Fong:1990nt}. Replacing the expressions (\ref{eq:humpNQ2}-\ref{eq:humpNQ3})
into (\ref{eq:sol3partg},\ref{eq:sol3partgbis}) and (\ref{eq:sol3partq},\ref{eq:sol3partqbis}), one finds 
those for the three-particle correlators in the Fong-Webber approximation \cite{Fong:1990ph,Fong:1990nt}.
This solution will be compared with that from (\ref{eq:tildef123g}) and (\ref{eq:tildef123}) after making 
use of (\ref{eq:psiell}) and (\ref{eq:psiy}) in subsection \ref{subsec:pheno}. 

\subsection{From two to three-particle correlations in the small $\boldsymbol{x}$ region}
\label{subsec:smallx}
In \cite{Ramos:2006dx}, the sign of the two-particle correlator (${\cal C}_A^{(2)}-1\geq0$) 
was studied as a function of $x$ in the region of the phase space where the two partons 
(hadrons after assuming the LPHD) are strongly correlated. From the previous inequality, 
it turned out that two partons with $\ell_i\gtrsim 2.6$ ($x_i\lesssim0.07$) at LHC energy scales 
(i.e. $Q=450$ GeV, see subsection \ref{subsec:pheno}) are correlated as they are emitted from the same cascade 
following the QCD AO. Asymptotically $Y\to\infty$, one has $\ell_i\gtrsim 4.5$ ($x_i\lesssim0.011$).

For three-particle correlations we study the sign of the 
cumulant of the genuine correlator $F^{(3)}_{123}>0$ and determine the approximate region in $x$ where 
diagrams displayed in Fig.\ref{fig:three-part} and Fig.\ref{fig:physpic3}d become dominant. One has,
$$
1-c\left(\psi_{1,\ell}+\psi_{2,\ell}+\psi_{3,\ell}\right)
+\xi_1^{12}+\xi_1^{13}+\xi_1^{23}-\epsilon_1>0.
$$
However, corrections $\xi_1^{ij}$, $\epsilon_1$ have been shown to be negligible and to 
vanish for particles having the same energy momentum. Thus, we rather study the sign of
$$
1-c\left(\psi_{1,\ell}+\psi_{2,\ell}+\psi_{3,\ell}\right)>0.
$$
Making use of $\psi_\ell=\gamma_0\sqrt{\frac{y}{\ell}}=\gamma_0\sqrt{\frac{Y-\ell}{\ell}}$
for the sake of simplicity, one has,
$$
1-3c\gamma_0\sqrt{\frac{Y-\ell}{\ell}}>0\Leftrightarrow\ell>\frac{M}{1+\frac{M}{Y}},
\quad M=\frac{9c^2}{\beta_0}=10.1.
$$
Thus, for LHC energy $Y=7.5$, the value of $\ell(x)$ where the cumulant becomes 
positive turns out to be $\ell\gtrsim4.3$, which in $x$ corresponds to $x\lesssim0.014$.
Asymptotically $Y\to\infty$, one has $\ell_i\gtrsim 10.1$ ($x_i\lesssim4.1\times10^{-5}$).
Therefore, there exists a range in $x$ where the observable ${\cal C}^{(3)}_{123}$ is 
dominated by the emission of two correlated partons emitted independently from the third 
one, that is $0.014\lesssim x\lesssim0.07$ for diagrams Fig.\ref{fig:physpic3}b
and Fig.\ref{fig:physpic3}c; for $x\lesssim0.014$, the process will be dominated 
by three particles emitted from the same partonic cascade following the QCD AO described
in Fig.\ref{fig:physpic3}d. Asymptotically $Y\to\infty$, one has 
$4.1\times10^{-5}\lesssim x\lesssim0.011$ for diagrams Fig.\ref{fig:physpic3}b
and Fig.\ref{fig:physpic3}c, and $x\lesssim4.1\times10^{-5}$ for Fig.\ref{fig:physpic3}d.
These values will indeed justify our choices for the representation of the 
three-particle correlations as function of $(x_1,x_2,x_3)$ in subsection \ref{subsec:pheno}.

\subsection{Beyond three-particle correlations}
\label{subsec:beyond}
It is worth reminding that the LPHD hypothesis has also been 
confronted to multi-particle factorial moments up to the 5th order in 
the experimental studies of $ep$ and $e^+e^-$ collisions at HERA \cite{Chekanov:2001sj} and
LEP \cite{Abbiendi:2006qr} respectively, where it was 
found that the LPHD hypothesis faces difficulties when it is applied to soft multi-particle 
fluctuations. In this work the studies are carried out by using the momentum and transverse momentum
cuts in order to test the MLLA soft limit calculations \cite{Lupia:1998nc}. The theoretical computation of 
multiplicity correlators or multiplicity fluctuations $\left<n(n-1)\ldots(n-k+1)\right>$ was performed 
in \cite{Malaza:1985jd} at MLLA up to the rank $k=5$ of the correlator.

However, performing these calculations for higher rank 
differential inclusive correlators, related to the previous ones by the integral
$$
\left<n(n-1)\ldots(n-k+1)\right>_A=\int dx_1\ldots dx_k x_1\ldots x_kD_A^{(k)}(x_1,\ldots,x_k,Y)
$$
becomes rather cumbersome. As an example, in this subsection, we display the DLA equation and 
solution of the 4-particle correlator. The DLA equation reads,
\begin{equation}\label{eq:4partcorr}
\hat{A}^{(4)}_{1234}=\frac{C_A}{N_c}\gamma_0^2G^{(4)}_{1234},
\end{equation}
where $\hat{A}$ has been defined in the appendix \ref{sec:4partcorr} in (\ref{eq:4partAhat}).
The solution of (\ref{eq:4partcorr}) with the definition of $\hat{A}$ 
(\ref{eq:4partAhat}) reads,
\begin{eqnarray}\label{eq:sol4corr}
{\cal C}^{(4)}_A\!-\!1\!=\!\frac{N_c}{C_A}H_1\left(\dot{{\cal C}}^{(2)}\right)
\!+\!\frac{N_c^2}{C_A^2}H_2\left(\dot{{\cal C}}^{(3)},\dot{{\cal C}}^{(2)}\right)
\!+\!\frac{N_c^3}{C_A^3}\frac{H_3\left(\dot{{\cal C}}^{(3)},\dot{{\cal C}}^{(2)}\right)}
{3+\Delta_{12}+\Delta_{13}+\Delta_{14}+\Delta_{23}+\Delta_{24}+\Delta_{34}},
\end{eqnarray}
where the functions $H_1$, $H_2$ and $H_3$ are written in the appendix \ref{sec:4partcorr}
in (\ref{eq:H1}), (\ref{eq:H2}) and (\ref{eq:H3}) respectively. The solution (\ref{eq:sol4corr})
can also be interpreted in terms of Feynman diagrams contributing to the emission of four hadrons 
inside the jet. Accordingly, the term $\propto\frac{N_c}{C_A}$ correspond to the case $A\to12(34)$
where two offspring are correlated while the other two are emitted independently; as a consequence it depends
only on the two-particle correlator. The second term $\propto\frac{N_c^2}{C_A^2}$ is associated to the 
cases $A\to(12)(34)$ and $A\to(123)4$, which translates into either emitting two sub-jets with two-particles
correlated within each, or emitting three correlated partons like in Fig.\ref{fig:three-part} 
with another independent emission. Finally, the term $\propto\frac{N_c^3}{C_A^3}$ after setting $H_3=1+\ldots$ 
corresponds to the full correlated emission of four offspring inside the same shower.
The inclusion of SLs corrections to (\ref{eq:sol4corr}) would be cumbersome and
stays beyond the scope of this paper. On the other hand, the computation of differential higher order 
rank ($k$) correlators at MLLA would imply the failure of the perturbative approach because of the increasing
size of higher order corrections $\propto(\psi_{1,\ell}+\ldots\psi_{k,\ell})={\cal O}(\sqrt{\alpha_s})$.
Hence, for higher order $k$ correlators, the small $x$ range
where MLLA predictions stay valid gets reduced even at high energy scales, such that 
(see subsection \ref{subsec:smallx})
$$
M_k=\frac{k^2c_k}{\beta_0},\qquad \ell_k>\frac{M_k}{1+\frac{M_k}{Y}}
$$
with
$$
c_k=\frac{1}{4N_c}\left[\frac{11}3N_c+(-1)^k\frac43n_fT_R\left(1-2\frac{C_F}{N_c}\right)^k\right].
$$

\section{Predictions for the LHC and phenomenological consequences}
\label{subsec:pheno}
In this section, we perform theoretical predictions for three-particle correlations 
for the LHC. We display the MLLA solutions (\ref{eq:sol3partg}) and (\ref{eq:sol3partq}) 
of the evolution equations (\ref{eq:mlla3correlgdiff}) and (\ref{eq:mlla3correlqdiff}) 
respectively. We compare the DLA solution of the evolutions equations 
from section \ref{subsec:dlasol} with the MLLA solution from the steepest descent 
evaluation of the one-particle distribution in subsection \ref{subsec:MLLAsteepdesc}
and the solution from the hump approximation in \ref{subsec:humpapp}. Thus,
\begin{itemize}
\item the DLA solution is computed by plugging (\ref{eq:simpledla}) into (\ref{eq:dlasolmunuA});
\item the MLLA solution from the steepest descent will be displayed
by substituting the MLLA two-particle correlators (\ref{eq:sol2partcorrstdescg}), 
(\ref{eq:sol2partcorrstdescq}) and the functions  (\ref{eq:fij}), (\ref{eq:tildef123g}), 
(\ref{eq:tildefij}) and (\ref{eq:tildef123}) into (\ref{eq:sol3partg}) and (\ref{eq:sol3partq}) 
for gluon and quark jets respectively;
\item the MLLA hump approximation will be displayed by plugging (\ref{eq:humpNQ2})-(\ref{eq:humpNQ3}) into 
(\ref{eq:sol3partgbis}) and (\ref{eq:sol3partqbis})
and finally (\ref{eq:sol3partg}) and (\ref{eq:sol3partq}).
\end{itemize}
In particular, the computation of the DLA and MLLA solutions from the steepest descent needs the 
prior inversion of the system of equations (\ref{eq:systemunubis}) in order to obtain $(\mu_i,\nu_i)$
as functions of the original kinematical variables ($\ell_i,y_i$). The correlators are functions of 
the variables $(\ell_i$, $y_i)$ and the virtuality of the jet $Q=E\Theta_0$. After setting 
$y_i=Y-\ell_i$ with fixed $Y=\ln(Q/Q_0)$ in the arguments of the solutions (\ref{eq:sol3partg}) and (\ref{eq:sol3partq})
the dependence can be reduced to the following: ${\cal C}^{(3)}_{G_{123}}(\ell_1,\ell_2,\ell_3,Y)$ 
and ${\cal C}^{(3)}_{Q_{123}}(\ell_1,\ell_2,\ell_3,Y)$.
\subsection{Predictions for the limiting spectrum $\boldsymbol{\lambda\approx0}$}
In this subsection we give predictions within the limiting spectrum $\lambda\lesssim0.5$
for charged hadrons mostly composed by pions and kaons.

\begin{figure}
\begin{center}
\epsfig{file=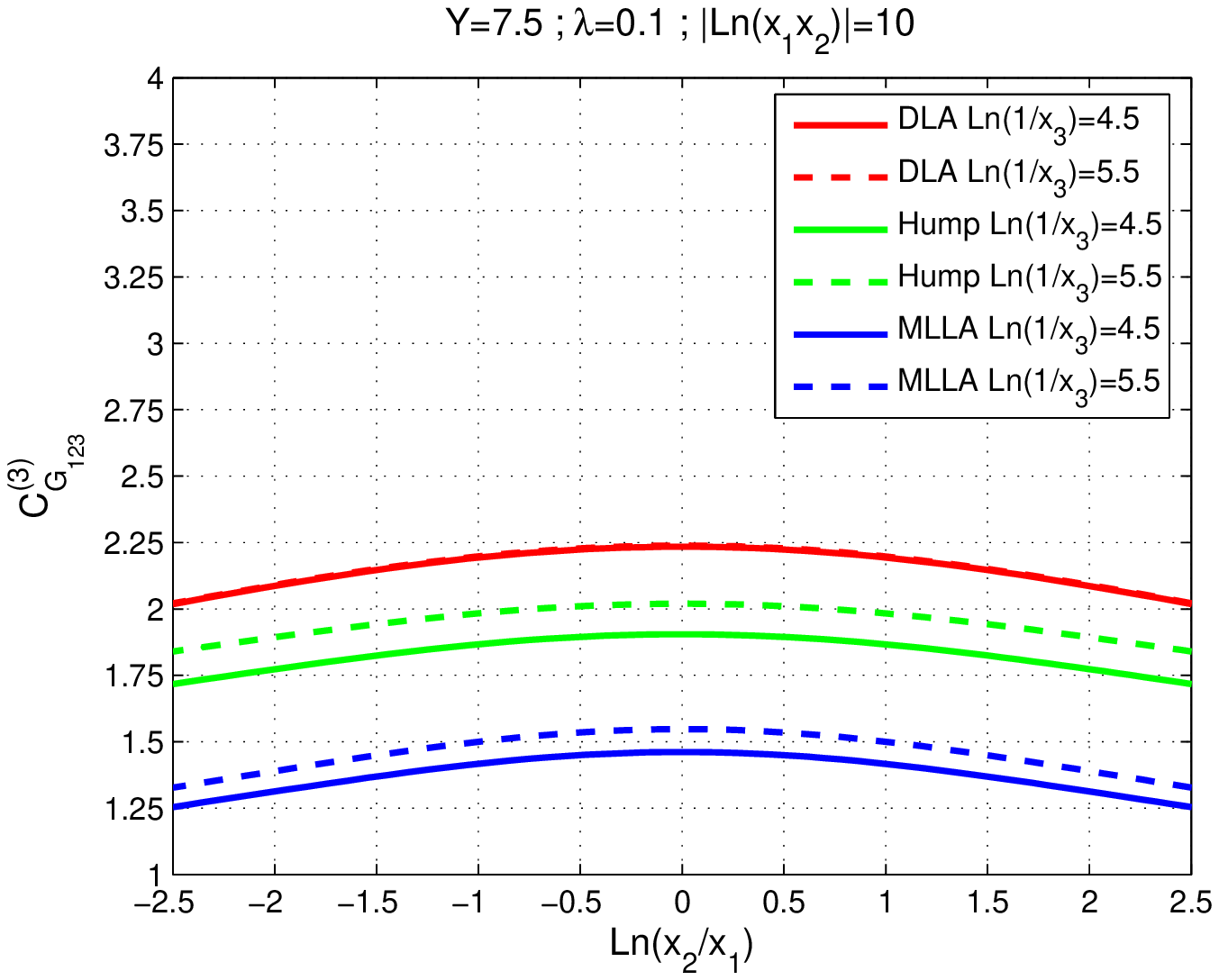, height=6.5truecm,width=7.5truecm}
\epsfig{file=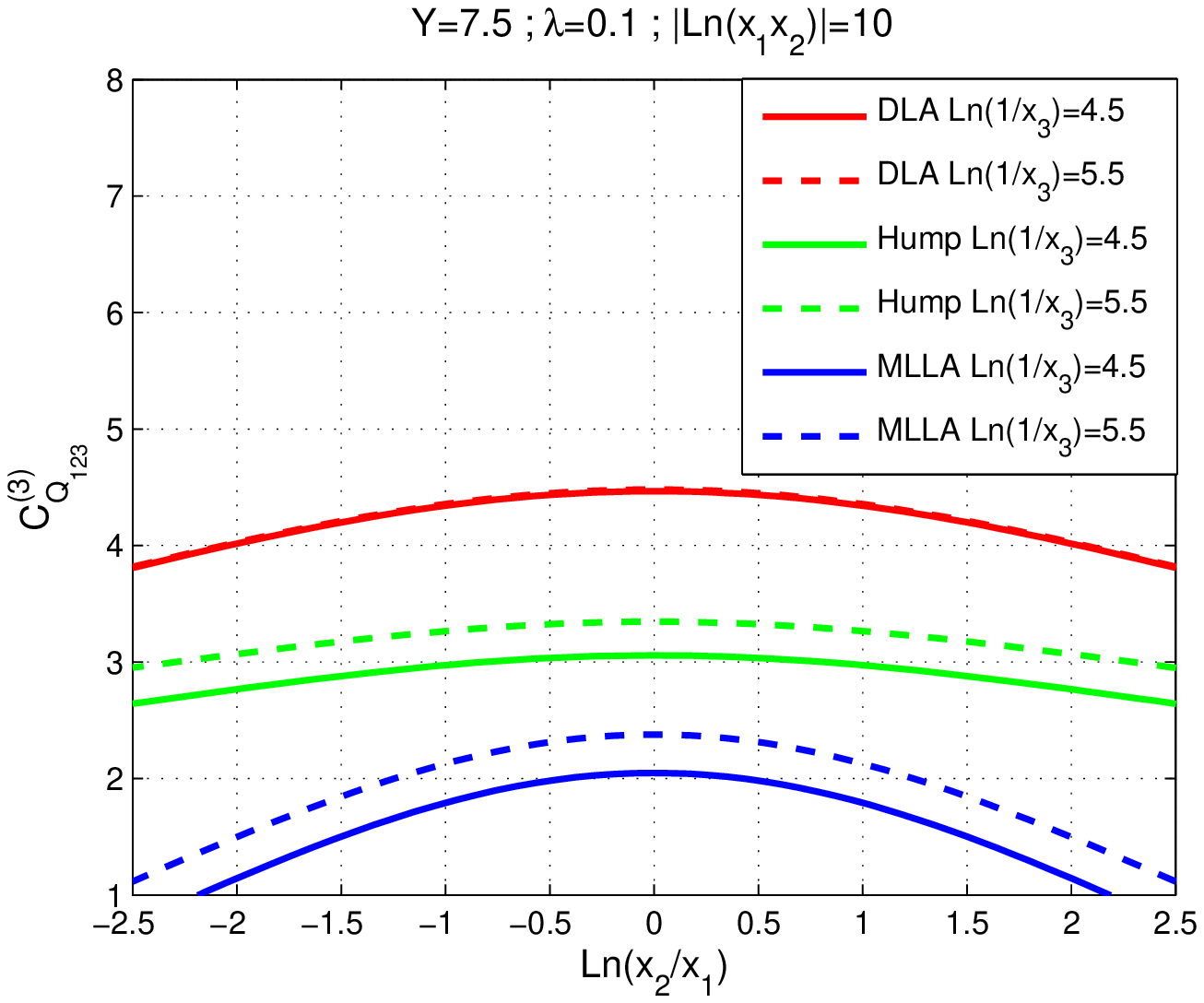, height=6.5truecm,width=7.5truecm}
\caption{\label{fig:corr3gqdiff} Three-particle correlations inside a gluon jet (left) and
a quark jet (right) as a function of $\ell_1-\ell_2=\ln(x_2/x_1)$ for $\ell_1+\ell_2=|\ln(x_1x_2)|=10$, 
$\ell_3=\ln(1/x_3)=4.5,\,5.5$, fixed $Y=7.5$ in the limiting spectrum approximation $\lambda\approx0$.}  
\end{center}
\end{figure}

In Fig.\ref{fig:corr3gqdiff}, the DLA (\ref{eq:dla3partsol}), 
MLLA hump approximation from subsection \ref{subsec:humpapp} and 
MLLA (\ref{eq:sol3partg}) three-particle correlators are displayed, as a function of the difference 
$(\ell_1-\ell_2)=\ln(x_2/x_1)$ for two fixed values of $\ell_3=\ln(1/x_3)=4.5,\,5.5$, fixed sum $(\ell_1+\ell_2)=|\ln(x_1x_2)|=10$ 
and finally fixed $Y=7.5$ (virtuality $Q=450$ GeV and $\Lambda_{QCD}=250$ MeV), which is realistic for the LHC 
phenomenology \cite{PerezRamos:2005nh}. The values $\ell_3=\ln(1/x_3)=4.5,\,5.5$ ($x_3=0.011,\,x_3=0.004$) 
have been chosen according to the range of the energy fraction $x_i\ll0.1$, 
where the MLLA scheme can only be applied and in particular, the range $x\lesssim0.014$, where
the cumulant correlator $F^{(3)}_{123}$ is dominant (see subsection \ref{subsec:smallx}). 

\begin{figure}
\begin{center}
\epsfig{file=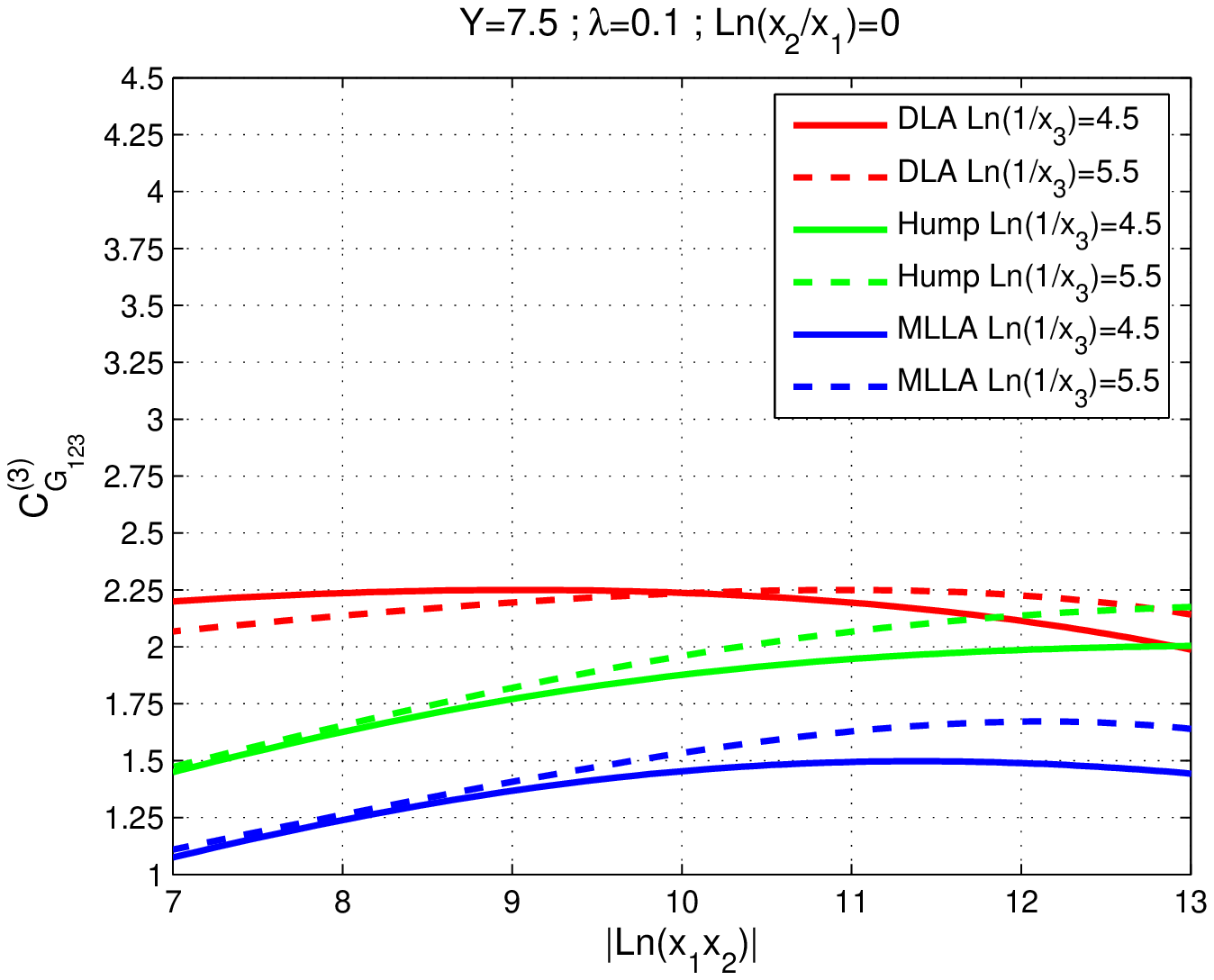, height=6.5truecm,width=7.5truecm}
\epsfig{file=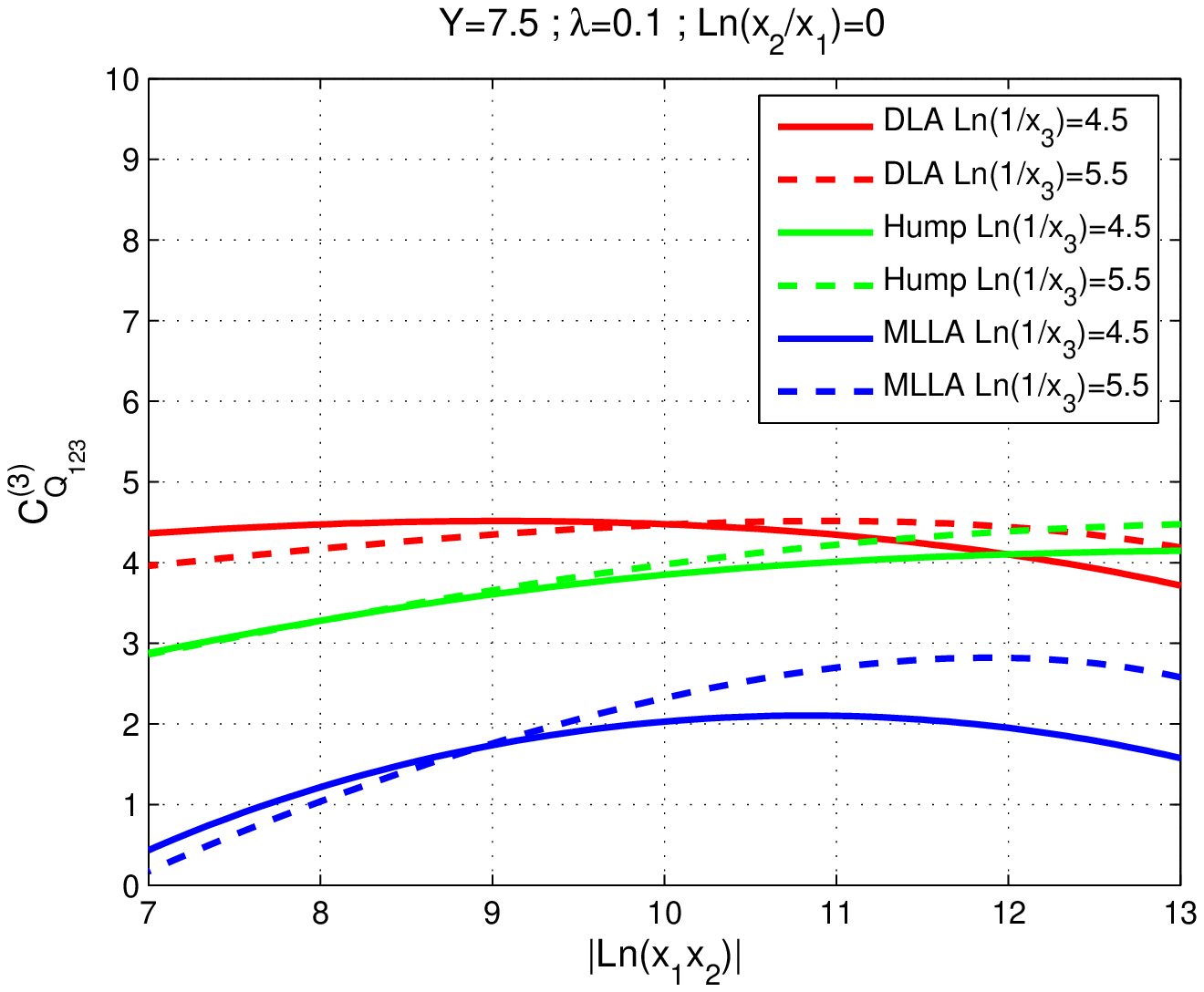, height=6.5truecm,width=7.5truecm}
\caption{\label{fig:corr3gqsum} Three-particle correlations inside a gluon jet (left) and
a quark jet (right) as a function of $\ell_1+\ell_2=|\ln(x_1x_2)|$ for $x_1=x_2$, 
$\ell_3=\ln(1/x_3)=4.5,\,5.5$, fixed $Y=7.5$ in the limiting spectrum approximation $\lambda\approx0$.}  
\end{center}
\end{figure}

In Fig.\ref{fig:corr3gqsum}, 
the DLA (\ref{eq:dla3partsol}), MLLA hump approximation from subsection 
\ref{subsec:humpapp} and MLLA (\ref{eq:sol3partg}) three-particle correlators are displayed,
in this case, as a function of the sum 
$(\ell_1+\ell_2)=|\ln(x_1x_2)|$ for the same values of $\ell_3=\ln(1/x_3)=4.5,\,5.5$, for $x_1=x_2$ and 
$Y=7.5$. The range $7.0\leq|\ln(x_1x_2)|\leq13.0$ has been chosen according to the condition
$x\lesssim0.014$ discussed in \ref{subsec:smallx}.

\begin{figure}
\begin{center}
\epsfig{file=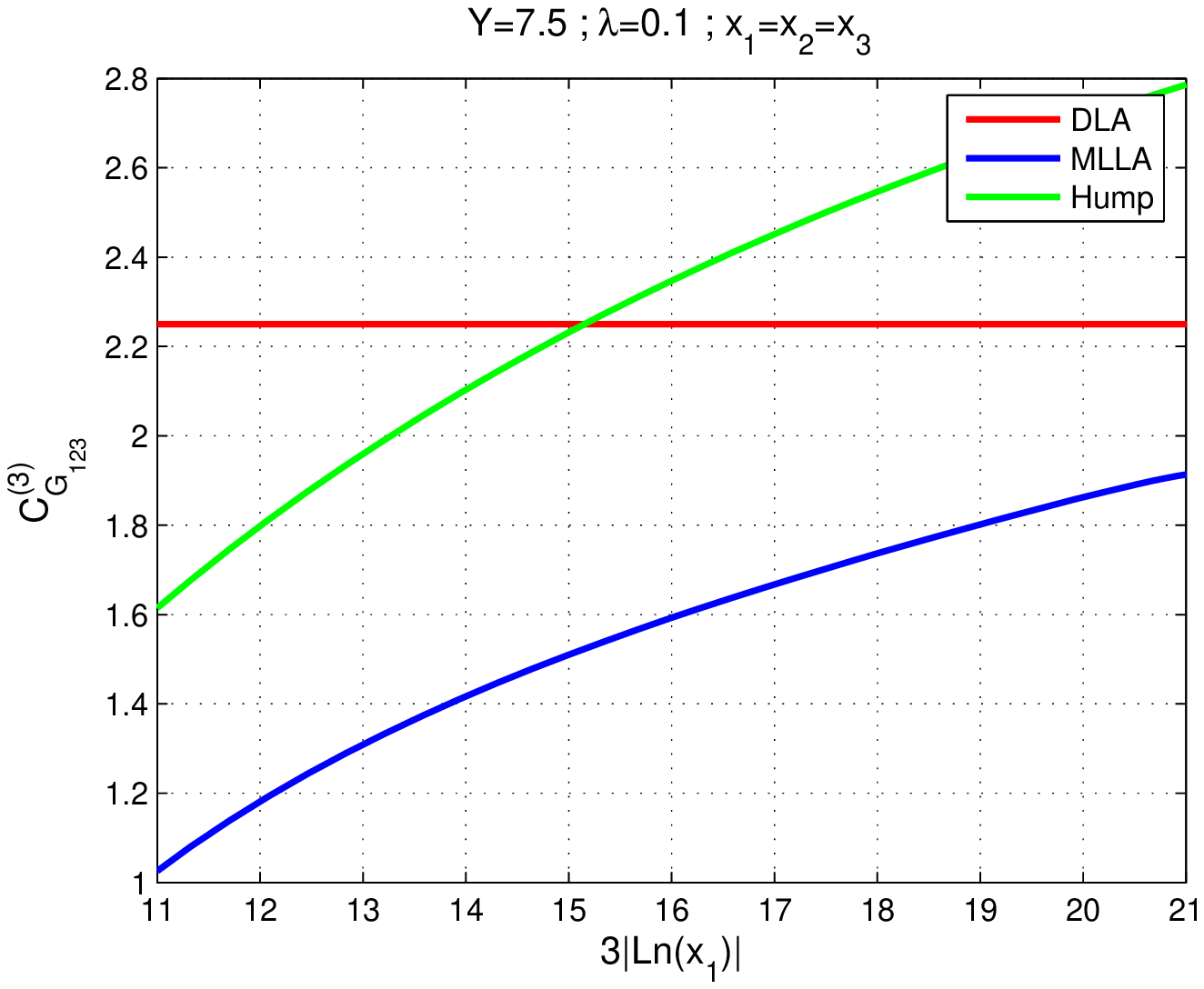, height=6.5truecm,width=7.5truecm}
\epsfig{file=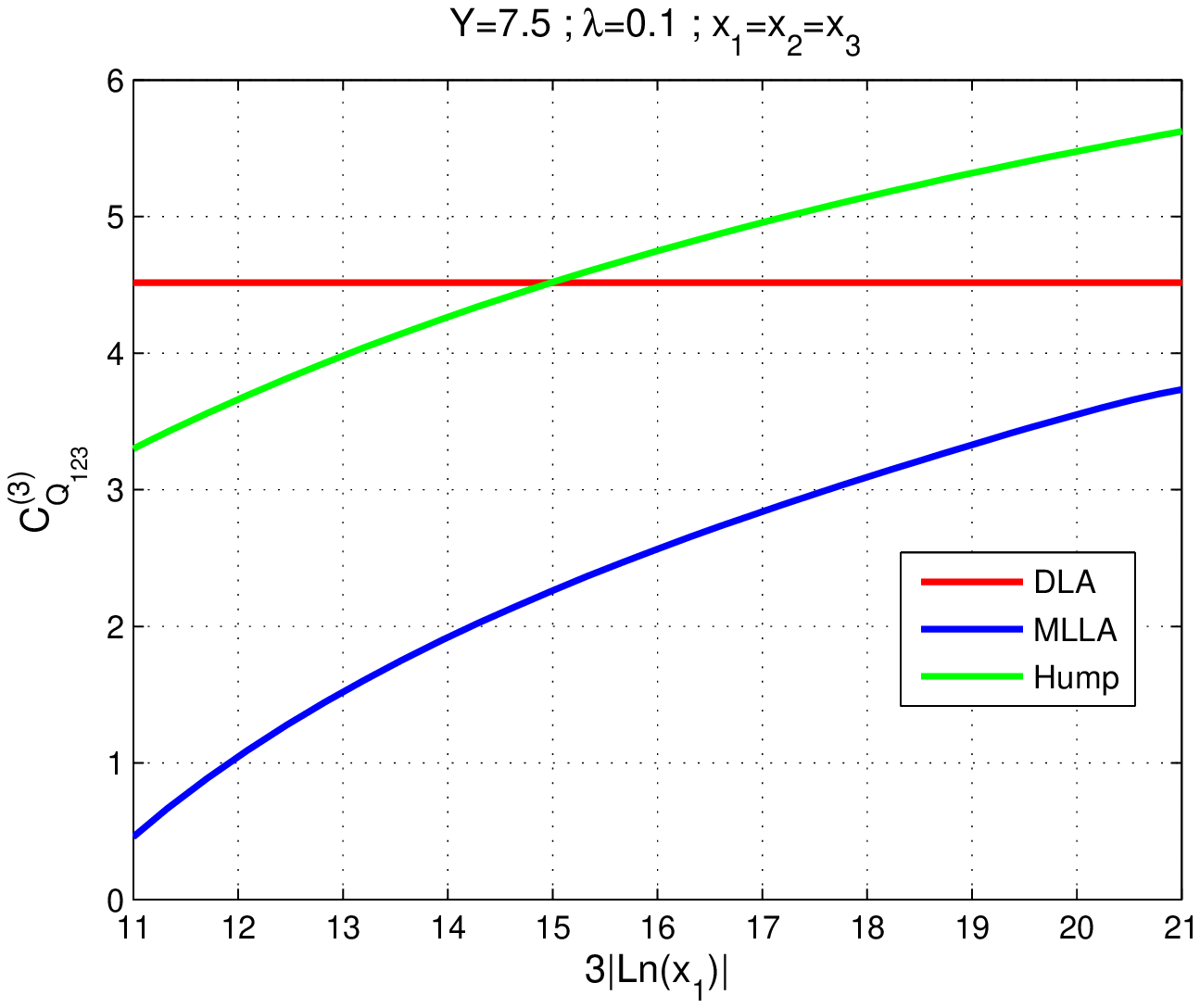, height=6.5truecm,width=7.5truecm}
\caption{\label{fig:corr3gq3sum} Three-particle correlations inside a gluon jet (left) and
a quark jet (right) as a function of $\ell_1+\ell_2+\ell_3=|\ln(x_1x_2x_3)|$ for $x_1=x_2=x_3$, 
fixed $Y=7.5$ in the limiting spectrum approximation $\lambda\approx0$.}  
\end{center}
\end{figure}

As expected in both cases, the DLA and MLLA three-particle correlators are larger inside a 
quark than in a gluon jet. Of course, these plots will be the same and the interpretation will
apply to all possible permutations of three particles (123). As observed and written above, 
the difference between the DLA and MLLA results is quite important pointing out 
that overall corrections in ${\cal O}(\sqrt{\alpha_s})$ are quite large. Indeed, the last 
behavior is not surprising as was already observed on the treatment of 
multiplicity fluctuations of the third kind, where \cite{Malaza:1985jd}
\begin{eqnarray}
\frac{\left<n(n-1)(n-2)\right>_G}{\left<n\right>^3_G}\!\!&\!\!=\!\!&\!\!
2.25\left[1-(1.425-0.021n_f)\sqrt{\alpha_s}\right],\cr
\frac{\left<n(n-1)(n-2)\right>_Q}{\left<n\right>^3_Q}\!\!&\!\!=\!\!&\!\!
4.52\left[1-(2.280-0.018n_f)\sqrt{\alpha_s}\right].
\notag
\end{eqnarray}
For instance, for one quark jet produced at the $Z^0$ peak of the $e^+e^-$ annihilation 
($Q=45.6$ GeV), one has $\alpha_s=0.134$. Replacing this value into the previous 
formula for a quark jet multiplicity correlator, one obtains a variation 
from 4.52 (DLA) to 0.83 (MLLA). That is one of the reasons for 
DLA has been known to provide unreliable predictions which should not be compared
with experiments. 
From Fig.\ref{fig:corr3gqdiff}, the 
correlation are observed to be the strongest when particles have the same energy $x_i=x_j$ for fixed
$x_k$ and to decrease when one parton is much harder the others. Indeed, in this region of the phase 
space two competing effects should be satisfied: on one hand, as a consequence of gluon coherence and AO, 
gluon emission angles should decrease and on the other hand, the convergence of the perturbative series 
$k_\perp=x_iE\Theta_i\geq Q_0$ should be guaranteed. That is why, as the collinear cut-off 
parameter $Q_0$ is reached, gluons are emitted at larger angles and destructive interferences 
with previous emissions occur. Moreover, the observable increases for softer partons with 
$x_3$ decreasing, which is for partons less sensitive to the energy balance. In 
Fig.\ref{fig:corr3gqsum} the MLLA correlations increase 
for softer partons, then flatten and decrease as a consequence of soft gluon coherence, 
reproducing for three-particle correlations, the hump-backed shape of the one-particle
distribution. 
Because of the limitation of the phase space, one has ${\cal C}^{(3)}\leq1$ for 
harder partons. Finally, in Fig.\ref{fig:corr3gq3sum}, we display the three-particle
correlators as function of the sum $|\ln(x_1x_2x_3)|$, for $x_1=x_2=x_3$; when 
compared with Fig.\ref{fig:corr3gqsum} and Fig.\ref{fig:corr3gqdiff}, the correlators 
are shown to be larger. That is why, and as expected, the correlations are the strongest
for particles having the same energy-momentum $x_1=x_2=x_3$. In these figures, 
the MLLA hump approximation is seen to become larger than the DLA correlator for
smaller values of $x$ than those close to the hump region, which is unphysical. This is
due to the fact that this approximation should not be trusted beyond the hump region
$\mid\ell-Y/2\mid\ll\sigma\propto Y^{3/2}$, $3Y/2=11.25$ in this case.

The MLLA hump approximation from subsection \ref{subsec:humpapp} is observed to be larger 
than the MLLA solution from the steepest descent of the one-particle distribution but one 
should bear in mind that this is only an approximation made for the sake of clarity in the
interpretation of the solutions. In particular, from Fig.\ref{fig:corr3gqdiff} one can 
observe a smoother descent for the slope of the correlators in this case than that 
given from the more exact steepest descent. This difference comes from the role played 
by the iterative corrections displayed in Fig.\ref{fig:epsilon1}, which decrease the 
correlators away from the hump region when one of the partons becomes harder than 
the others. Near the maximum $x_i=x_j$ of the correlators, the difference between the two approaches is
${\cal O}\left(\frac{\ell_k^2}{Y^2}\gamma_0\right)$ and should decrease for $x_i\to1$, 
according to (\ref{eq:humpepsilon1}). 
\subsection{Predictions beyond the limiting spectrum $\boldsymbol{\lambda\ne0}$}
\begin{figure}
\begin{center}
\epsfig{file=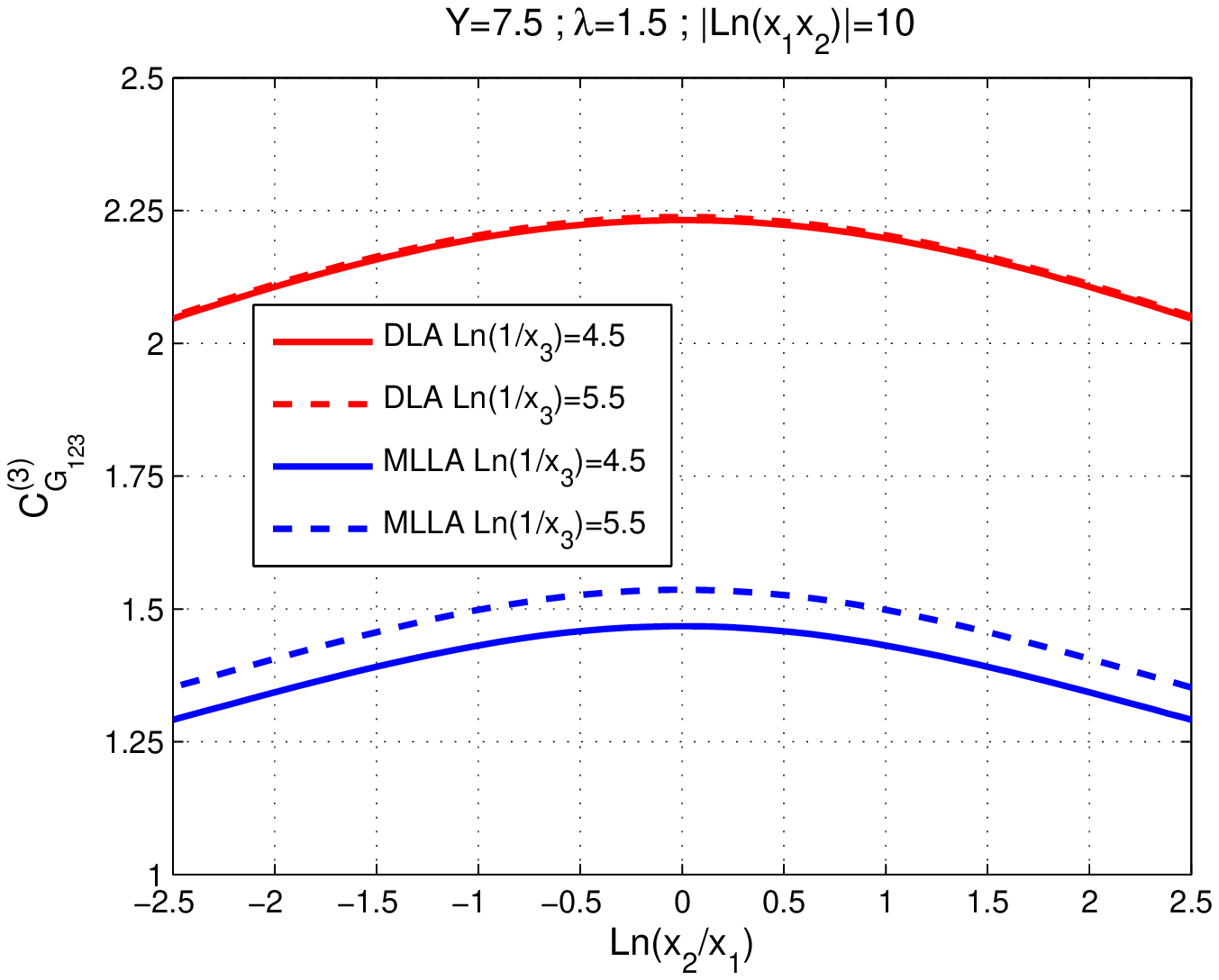, height=6.5truecm,width=7.5truecm}
\epsfig{file=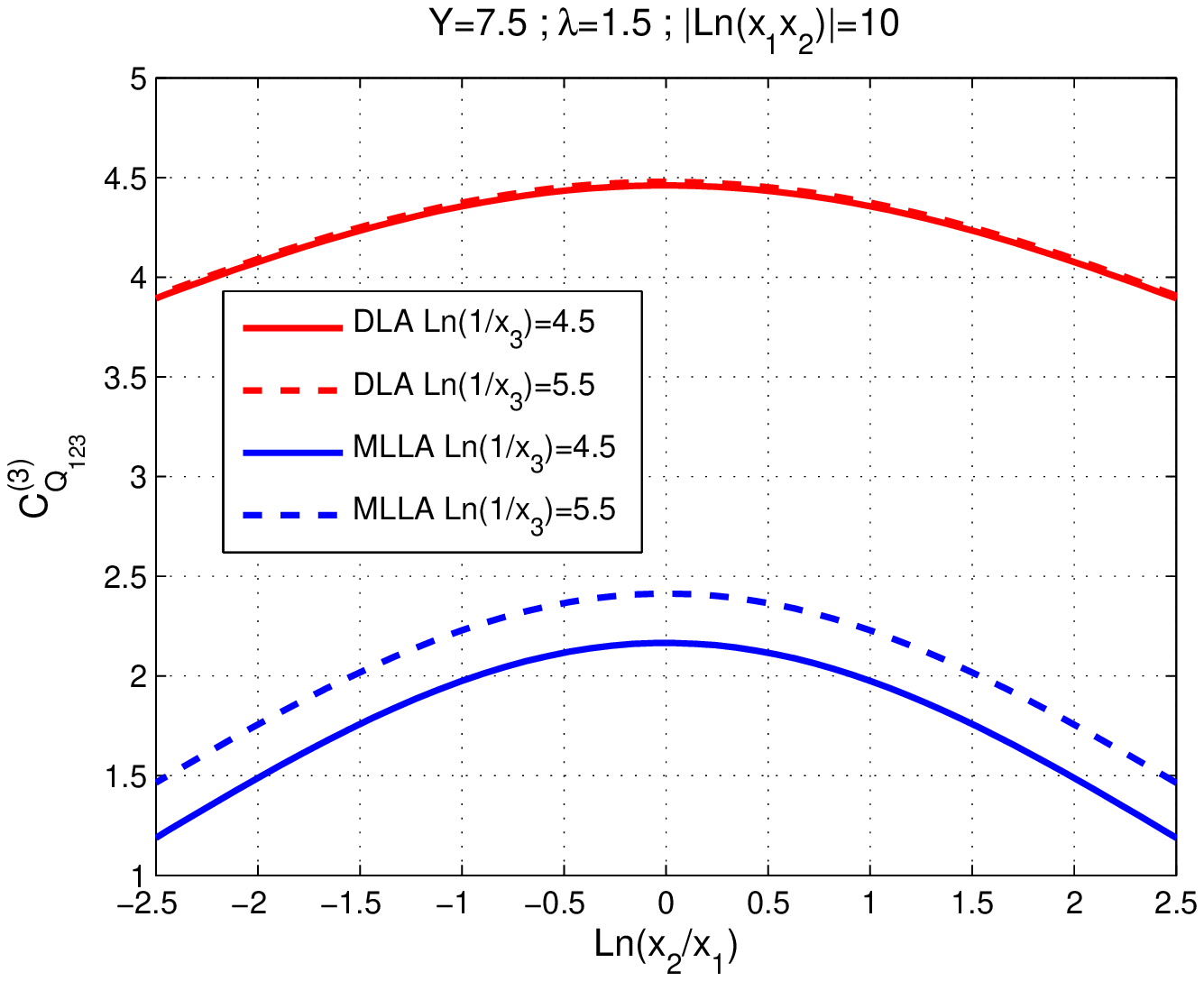, height=6.5truecm,width=7.5truecm}
\caption{\label{fig:corr3gqgdiffB} Three-particle correlations inside a gluon jet (left) and
a quark jet (right) as a function of $\ell_1-\ell_2=\ln(x_2/x_1)$ for $\ell_1+\ell_2=|\ln(x_1x_2)|=10$, 
$\ell_3=\ln(1/x_3)=4.5,\,5.5$, fixed $Y=7.5$ in the limiting spectrum approximation $\lambda=1.5$.}  
\end{center}
\end{figure}
\begin{figure}
\begin{center}
\epsfig{file=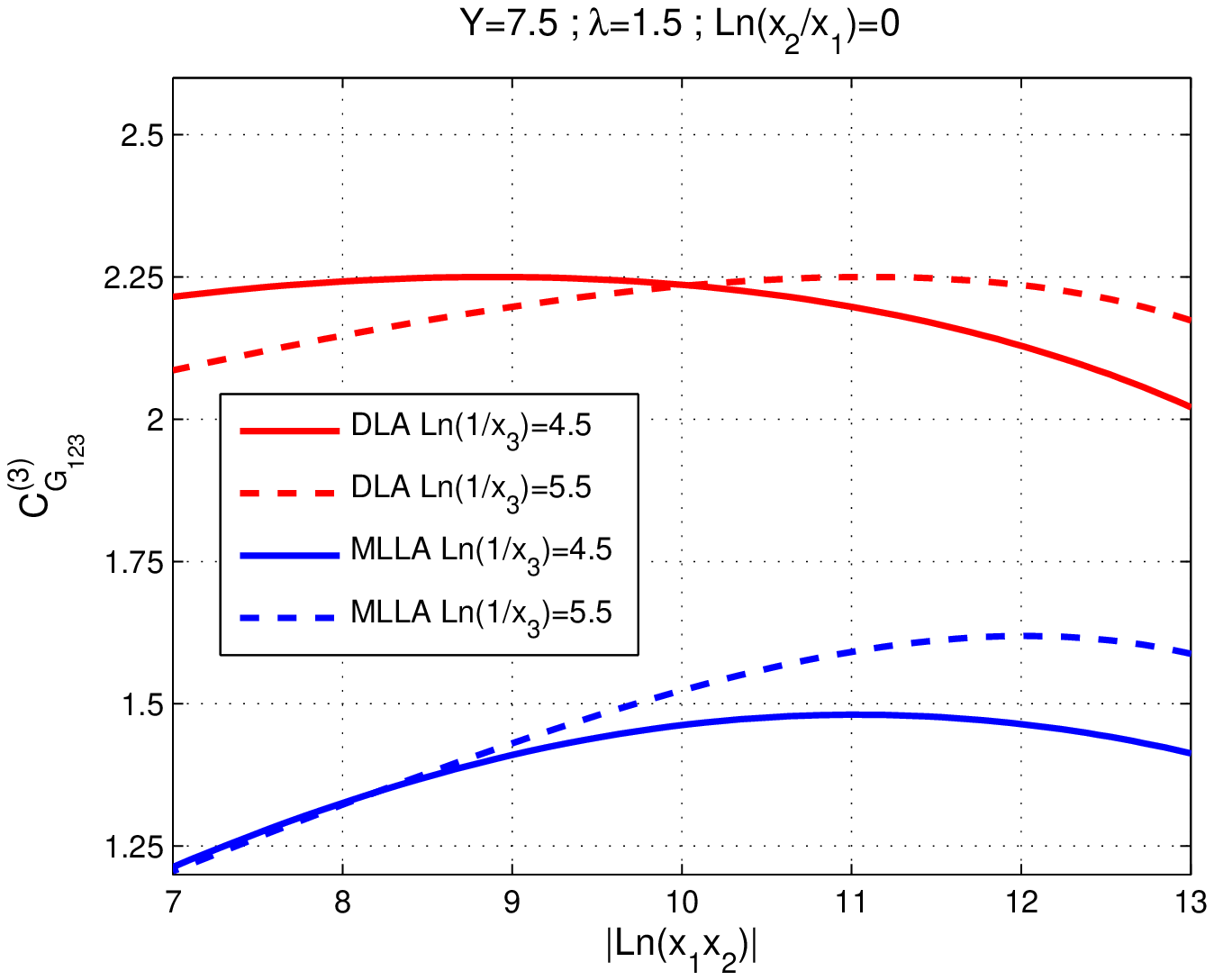, height=6.5truecm,width=7.5truecm}
\epsfig{file=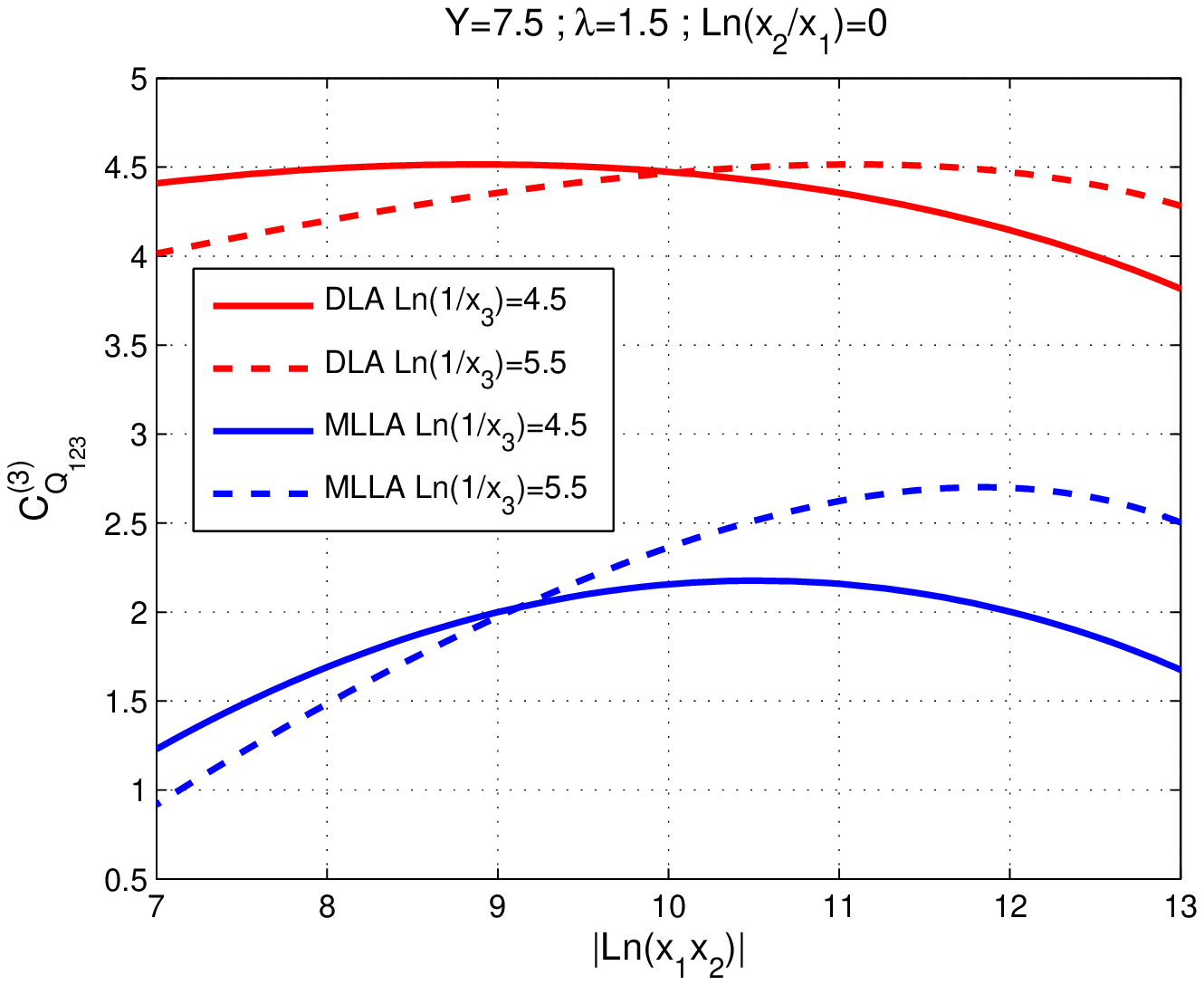, height=6.5truecm,width=7.5truecm}
\caption{\label{fig:corr3gqsumB} Three-particle correlations inside a gluon jet (left) and
a quark jet (right) as a function of $\ell_1+\ell_2=|\ln(x_1x_2)|$ for $x_1=x_2$, 
$\ell_3=\ln(1/x_3)=4.5,\,5.5$, fixed $Y=7.5$ in the limiting spectrum approximation $\lambda=1.5$.}  
\end{center}
\end{figure}

The approximated evaluation of the one-particle distribution from the steepest descent method made 
possible the evaluation of the two-particle correlations beyond the limiting spectrum 
approximation, that is for $Q_0\ne\Lambda_{QCD}$. Accordingly, it makes also possible the 
evaluation of the three-particle correlators ${\cal C}^{(3)}_{G_{123}}(\ell_1,\ell_2,\ell_3,Y)$ 
and ${\cal C}^{(3)}_{Q_{123}}(\ell_1,\ell_2,\ell_3,Y)$ beyond this limit $\lambda\ne0$. This parameter, 
also known as hadronization parameter, guarantees in particular the convergence of the perturbative 
approach $\alpha_s\ll1$. In Fig.\ref{fig:corr3gqgdiffB} and Fig.\ref{fig:corr3gqsumB} we display the same set of 
curves beyond the limiting spectrum ($\lambda=1.5$) as in Fig.\ref{fig:corr3gqdiff} and 
Fig.\ref{fig:corr3gqsum} in the limiting spectrum ($\lambda\sim0$), with the 
exception of curves coming from the hump approximation. The value of $\lambda$
in this case was evaluated for $Q_0\sim1$ GeV, which corresponds to the proton mass, and 
$\Lambda_{QCD}=250$ MeV. As observed the correlation increases with $\lambda$
and the range where ${\cal C}^{(3)}\geq1$ becomes larger in this case.

\section{Conclusions}

In this paper we provide the first full pQCD treatment of three-particle
correlations in parton showers and a further refined test of the LPHD within the
limiting spectrum approximation and beyond. The evolution equations
satisfied by this differential observable have been
obtained for the first time and the differential
version of the equations has been solved iteratively. It has been possible to interpret the analytical
solution in terms of Feynman diagrams describing the process and to evaluate it
from the steepest descent method applied to the single inclusive distribution. 
The correlations have been displayed in the range $x\lesssim0.014$, where the process is 
dominated by three particles emitted from the same partonic cascade following the QCD AO described
in Fig.\ref{fig:three-part} and Fig.\ref{fig:physpic3}d. Furthermore, four-particle correlations have been computed at DLA
so as to show that the inclusion of higher order corrections for more than three particles would
rather be a cumbersome task. The correlations have been shown to
be strongest for the softest hadrons having the same energy $x_1=x_2=x_3$ in both quark and gluon jets,
increasing as a function of $\ln(x_i/x_j)$ and $|\ln(x_ix_j)|$ when $x_k$ softens, that is for partons
being less sensitive to the energy balance. 

Coherence effects appear when one or two of the partons involved in the process 
is harder than the others, thus reproducing for this observable the hump-backed shape of the
one particle distribution. Away from the maximum at $x_i=x_i$, because of limitation of the
phase space, one has ${\cal C}^{(3)}\leq1$. Predictions beyond the limiting spectrum for heavier charged hadrons as 
compared with pions and kaons show that the correlations should increase as the parameter 
$Q_0$ equals the mass of such hadrons and the range where ${\cal C}^{(3)}\geq1$ has been 
enlarged beyond this limit. The last statement is not surprising because
soft gluon emission gets suppressed between the two scales $Q_0$ and $\Lambda_{QCD}$ for 
$\lambda\ne0$, thus decreasing the particle yield inside the whole jet. This measurement would in particular
provide an additional and independent check of the LPHD for massive charged hadrons.
As was shown in \ref{subsec:dlasol}, the DLA solution of the evolution equations provide general features of the 
observable showing its unreliability to be compared with the experiment. That is why, the MLLA shape and overall 
normalization of this observable should be compared with the data. In the case of $p\bar p$ collisions at the
Tevatron, since diet events consist of both gluon and quark jets, in order
to compare data to theory, a parameter $f_g$ for mixed samples of quark and gluon jets was chosen \cite{:2008ec}.
In $pp$ collisions at the LHC, the same procedure can be applied so as to measure the two- and
three-particle correlations. Furthermore, MLLA corrections have been shown to be larger for three 
than for two particles, that is to increase as the number of particles increases. 

As was the case for two particles, the three-particle correlations are larger inside a quark than in a gluon 
jet. Same trends have been observed in HERA and LEP data for soft multi-particle fluctuations in 
\cite{Chekanov:2001sj,Abbiendi:2006qr}.

Finally, we give the first analytical predictions for intra-jet
three-particle correlations in view of forthcoming measurements by ATLAS, CMS and ALICE at the LHC. 

\section*{Acknowledgements}

We gratefully acknowledge enlightening discussions with W. Ochs and 
E. Sarkisyan-Grinbau as well as 
support from Generalitat Valenciana under grant PROMETEO/2008/004
and M.A.S. from FPA2008-02878 and GVPROMETEO2010-056. V.M acklowledges support from
the grant HadronPhysics2, a FP7-Integrating 
Activities and Infrastructure Program of the
European Commission under Grant 227431, by UE (Feder) and the MICINN 
(Spain) grant FPA 2010-21750-C02-01.

\appendix
\section{MLLA approximation}
\label{appendix:taylorseries}
In (\ref{eq:correl3q}), for $\ln(1-z)\ll\ln x$ and $\ln z\ll\ln x$, 
we perform the following Taylor expansions:
\begin{equation}
Q^{(3)}(1-z)-Q^{(3)}\approx\ln(1-z)\frac{dQ^{(3)}}{d\ell_1}+{\cal O}(\alpha_s),
\end{equation}
\begin{eqnarray}
&&\left(Q^{(2)}_{ij}(1-z)-Q^{(2)}_{ij}\right)(G_k(z)-Q_k)
+\left(G^{(2)}_{ij}(z)-Q^{(2)}_{ij}\right)\left(Q_k(1-z)-Q_k\right)\cr
&&=\ln(1-z)\left[\frac{dQ^{(2)}_{ij}}{d\ell_1}(G_k-Q_k)
+\left(G^{(2)}_{ij}-Q^{(2)}_{ij}\right)\frac{dQ_k}{d\ell_1}\right]+{\cal O}(\alpha_s),
\end{eqnarray}
\begin{equation}
\left(Q_i-G_i(z)\right)\left(Q_j(1-z)-Q_j\right)Q_k\approx\ln(1-z)(Q_i-G_i)\frac{dQ_j}{d\ell_1}Q_k+{\cal O}(\alpha_s).
\end{equation}
Since none of these terms contribute to MLLA ${\cal O}(\sqrt{\alpha_s})$, they will be dropped hereafter. In equation 
(\ref{eq:correl3g}), we perform the following approximations in the hard fragmentation region,
\begin{equation}\label{eq:tranfs1}
\left(G^{(2)}_{ij}(z)-G^{(2)}_{ij}\right)\left(G_k(1-z)-G_k\right)\approx
\ln z\ln(1-z)\frac{dG^{(2)}_{ij}}{d\ell_1}\frac{G_k}{d\ell_1}+{\cal O}(\alpha_s),
\end{equation}
\begin{equation}\label{eq:tranfs2}
(G_i-G_i(z))(G_j(1-z)-G_j)G_k\approx-\ln z\ln(1-z)\frac{dG_i}{d\ell_1}\frac{dG_j}{d\ell_1}G_k+{\cal O}(\alpha_s).
\end{equation}
Neither (\ref{eq:tranfs1}) nor (\ref{eq:tranfs2}) contribute to MLLA. The other terms in (\ref{eq:correl3g}) can be written as,
\begin{equation}\label{eq:tranfs3}
2Q^{(3)}(z)-G^{(3)}\approx(2Q^{(3)}-G^{(3)})+2\ln z\frac{dQ^{(3)}}{d\ell_1}+{\cal O}(\alpha_s)
\end{equation}
\begin{eqnarray}\label{eq:tranfs4}
&&2\left(Q^{(2)}_{ij}(z)-G^{(2)}_{ij}\right)(Q_k(1-z)-G_k)\approx
2\left(Q^{(2)}_{ij}-G^{(2)}_{ij}\right)(Q_k-G_k)\\
&&+2\ln(1-z)\left(Q^{(2)}_{ij}-G^{(2)}_{ij}\right)\frac{dQ_k}{d\ell_1}
+2\ln z(Q_k-G_k)\frac{dQ^{(2)}_{ij}}{d\ell_1}+{\cal O}(\alpha_s),\notag
\end{eqnarray}
\begin{equation}\label{eq:tranfs5}
(2Q_i(z)Q_j(z)-G_iG_j)G_k\approx (2Q_iQ_j-G_iG_j)G_k+\ln z\left(Q_i\frac{dQ_j}{d\ell_1}
+\frac{dQ_i}{d\ell_1}Q_j\right)+{\cal O}(\alpha_s),
\end{equation}
\begin{eqnarray}
(G_i-2Q_i(z))(2Q_j(1-z)-G_j)G_k\!\!&\!\!\approx\!\!&\!\!(G_i-2Q_i)(2Q_j-G_j)G_k-2(2Q_j-G_j)G_k\ln z\frac{dQ_i}{d\ell_1}G_k\cr
\!\!&\!\!+\!\!&\!\!2(G_i-2Q_i)\ln(1-z)\frac{dQ_j}{d\ell_1}G_k+{\cal O}(\alpha_s),
\label{eq:tranfs6}
\end{eqnarray}
such that only the first terms in (\ref{eq:tranfs3}), (\ref{eq:tranfs4}), (\ref{eq:tranfs5})
and (\ref{eq:tranfs6}) will be kept in the following. 
Furthermore, we make use of the identity \cite{Ramos:2006dx}
$$
\int^1dz\Phi_g^g(z)\left(G^{(3)}(z)-zG^{(3)}\right)=\int^1dz(1-z)\Phi_g^g(z)\left(G^{(3)}(z)+
\left(G^{(3)}(z)-G^{(3)}\right)\right),
$$
such that $G^{(n)}(z)-zG^{(n)}$ can be replaced by,
$$
G^{(n)}(z)-zG^{(n)}\to (1-z)\left[G^{(n)}(z)+\left(G^{(n)}(z)-G^{(n)}\right)\right]
\approx(1-z)\left[G^{(n)}(z)+\ln z\frac{dG^{(n)}}{d\ell_1}\right],
$$
($n=1,2,3$) in the r.h.s. of equations (\ref{eq:gpr}), (\ref{eq:G2prsub}) and (\ref{eq:correl3g}). 
Indeed, terms $\propto\ln z,\ln(1-z)$ provide NMLLA corrections ${\cal O}(\alpha_s)$ which 
improve energy conservation; however, their inclusion goes beyond the scope of the present paper.
\subsection{One and two particle distributions at small $\boldsymbol{x}$}
\label{subsec:oneandtwopartdist}

The MLLA integro-differential version of equations 
(\ref{eq:qpr},\ref{eq:gpr}) and (\ref{eq:G2prsub},\ref{eq:Q2prsub}) is obtained
after integrating over the regular part of the splitting functions, 
such that \cite{Dokshitzer:1991wu,Ramos:2006dx,Fong:1990nt}
\begin{eqnarray}
Q_{i,y}\!\!&\!\!=\!\!&\!\!\frac{C_F}{N_c}\int_0^\ell d\ell'
\gamma_0^2(\ell'+y)G_i(\ell',y)-\frac34\frac{C_F}{N_c}\gamma_0^2(\ell+y) G_i(\ell,y),
\label{eq:solq}\\
G_{i,y}\!\!&\!\!=\!\!&\!\!\int_0^{\ell} d\ell'\gamma_0^2(\ell'+y)G_i(\ell',y)-a\gamma_0^2(\ell+y)G_i(\ell,y),
\label{eq:solg}
\end{eqnarray}
with $\gamma_0^2(\ell+y)=\frac{1}{\beta_0(\ell+y+\lambda)}$, 
and the two-particle correlations ($\hat A^{(2)}_{ij}=A^{(2)}_{ij}-A_iA_j$) \cite{Ramos:2006dx,Fong:1990nt},
\begin{eqnarray}
\label{eq:eveeqq}
\hat Q^{(2)}_{ij,y}
\!\!&\!\!=\!\!&\!\!\frac{C_F}{N_c}
\int_0^{\ell_i}d\ell
\gamma_0^2(\ell+y_j)G^{(2)}_{ij}(\ell,y_j,\eta_{ij})-\frac34\frac{C_F}{N_c}\gamma_0^2(\ell_i+y_j)
G^{(2)}_{ij}(\ell_i,y_j,\eta_{ij}),\\
\label{eq:eveeqg}
\hat G^{(2)}_{ij,y}
\!\!&\!\!=\!\!&\!\!\int_0^{\ell_i}\gamma_0^2(\ell+y_j)G^{(2)}(\ell,y_j,\eta_{ij})
- a\gamma_0^2(\ell_i+y_j) G^{(2)}_{ij}(\ell_i,y_j,\eta_{ij})\cr
\!\!&\!\!+\!\!&\!\! (a-b)\gamma_0^2(\ell_i+y_j)
G(\ell_i,y_j+\eta_{ij})G(\ell_i+\eta_{ij},y_j),
\end{eqnarray}
with $\gamma_0^2(\ell_i+y_j)=\frac{1}{\beta_0(\ell_i+y_j+\eta_{ij}+\lambda)}$, 
after accounting for hard corrections ${\cal O}(\sqrt{\alpha_s})$.
After differentiating (\ref{eq:solq},\ref{eq:solg}) and (\ref{eq:eveeqq},\ref{eq:eveeqg}) with respect to $``\ell"$, 
one has  \cite{Ramos:2006dx}
\begin{eqnarray}\label{eq:qdiffeqspec}
Q_{i,\ell y}\!\!&\!\!=\!\!&\!\!\frac{C_F}{N_c}\gamma_0^2G_i-\frac{C_F}{N_c}\frac34\gamma_0^2(G_{i,\ell}-\beta_0\gamma_0^2G_i),\\
G_{i,\ell y}\!\!&\!\!=\!\!&\!\!\gamma_0^2G_i-a\gamma_0^2(G_{i,\ell}-\beta_0\gamma_0^2G_i),
\label{eq:gdiffeqspec}
\end{eqnarray}
from where the following useful relations hold in MLLA \cite{Ramos:2006dx},
\begin{eqnarray}
\frac{Q_{i,\ell y}}{\gamma_0^2Q_i}\!\!&\!\!=\!\!&\!\!
\left[1-\frac34\psi_{i,\ell}\right]\frac{C_F}{N_c}\frac{G_i}{Q_i}+{\cal O}(\gamma_0^2),\\
\label{eq:ratioGQ}
\frac{G_i}{Q_i}\!\!&\!\!=\!\!&\!\!\frac{N_c}{C_F}\left[1-\left(a-\frac34\right)\psi_{i,\ell}\right]+{\cal O}(\gamma_0^2),\\
\frac{Q_{i,\ell y}}{\gamma_0^2Q_i}\!\!&\!\!=\!\!&\!\!1-a\psi_{i,\ell}+{\cal O}(\gamma_0^2).
\label{eq:QilyQ}
\end{eqnarray}
Corrections $\propto\beta_0$ in (\ref{eq:qdiffeqspec}) and (\ref{eq:gdiffeqspec}), which are NMLLA, 
account for the running of the coupling constant $\alpha_s$ and those $\propto \frac34,a,(a-b)$ account 
for energy conservation in the hard parton splitting region.
The MLLA gluon inclusive spectrum is given by the solution of (\ref{eq:gdiffeqspec}) 
\cite{Dokshitzer:1991wu} and can be written in the form \cite{PerezRamos:2007cr}:
\begin{equation}
G_i(\ell,y) = 2\ \frac{\Gamma(B)}{\beta_0}\
 \int_0^\frac{\pi}{2}\
  \frac{d\tau}{\pi}\,e^{-B\alpha}
\  {\cal F}_B(\tau,y,\ell),
\label{eq:ifD}
\end{equation}
where the integration
is performed with respect to $\tau$ defined by
$\displaystyle \alpha = \frac{1}{2}\ln\frac{y}{\ell}  + i\tau$ and with
\begin{eqnarray*}
{\cal F}_B(\tau,y,\ell) &=& \left[ \frac{\cosh\alpha
-\displaystyle{\frac{y-\ell}{y+\ell}}
\sinh\alpha} 
 {\displaystyle \frac{\ell +
y}{\beta_0}\,\frac{\alpha}{\sinh\alpha}} \right]^{B/2}
  I_B(2\sqrt{Z(\tau,y,\ell)}), \cr
&& \cr
&& \cr
 Z(\tau,y,\ell) &=&
\frac{\ell + y}{\beta_0}\,
\frac{\alpha}{\sinh\alpha}\,
 \left(\cosh\alpha
%+ (1-2\zeta)
-\frac{y-\ell}{y+\ell}
\sinh\alpha\right),
\label{eq:calFdef}
\end{eqnarray*}
$B=a/\beta_0$ and $I_B$ is the modified Bessel function of the first kind. The formula in (\ref{eq:ifD})
corresponds indeed to the so-called hump-backed plateau, which describes the energy spectrum of soft hadrons
in the limiting spectrum approximation $Q_0=\Lambda_{QCD}$ \cite{Dokshitzer:1991wu,Khoze:1996dn}. This result
is well known and constitutes one of the strikest predictions of pQCD. The corresponding solution
of (\ref{eq:qdiffeqspec}) for $Q_i(\ell,y)$ can be obtained from (\ref{eq:ratioGQ}) with accuracy 
${\cal O}(\sqrt{\alpha_s})$.
The system of differential evolution equations for two-particle correlations follows 
from (\ref{eq:eveeqq}) and (\ref{eq:eveeqg}), such that \cite{Ramos:2006dx}
\begin{eqnarray}\label{eq:diffcorrel2q}
\left[Q^{(2)}_{ij}-Q_iQ_j\right]_{\ell y}\!\!&\!\!=\!\!&\!\!\frac{C_F}{N_c}\gamma_0^2G^{(2)}_{ij}
-\frac34\frac{C_F}{N_c}\gamma_0^2\left(G^{(2)}_{ij,\ell}-\beta_0\gamma_0^2G^{(2)}_{ij}\right),\\
\left[G^{(2)}_{ij}-G_iG_j\right]_{\ell y}\!\!&\!\!=\!\!&\!\!\gamma_0^2G^{(2)}_{ij}
\!-\!a\gamma_0^2\left(G^{(2)}_{ij,\ell}\!-\!\beta_0\gamma_0^2G^{(2)}_{ij}\right)\!+\!(a-b)\gamma_0^2
\left[\left(G_iG_j\right)_\ell-\beta_0\gamma_0^2G_iG_j\right].\label{eq:diffcorrel2g}
\end{eqnarray}
In \cite{Ramos:2006dx}, the system (\ref{eq:eveeqq},\ref{eq:eveeqg}) was solved iteratively 
after replacing $G^{(2)}_{ij}=C^{(2)}_{G,ij}G_iG_j$ and $Q^{(2)}_{ij}=C^{(2)}_{Q,ij}Q_iQ_j$
in (\ref{eq:diffcorrel2g}) and (\ref{eq:diffcorrel2q}) respectively. 
The MLLA solutions of (\ref{eq:diffcorrel2q}) and (\ref{eq:diffcorrel2g}), 
which are to be used in the present paper read \cite{Ramos:2006dx}
\begin{eqnarray}
 {\cal C}^{(2)}_{G_{ij}} -1\!\!&\!\!=\!\!&\!\!\frac{1 -\delta_1^{ij} -b\left(\psi_{i,\ell} 
+\psi_{j,\ell}\right)}
{1+ \Delta_{ij}
+\delta_1^{ij}}
\label{eq:CGfull},\\
\frac{ {\cal C}^{(2)}_{Q_{ij}} -1}{ {\cal C}^{(2)}_{G_{ij}} -1}\!\!&\!\!=\!\!&\!\!
\frac{N_c}{C_F}\left[1+(b-a)(\psi_{i,\ell} +\psi_{j,\ell})\frac{1+\Delta_{ij}}{2+\Delta_{ij}}\right],
\label{eq:CQfull}
\end{eqnarray}
which were evaluated by the steepest descent method over the single inclusive distribution 
in \cite{Ramos:2006mk}. We have introduced the following notations and functions \cite{Ramos:2006dx},
\begin{eqnarray}
&&\Delta_{ij}=\gamma_0^{-2}\left(\psi_{i,\ell}\psi_{j,y}+\psi_{i,y}\psi_{j,\ell}\right)={\cal O}(1);\\
\label{eq:nota4bis}
&& \quad \chi^{ij} =  \ln \dot{{\cal C}}_{G_{ij}}^{(2)}={\cal O}(1),\quad
\chi_{\ell}^{ij} = \frac{\partial\chi^{ij}}{\partial\ell}={\cal O}(\gamma_0^2),\qquad 
\chi_{y}=\frac{\partial\chi^{ij}}{\partial y}={\cal O}(\gamma_0^2);\\
&&\delta_1^{ij} = \gamma_0^{-2}\Big[\chi_{\ell}^{ij}(\psi_{i,y}+\psi_{j,y}) +
   \chi_{y}^{ij}(\psi_{j,\ell}+\psi_{i,\ell})\Big]={\cal O}(\gamma_0)\label{eq:delta1},
\end{eqnarray}
where, following from (\ref{eq:ordermag1}) and (\ref{eq:ordermag2}), we have evaluated the 
corresponding order of magnitude of these quantities in powers
of the anomalous dimension $\gamma_0\propto\sqrt{\alpha_s}$. The solution is iterative 
with respect to corrections $\chi$ and $\delta_1$, which need the prior evaluation of the
DLA solution $\dot{{\cal C}}_{G_{ij}}^{(2)}$ of the equations.

\section{Iterative solution of the evolution equations}
\label{appendix:iteratsolgq}
Let us first solve the equation (\ref{eq:mlla3correlgdiff}). For the sake of simplicity,
it is much easier to solve the equivalent equation:
\begin{eqnarray}
\label{eq:otherG3}
\hat{G}_{\ell y}^{(3)}\!\!&\!\!=\!\!&\!\!\gamma_0^2G^{(3)}
\!-\!a\gamma_0^2\left(G^{(3)}_\ell\!-\!\beta_0\gamma_0^2G^{(3)}\right)\!+\!(a-b)\gamma_0^2\left\{
\left[G^{(2)}_{12}G_3+G^{(2)}_{13}G_2+G^{(2)}_{23}G_1\right]_\ell\right.\\
\!\!&\!\!-\!\!&\!\!\left.\beta_0\gamma_0^2\left[G^{(2)}_{12}G_3+G^{(2)}_{13}G_2+G^{(2)}_{23}G_1\right]\right\}+
(2a-3b+c)\gamma_0^2\left[(G_1G_2G_3)_\ell\!-\!\beta_0\gamma_0^2G_1G_2G_3\right].\notag
\end{eqnarray}
One has to substitute the following in the l.h.s. of the equation (\ref{eq:otherG3}):
$$
G^{(3)}={\cal C}^{(3)}_{G_{123}}G_1G_2G_3,\quad G^{(2)}_{ij}={\cal C}^{(2)}_{G_{ij}}G_iG_j.
$$
Thus, after normalizing by $\gamma_0^2G_1G_2G_3$, one finds,
\begin{eqnarray}\label{eq:rhswithepsilons}
\frac{\left[({\cal C}^{(3)}_{G_{123}}-1)G_1G_2G_3\right]_{\ell y}}{\gamma_0^2G_1G_2G_3}\!\!&\!\!=\!\!&\!\!
{\cal C}^{(3)}_{G_{123}}(\epsilon_1+\epsilon_2)+({\cal C}^{(3)}_{G_{123}}-1)
\left[3+\Delta_{12}+\Delta_{13}+\Delta_{23}\right.\\
\!\!&\!\!-\!\!&\!\!\left.a(\psi_{1,\ell}+\psi_{2,\ell}+\psi_{3,\ell})+3a\beta_0\gamma_0^2\right],\notag
\end{eqnarray}
while for the other terms in the r.h.s. of the same equation one finds,
\begin{eqnarray}
\frac{\left[({\cal C}^{(2)}_{G_{ij}}-1)G_1G_2G_3\right]_{\ell y}}{\gamma_0^2G_1G_2G_3}
\!\!&\!\!=\!\!&\!\!({\cal C}^{(2)}_{G_{ij}}-1)\!\left(\!\sum_{i=1}^{3}\frac{G_{i,\ell y}}{\gamma_0^2G_i}
+\Delta_{12}+\Delta_{13}+\Delta_{23}\right)\!\!+{\cal C}^{(2)}_{G_{ij}}\xi_1^{ij}+
{\cal C}^{(2)}_{G_{ij}}\delta_2^{ij}\cr
\!\!&\!\!=\!\!&\!\!({\cal C}^{(2)}_{G_{ij}}-1)\left[3+\Delta_{12}+\Delta_{13}+\Delta_{23}-
a(\psi_{1,\ell}+\psi_{2,\ell}+\psi_{3,\ell})+3a\beta_0\gamma_0^2\right.\cr
\!\!&\!\!+\!\!&\!\!\left.\xi_1^{ij}+\delta_2^{ij}\right]
+\xi_1^{ij}+\delta_2^{ij}.\label{eq:lhsG3}
\end{eqnarray}
The r.h.s. provides the following contribution
\begin{eqnarray}
\frac{r.h.s.}{\gamma_0^2G_1G_2G_3}\!\!&\!\!=\!\!&\!\!{\cal C}^{(3)}_{G_{123}}-a{\cal C}^{(3)}_{G_{123}}
(\psi_{1,\ell}+\psi_{2,\ell}+\psi_{3,\ell}+\zeta_\ell-\beta_0\gamma_0^2)+(a-b)
\left[{\cal C}^{(2)}_{G_{12}}(\chi_\ell^{12}+\psi_{1,\ell}+\psi_{2,\ell}\right.\cr
\!\!&\!\!+\!\!&\!\!\left.\psi_{3,\ell})+{\cal C}^{(2)}_{G_{13}}(\chi_\ell^{13}+\psi_{1,\ell}
+\psi_{2,\ell}+\psi_{3,\ell})+{\cal C}^{(2)}_{G_{23}}(\chi_\ell^{23}+\psi_{1,\ell}
+\psi_{2,\ell}+\psi_{3,\ell})\right.\cr
\!\!&\!\!-\!\!&\!\!\left.\beta_0\gamma_0^2({\cal C}^{(2)}_{G_{12}}+{\cal C}^{(2)}_{G_{13}}
+{\cal C}^{(2)}_{G_{23}})\right]+(3b-2a-c)(\psi_{1,\ell}
+\psi_{2,\ell}+\psi_{3,\ell}-\beta_0\gamma_0^2).\label{eq:lhsQ3}
\end{eqnarray}
After adding (\ref{eq:rhswithepsilons}) and (\ref{eq:lhsG3}) and equating with 
(\ref{eq:lhsQ3}) together with some algebra in between, one finds the 
solution written in (\ref{eq:sol3partg}). Following the same iterative procedure 
$$
Q^{(3)}={\cal C}^{(3)}_{Q_{123}}Q_1Q_2Q_3,\quad Q^{(2)}_{ij}={\cal C}^{(2)}_{Q_{ij}}Q_iQ_j;
\quad G^{(3)}={\cal C}^{(3)}_{G_{123}}G_1G_2G_3,
$$
for the quark jet evolution equation written in (\ref{eq:mlla3correlqdiff}), one has,
\begin{eqnarray}\label{eq:procedquark}
&&\left({\cal C}^{(3)}_{Q_{123}}-1\right)
\left(\tilde\Delta_{12}+\tilde\Delta_{13}+\tilde\Delta_{23}+\sum_{i=1}^{3}\frac{Q_{i,\ell y}}{\gamma_0^2Q_i}
+\tilde\epsilon_1+\tilde\epsilon_2\right)\\
&&-\left({\cal C}^{(2)}_{Q_{12}}-1\right)\left(\tilde\Delta_{12}+\tilde\Delta_{13}+
\tilde\Delta_{23}+\sum_{i=1}^{3}\frac{Q_{i,\ell y}}{\gamma_0^2Q_i}
+\tilde\xi_1^{12}+\tilde\delta_2^{12}\right)\cr
&&-\left({\cal C}^{(2)}_{Q_{13}}-1\right)\left(\tilde\Delta_{12}+\tilde\Delta_{13}+
\tilde\Delta_{23}+\sum_{i=1}^{3}\frac{Q_{i,\ell y}}{\gamma_0^2Q_i}
+\tilde\xi_1^{13}+\tilde\delta_2^{13}\right)\cr
&&-\left({\cal C}^{(2)}_{Q_{23}}-1\right)\left(\tilde\Delta_{12}+\tilde\Delta_{13}+
\tilde\Delta_{23}+\sum_{i=1}^{3}\frac{Q_{i,\ell y}}{\gamma_0^2Q_i}
+\tilde\xi_1^{23}+\tilde\delta_2^{23}\right)\cr
&&=\frac{C_F}{N_c}{\cal C}^{(3)}_{G_{123}}\left[1-\frac34(\psi_{1,\ell}
+\psi_{2,\ell}+\psi_{3,\ell}+\zeta_\ell-\beta_0\gamma_0^2)\right]\frac{G_1G_2G_3}{Q_1Q_2Q_3}\cr
&&+(\tilde\xi_1^{12}+\tilde\delta_2^{12})+(\tilde\xi_1^{13}+\tilde\delta_2^{13})+(\tilde\xi_1^{23}+
\tilde\delta_2^{23})-\tilde\epsilon_1-\tilde\epsilon_2.\notag
\end{eqnarray}
Finally by adding and subtracting $(\tilde\epsilon_1+\tilde\epsilon_2)$ in every term 
$\propto\left({\cal C}^{(2)}_{Q_{ij}}-1\right)$ in the l.h.s. of (\ref{eq:procedquark}) one 
finds (\ref{eq:sol3partq}).
\section{Steepest descent evaluation: reminder from \cite{Ramos:2006mk}}
\label{appendix:steepdesceval}
The evaluation of the integral representation by the steepest descent method at small 
$x\ll1$ (or large $\ell\gg1$) and very high energy $Y\gg1$ leads to the result,
\begin{equation}\label{eq:speepG}
G(\ell,y)\approx{\cal N}(\mu,\nu,\lambda)\exp\left[\frac2{\beta_0}
\left(\sqrt{\ell+y+\lambda}-\sqrt{\lambda}\right)\frac{\mu-\nu}{\sinh\mu-\sinh\nu}+\nu-
\frac{a}{\beta_0}(\mu-\nu)\right],
\end{equation}
where
$$
{\cal N}(\mu,\nu,\lambda)=\frac12(\ell+y+\lambda)
\frac{\left(\frac{\beta_0}{\lambda}\right)^{1/4}}{\sqrt{\pi\cosh\nu Det A(\mu,\nu)}},
$$
with
$$
Det A(\mu,\nu)=\beta_0(\ell+y+\lambda)^3\left[\frac{(\mu-\nu)\cosh\mu\cosh\nu
+\cosh\mu\sinh\nu-\sinh\mu\sinh\nu}{\sinh^3\mu\cosh\nu}\right].
$$
The logarithmic derivatives of the spectrum given in (\ref{eq:psiell}) and (\ref{eq:psiy}) were derived 
from (\ref{eq:speepG}) and it was also shown that (\ref{eq:speepG}) reproduces the Gaussian shape of the
inclusive distribution near the hump $\ell_{max}\approx Y/2$. From (\ref{eq:speepG}), one has indeed,
\begin{equation}
G(\ell,y)\approx\left(\frac3{\pi\sqrt{\beta_0}[(\ell+y+\lambda)^{3/2}-\lambda^{3/2}]}\right)^{1/2}
\exp\left(-\frac2{\sqrt{\beta_0}}\frac{3}{(\ell+y+\lambda)^{3/2}-\lambda^{3/2}}
\frac{(\ell-Y/2)^2}{2}\right),
\end{equation}
where the MLLA $\ell_{max}$ reads,
$$
\ell_{max}\approx\frac{Y}2+\frac12\frac{a}{\beta_0}\left(\sqrt{Y+\lambda}-\sqrt{\lambda}\right).
$$
Setting $a=0$ and $\lambda=0$ in the previous expressions one recovers the DLA results,
which are needed for subsection \ref{subsec:dlasol}.
The functions entering as a function of ($\mu,\nu$) in (\ref{eq:psiell}) and (\ref{eq:psiy}) 
are the following,
\begin{eqnarray}
\tilde Q(\mu,\nu)\!\!&\!\!=\!\!&\!\!\frac{\cosh\mu\sinh\mu\cosh\nu-(\mu-\nu)\cosh\nu-\sinh\nu}
{(\mu-\nu)\cosh\mu\cosh\nu+\cosh\mu\sinh\nu-\sinh\mu\cosh\nu},\\
K(\mu,\nu)\!\!&\!\!=\!\!&\!\!-\frac12\sinh\nu\frac{(\mu-\nu)\cosh\mu-\sinh\mu}
{(\mu-\nu)\cosh\mu\cosh\nu+\cosh\mu\sinh\nu-\sinh\mu\cosh\nu},\\
L(\mu,\nu)\!\!&\!\!=\!\!&\!\!\frac32\coth\mu-\frac12\frac{(\mu-\nu)\cosh\nu\sinh\mu+\sinh\nu\sinh\mu}
{(\mu-\nu)\cosh\mu\cosh\nu+\cosh\mu\sinh\nu-\sinh\mu\cosh\nu},\\
C(\mu,\nu)\!\!&\!\!=\!\!&\!\!L(\mu,\nu)+\tanh\nu\coth\mu\left(1+K(\mu,\nu)\right).
\end{eqnarray}
The expressions for the two particle correlations follow from (\ref{eq:CGfull}) and (\ref{eq:CQfull}) \cite{Ramos:2006mk},
\begin{eqnarray}\label{eq:sol2partcorrstdescg}
{\cal C}_{G_{ij}}^{(2)}\!\!&\!\!=\!\!&\!\!1+\frac{1-b\gamma_0(e^{\mu_i}+e^{\mu_j})-\delta_1^{ij}}
{1+2\cosh(\mu_i-\mu_j)+\Delta'(\mu_i,\nu_i,\mu_j,\nu_j)+\delta_1^{ij}},\\
{\cal C}_{Q_{ij}}^{(2)}\!\!&\!\!=\!\!&\!\!1+\frac{N_c}{C_F}
\left[{\cal C}_{G_{ij}}^{(2)}-1+
\frac12(b-a)\gamma_0\frac{e^{\mu_i}+e^{\mu_j}}{1+\cosh(\mu_i-\mu_j)}\right],
\label{eq:sol2partcorrstdescq}
\end{eqnarray}
where,
\begin{equation}
\delta_1^{ij}=\beta_0\gamma_0\frac{2\sinh^2\left(\frac{\mu_i-\mu_j}{2}\right)}
{3+4\sinh^2\left(\frac{\mu_i-\mu_j}{2}\right)}\Big(\tilde Q(\mu_i,\nu_i)+\tilde Q(\mu_j,\nu_j)\Big),
\end{equation}
and
\begin{eqnarray}
\Delta'(\mu_i,\nu_i,\mu_j,\nu_j)\!\!&\!\!=\!\!&\!\!-a\gamma_0\left[e^{\mu_i}+e^{\mu_j}-\sinh(\mu_i-\mu_j)
(\tilde Q_i-\tilde Q_j)+\cosh\mu_1\tanh\nu_2+\cosh\mu_2\tanh\nu_1\right.\cr
\!\!&\!\!-\!\!&\!\!\left.\sinh\mu_i\tanh\nu_j\coth\mu_j-\sinh\mu_j\tanh\nu_i\coth\mu_i\right.\cr
\!\!&\!\!+\!\!&\!\!\left.\sinh(\mu_i-\mu_j)\Big(\tanh\nu_i\coth\mu_i\tilde Q_i-\tanh\nu_j\coth\mu_j\tilde Q_j\Big)\right]\cr
\!\!&\!\!-\!\!&\!\!\beta_0\gamma_0\!\left[\cosh\mu_i-\sinh\mu_iC_j+\cosh\mu_j-\sinh\mu_jC_i+
\sinh(\mu_i-\mu_j)(C_i\tilde Q_i-C_j\tilde Q_j)\right.\cr
\!\!&\!\!+\!\!&\!\!\left.\cosh\mu_i\tanh\nu_j(1+K_j)+\cosh\mu_j\tanh\nu_i(1+K_i)\right].
\end{eqnarray}
The solutions (\ref{eq:sol2partcorrstdescg}) and (\ref{eq:sol2partcorrstdescq}) are the ones to be
used in this paper for the evaluations of the three-particle correlations and will be directly 
inserted in the solutions (\ref{eq:sol3partg}) and (\ref{eq:sol3partq}) respectively.
\subsection{Corrections $\boldsymbol{\xi_1^{ij}},\boldsymbol{\tilde\xi_1^{ij}}$ 
\label{subappex:corrxis}
and $\boldsymbol{\epsilon_1},\boldsymbol{\tilde\epsilon_1}$}
For the computation of these corrections, one only needs to take the DLA part of the logarithmic derivatives
of the one-particle distribution $\psi_{i,\ell}=\gamma_0e^{\mu_i}$ and $\psi_{i,y}=\gamma_0e^{-\mu_i}$, such 
that after replacement in (\ref{eq:xiell}) and (\ref{eq:tildexiell}) one finds,
\begin{eqnarray}
\xi_1^{ij}&=&\frac1{\gamma_0}\left[\chi_{\ell}^{ij}\left(e^{-\mu_1}+e^{-\mu_2}+e^{-\mu_3}\right)+
\chi_y^{ij}\left(e^{\mu_1}+e^{\mu_2}+e^{\mu_3}\right)\right],\\
\tilde\xi_1^{ij}&=&\frac1{\gamma_0}\left[\tilde\chi_{\ell}^{ij}\left(e^{-\mu_1}+e^{-\mu_2}+e^{-\mu_3}\right)+
\tilde\chi_y^{ij}\left(e^{\mu_1}+e^{\mu_2}+e^{\mu_3}\right)\right],
\end{eqnarray}
where
\begin{equation}
\chi_{\ell}^{ij}=\beta_0\gamma_0^2\frac{\tanh\frac{\mu_i-\mu_j}{2}}{1+2\cosh(\mu_i-\mu_j)}
\frac{e^{\mu_i}\tilde Q_i-e^{\mu_j}\tilde Q_j}{2},\quad\tilde\chi_\ell^{ij}=-\frac{N_c}{C_F}
\frac{\dot{{\cal C}}_{G_{ij}}^{(2)}}{\dot{{\cal C}}_{Q_{ij}}^{(2)}}\chi_{\ell}^{ij},
\end{equation}
\begin{equation}
\chi_{y}^{ij}=-\beta_0\gamma_0^2\frac{\tanh\frac{\mu_i-\mu_j}{2}}{1+2\cosh(\mu_i-\mu_j)}
\frac{e^{-\mu_i}\tilde Q_i-e^{-\mu_j}\tilde Q_j}{2},\quad\tilde\chi_y^{ij}=-\frac{N_c}{C_F}
\frac{\dot{{\cal C}}_{G_{ij}}^{(2)}}{\dot{{\cal C}}_{Q_{ij}}^{(2)}}\chi_{y}^{ij};
\end{equation}
with
\begin{equation}
\dot{{\cal C}}_{G_{ij}}^{(2)}=1+\frac1{1+2\cosh(\mu_i-\mu_j)},\quad
\dot{{\cal C}}_{Q_{ij}}^{(2)}=1+\frac{N_c}{C_F}\frac1{1+2\cosh(\mu_i-\mu_j)}.
\end{equation}
Accordingly, replacing $\psi_{i,\ell}=\gamma_0e^{\mu_i}$ and $\psi_{i,y}=\gamma_0e^{-\mu_i}$ in (\ref{eq:epsilon1})
and (\ref{eq:tildeepsilon1}), one has
\begin{eqnarray}
\epsilon_1\!\!&\!\!=\!\!&\!\!\frac1{\gamma_0}\left[\zeta_{\ell}\left(e^{-\mu_1}+e^{-\mu_2}+e^{-\mu_3}\right)+
\zeta_y\left(e^{\mu_1}+e^{\mu_2}+e^{\mu_3}\right)\right],\\
\tilde\epsilon_1\!\!&\!\!=\!\!&\!\!\frac1{\gamma_0}
\left[\tilde\zeta_{\ell}\left(e^{-\mu_1}+e^{-\mu_2}+e^{-\mu_3}\right)+
\tilde\zeta_y\left(e^{\mu_1}+e^{\mu_2}+e^{\mu_3}\right)\right],
\end{eqnarray}
where $\zeta_\ell,\tilde\zeta_\ell$ and $\zeta_y,\tilde\zeta_y$ 
should be found from the DLA expression of ${\cal C}^{(3)}$ written 
in (\ref{eq:dla3partsol}), for $C_A=N_c$ in a gluon jet and $C_A=C_F$ in a quark jet. 
Introducing the parametrization in ($\mu,\nu$), one has respectively,
\begin{eqnarray}\label{eq:dlasolmunug}
\dot{{\cal C}}_{G_{123}}^{(3)}\!\!&\!\!=\!\!&\!\!1+\left(\dot{{\cal C}}_{G_{12}}^{(2)}-1\right)
+\left(\dot{{\cal C}}_{G_{13}}^{(2)}-1\right)+\left(\dot{{\cal C}}_{G_{23}}^{(2)}-1\right)\\
\!\!&\!\!+\!\!&\!\!\frac{1}{2}\frac{\left(\dot{{\cal C}}_{G_{12}}^{(2)}-1\right)
+\left(\dot{{\cal C}}_{G_{13}}^{(2)}-1\right)
+\left(\dot{{\cal C}}_{G_{23}}^{(2)}-1\right)}{1+\cosh(\mu_1-\mu_2)+\cosh(\mu_1-\mu_3)+\cosh(\mu_2-\mu_3)}\cr
\!\!&\!\!+\!\!&\!\!\frac{1}{2}\frac1{1+\cosh(\mu_1-\mu_2)+\cosh(\mu_1-\mu_3)+\cosh(\mu_2-\mu_3)},\notag
\end{eqnarray}
and
\begin{eqnarray}\label{eq:dlasolmunuq}
\dot{{\cal C}}_{Q_{123}}^{(3)}\!\!&\!\!=\!\!&\!\!1+\left(\dot{{\cal C}}_{Q_{12}}^{(2)}-1\right)
+\left(\dot{{\cal C}}_{Q_{13}}^{(2)}-1\right)+\left(\dot{{\cal C}}_{Q_{23}}^{(2)}-1\right)\\
\!\!&\!\!+\!\!&\!\!\frac{N_c}{2C_F}\frac{\left(\dot{{\cal C}}_{Q_{12}}^{(2)}-1\right)+\left(\dot{{\cal C}}_{Q_{13}}^{(2)}-1\right)
+\left(\dot{{\cal C}}_{Q_{23}}^{(2)}-1\right)}{1+\cosh(\mu_1-\mu_2)+\cosh(\mu_1-\mu_3)+\cosh(\mu_2-\mu_3)}\cr
\!\!&\!\!+\!\!&\!\!\frac{N_c^2}{2C_F^2}\frac1{1+\cosh(\mu_1-\mu_2)+\cosh(\mu_1-\mu_3)+\cosh(\mu_2-\mu_3)}.
\notag
\end{eqnarray}
Thus, in order to get $\zeta_\ell$ and $\zeta_y$, one should start from 
(\ref{eq:dlasolmunug},\ref{eq:dlasolmunuq}) and make use of
$$
\frac{\partial\mu_i}{\partial\ell}-\frac{\partial\mu_j}{\partial\ell}=
-\beta_0\gamma_0^2\frac{e^{\mu_i}\tilde Q_i-e^{\mu_j}\tilde Q_j}2,\quad
\frac{\partial\mu_i}{\partial y}-\frac{\partial\mu_j}{\partial y}=
\beta_0\gamma_0^2\frac{e^{-\mu_i}\tilde Q_i-e^{-\mu_j}\tilde Q_j}2.
$$
Therefore, everything is ready for the computation of
\begin{equation}
\zeta_\ell=\frac1{\dot{{\cal C}}_{G_{123}}^{(3)}}\dot{{\cal C}}_{G_{123},\ell}^{(3)},\quad
\zeta_y=\frac1{\dot{{\cal C}}_{G_{123}}^{(3)}}\dot{{\cal C}}_{G_{123},y}^{(3)};\quad
\tilde\zeta_\ell=\frac1{\dot{{\cal C}}_{Q_{123}}^{(3)}}\dot{{\cal C}}_{Q_{123},\ell}^{(3)},\quad
\tilde\zeta_y=\frac1{\dot{{\cal C}}_{Q_{123}}^{(3)}}\dot{{\cal C}}_{Q_{123},y}^{(3)}.
\end{equation}
For instance,
\begin{eqnarray}
\dot{{\cal C}}_{G_{123},\ell}^{(3)}\!\!&\!\!=\!\!&\!\!\chi_\ell^{12}\dot{{\cal C}}_{G_{12}}^{(2)}+
\chi_\ell^{13}\dot{{\cal C}}_{G_{13}}^{(2)}+\chi_\ell^{23}\dot{{\cal C}}_{G_{23}}^{(2)}+\frac12\frac{\chi_\ell^{12}\dot{{\cal C}}_{G_{12}}^{(2)}+\chi_\ell^{13}\dot{{\cal C}}_{G_{13}}^{(2)}+\chi_\ell^{23}\dot{{\cal C}}_{G_{23}}^{(2)}}{1+\cosh(\mu_1-\mu_2)+\cosh(\mu_1-\mu_3)+\cosh(\mu_2-\mu_3)}\cr
\!\!&\!\!-\!\!&\!\!\frac12\frac{\left(\dot{{\cal C}}_{G_{12}}^{(2)}-1\right)
+\left(\dot{{\cal C}}_{G_{13}}^{(2)}-1\right)+\left(\dot{{\cal C}}_{G_{23}}^{(2)}-1\right)}
{\left[1+\cosh(\mu_1-\mu_2)+\cosh(\mu_1-\mu_3)+\cosh(\mu_2-\mu_3)\right]^2}\left[\sinh(\mu_1-\mu_2)
\left(\frac{\partial\mu_1}{\partial\ell}-\frac{\partial\mu_2}{\partial\ell}\right)\right.\cr
\!\!&\!\!+\!\!&\!\!\left.\sinh(\mu_1-\mu_3)
\left(\frac{\partial\mu_1}{\partial\ell}-\frac{\partial\mu_3}{\partial\ell}\right)+
\sinh(\mu_2-\mu_3)
\left(\frac{\partial\mu_2}{\partial\ell}-\frac{\partial\mu_3}{\partial\ell}\right)\right]\cr
\!\!&\!\!-\!\!&\!\!\frac12\frac{1}
{\left[1+\cosh(\mu_1-\mu_2)+\cosh(\mu_1-\mu_3)+\cosh(\mu_2-\mu_3)\right]^2}\left[\sinh(\mu_1-\mu_2)
\left(\frac{\partial\mu_1}{\partial\ell}-\frac{\partial\mu_2}{\partial\ell}\right)\right.\cr
\!\!&\!\!+\!\!&\!\!\left.\sinh(\mu_1-\mu_3)
\left(\frac{\partial\mu_1}{\partial\ell}-\frac{\partial\mu_3}{\partial\ell}\right)+
\sinh(\mu_2-\mu_3)
\left(\frac{\partial\mu_2}{\partial\ell}-\frac{\partial\mu_3}{\partial\ell}\right)\right],
\end{eqnarray}
and
\begin{eqnarray}
\dot{{\cal C}}_{Q_{123},\ell}^{(3)}\!\!&\!\!=\!\!&\!\!\tilde\chi_\ell^{12}\dot{{\cal C}}_{Q_{12}}^{(2)}
+\tilde\chi_\ell^{13}\dot{{\cal C}}_{Q_{13}}^{(2)}+\tilde\chi_\ell^{23}\dot{{\cal C}}_{Q_{23}}^{(2)}
+\frac{N_c}{2C_F}\frac{\tilde\chi_\ell^{12}\dot{{\cal C}}_{Q_{12}}^{(2)}
+\tilde\chi_\ell^{13}\dot{{\cal C}}_{Q_{13}}^{(2)}
+\tilde\chi_\ell^{23}\dot{{\cal C}}_{Q_{23}}^{(2)}}{1+\cosh(\mu_1-\mu_2)+\cosh(\mu_1-\mu_3)+\cosh(\mu_2-\mu_3)}\cr
\!\!&\!\!-\!\!&\!\!\frac{N_c}{2C_F}\frac{\left(\dot{{\cal C}}_{Q_{12}}^{(2)}-1\right)
+\left(\dot{{\cal C}}_{Q_{13}}^{(2)}-1\right)+\left(\dot{{\cal C}}_{Q_{23}}^{(2)}-1\right)}
{\left[1+\cosh(\mu_1-\mu_2)+\cosh(\mu_1-\mu_3)+\cosh(\mu_2-\mu_3)\right]^2}\left[\sinh(\mu_1-\mu_2)
\left(\frac{\partial\mu_1}{\partial\ell}-\frac{\partial\mu_2}{\partial\ell}\right)\right.\cr
\!\!&\!\!+\!\!&\!\!\left.\sinh(\mu_1-\mu_3)
\left(\frac{\partial\mu_1}{\partial\ell}-\frac{\partial\mu_3}{\partial\ell}\right)+
\sinh(\mu_2-\mu_3)
\left(\frac{\partial\mu_2}{\partial\ell}-\frac{\partial\mu_3}{\partial\ell}\right)\right]\cr
\!\!&\!\!-\!\!&\!\!\frac{N_c^2}{2C_F^2}\frac{1}
{\left[1+\cosh(\mu_1-\mu_2)+\cosh(\mu_1-\mu_3)+\cosh(\mu_2-\mu_3)\right]^2}\left[\sinh(\mu_1-\mu_2)
\left(\frac{\partial\mu_1}{\partial\ell}-\frac{\partial\mu_2}{\partial\ell}\right)\right.\cr
\!\!&\!\!+\!\!&\!\!\left.\sinh(\mu_1-\mu_3)
\left(\frac{\partial\mu_1}{\partial\ell}-\frac{\partial\mu_3}{\partial\ell}\right)+
\sinh(\mu_2-\mu_3)
\left(\frac{\partial\mu_2}{\partial\ell}-\frac{\partial\mu_3}{\partial\ell}\right)\right].
\end{eqnarray}
For derivatives with respect to $y$, it is enough to replace $\ell$ by $y$ in the previous expressions.
\begin{figure}
\begin{center}
\epsfig{file=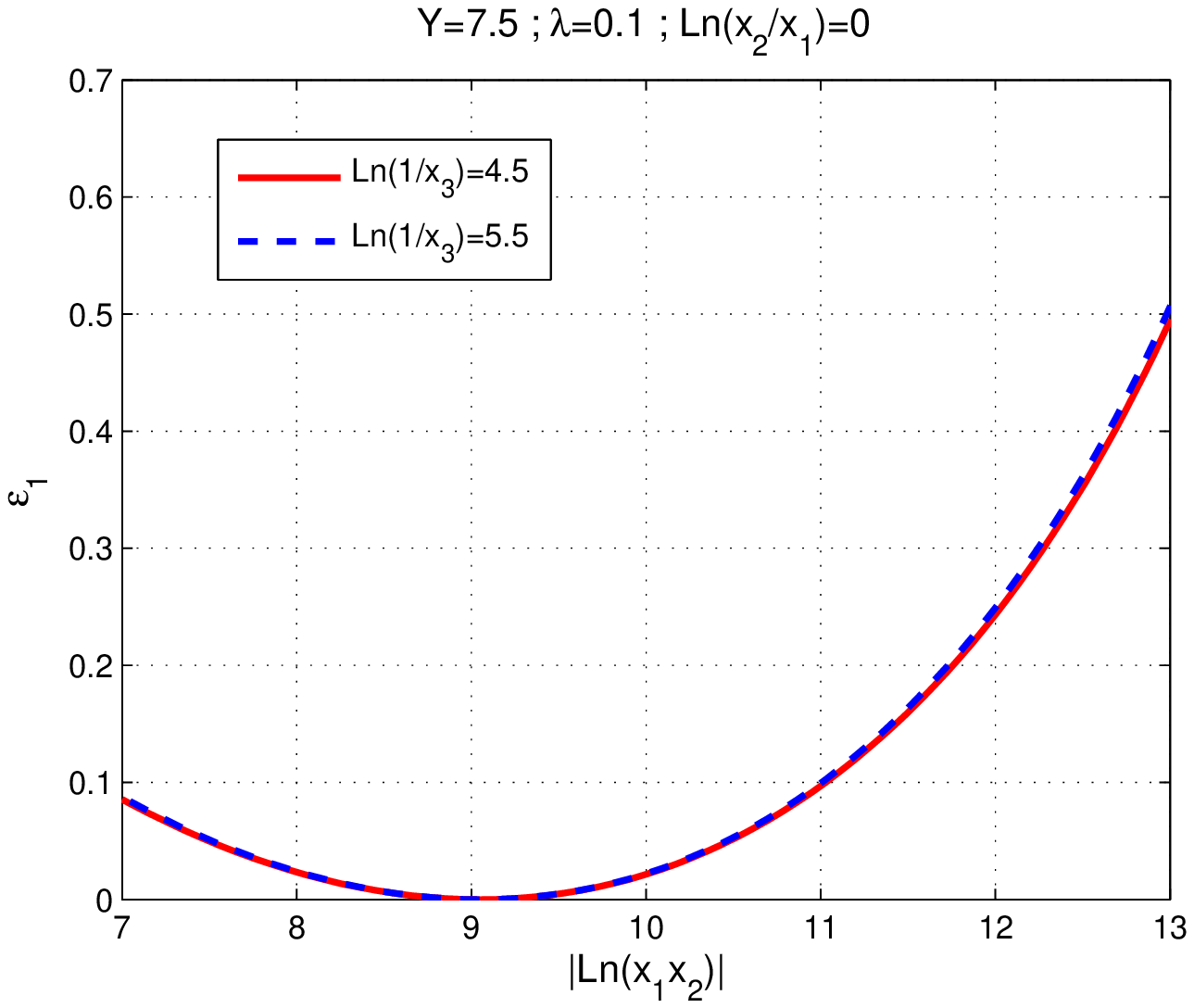, height=6.5truecm,width=7.5truecm}
\epsfig{file=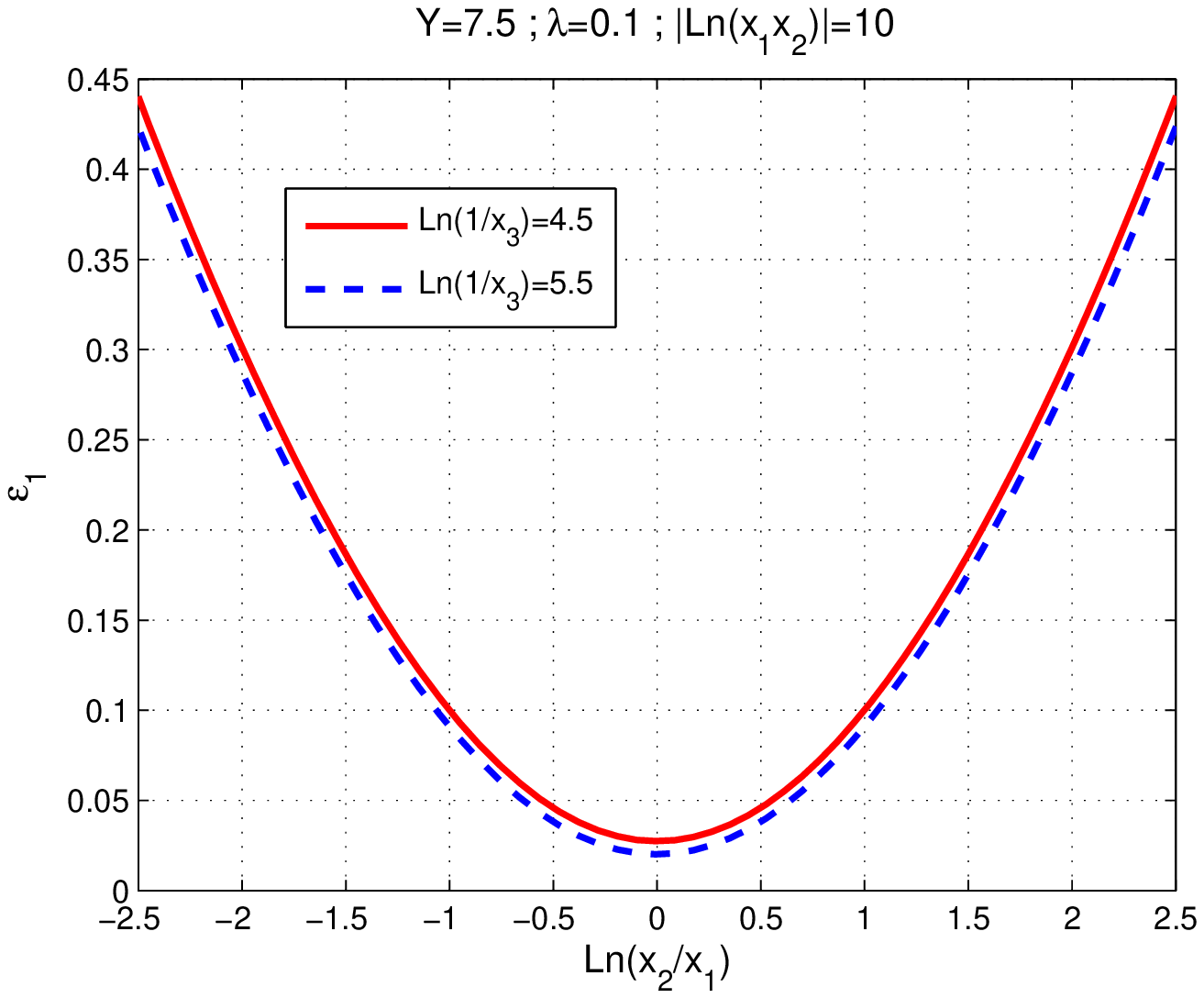, height=6.5truecm,width=7.5truecm}
\caption{\label{fig:epsilon1} Correction $\epsilon_1(\ell_1,\ell_2,\ell_3,Y)$ 
as a function of $\ell_1-\ell_2=\ln(x_2/x_1)$ for $\ell_1+\ell_2=|\ln(x_1x_2)|=10$, 
$\ell_3=\ln(1/x_3)=4.5,\,5.5$, fixed $Y=7.5$ in the limiting spectrum approximation $\lambda\approx0$.}  
\end{center}
\end{figure}
In Fig.\ref{fig:epsilon1}, we display $\epsilon_1(\ell_1,\ell_2,\ell_3,Y)$ as a function of the sum $|\ln(x_1x_2)|$ and 
the difference  $(\ell_1-\ell_2)=\ln(x_2/x_1)$ for two fixed values of $\ell_3=\ln(1/x_3)=4.5,\,5.5$, $x_1=x_2$ and
fixed sum $(\ell_1+\ell_2)=|\ln(x_1x_2)|=10$ and fixed $Y=7.5$. As expected, this correction decreases the correlations
away from the hum region and for harder particles. 

\subsection{Hump approximation}
\label{subappendix:humpapprox}
In this approximation, we consider the energy of the three partons to be close to the maximum of
the single inclusive distribution $\mid\ell-Y/2\mid\ll\sigma\propto Y^{3/2}$ for $i=1,2,3$. 
In \cite{Ramos:2006mk}, it was demonstrated that,
\begin{equation}
\psi_{i,\ell}\stackrel{\ell_{i}\sim Y/2}{\approx}\gamma_0(1+\mu_i+\frac12\mu_i^2),\quad \psi_{i,y}\stackrel{\ell_{i,j}\sim Y/2}
{\approx}\gamma_0(1-\mu_i+\frac12\mu_j^2),
\quad \mu_i\stackrel{\ell_i\sim Y/2}{\approx}\frac32\frac{y-\ell}{y+\ell},
\end{equation}
for $a,\beta_0,\lambda=0$, which is DLA. In the same approximation
one has the following for $a,\beta_0\ne0$ and $\lambda=0$,
\begin{equation}
\Delta_{ij}\stackrel{\ell_{i,j}\sim Y/2}{\approx}2+(\mu_i-\mu_j)^2-a\gamma_0(2+\mu_i+\mu_j)
-2\beta_0\gamma_0,
\end{equation}
where
\begin{equation}
(\mu_i-\mu_j)^2\stackrel{\ell_{i,j}\sim Y/2}{\approx}9\left(\frac{\ell_i-\ell_j}{Y}\right)^2,\quad 
\mu_i+\mu_j\stackrel{\ell_{i,j}\sim Y/2}{\approx}3\left(1-\frac{\ell_i+\ell_j}{Y}\right).
\end{equation}
Moreover, 
\begin{equation}
\delta_1^{ij}\stackrel{\ell_{i}\sim Y/2}{\approx}\frac29\beta_0\gamma_0(\mu_i-\mu_j)^2=2\beta_0\gamma_0\left(\frac{\ell_i-\ell_j}{Y}\right)^2,
\end{equation}
since $\gamma_0\left(\frac{\ell_i-\ell_j}{Y}\right)^2\ll\left(\frac{\ell_i-\ell_j}{Y}\right)^2$, $\delta_1$ was
neglected in this approximation.

Applying the previous expansions to (\ref{eq:NG2}-\ref{eq:DG3}) and (\ref{eq:NQ2}-\ref{eq:DQ3}), 
it is easy to find:
\begin{subequations}
\begin{eqnarray}
\label{eq:humpNQ2}
N_{Q_{ij}}^{(2)}\!\!&\!\!=\!\!&\!\!0,\\
N_{G}^{(3)}\!\!&\!\!=\!\!&\!\!1-\frac{3c}{\sqrt{\beta_0}}
\left(\frac52-\frac{|\ln(x_1x_2x_3)|}{\ln(Q/Q_0)}\right)\frac1{\sqrt{\ln(Q/Q_0)}}=
1-3c\left(\frac52-\frac{\ell_1+\ell_2+\ell_3}{Y}\right)\gamma_0,\\
D_{G}^{(3)}\!\!&\!\!=\!\!&\!\!D_{G}^{(2)}\!=\!8\!+\!9\!\left[\frac{\ln(x_2/x_1)}{\ln(Q/Q_0)}\right]^2\!
+9\!\left[\frac{\ln(x_3/x_1)}{\ln(Q/Q_0)}\right]^2\!+9\!\left[\frac{\ln(x_3/x_2)}{\ln(Q/Q_0)}\right]^2
\!-\!\frac{6\beta_0}{\sqrt{\beta_0\ln(Q/Q_0)}},\\
\!\!&\!\!-\!\!&\!\!\frac{3a}{\sqrt{\beta_0}}
\left(5-2\frac{|\ln(x_1x_2x_3)|}{\ln(Q/Q_0)}\right)\frac1{\sqrt{\ln(Q/Q_0)}},\cr
\!\!&\!\!=\!\!&\!\!8\!+\!9\!\left(\frac{\ell_1\!-\!\ell_2}{Y}\right)^2
\!\!+\!9\!\left(\frac{\ell_1\!-\!\ell_3}{Y}\right)^2\!\!+\!9\left(\frac{\ell_2\!-\!\ell_3}{Y}\right)^2\!
-\!6\beta_0\gamma_0\!-\!3a\!\left(5-2\frac{\ell_1\!+\!\ell_2\!+\!\ell_3}{Y}\right)\!\gamma_0,\cr
N_{G_{ij}}^{(2)}\!\!&\!\!=\!\!&\!\!1-\frac{3b}{\sqrt{\beta_0}}
\left(\frac52-\frac{|\ln(x_1x_2x_3)|}{\ln(Q/Q_0)}\right)\frac1{\sqrt{\ln(Q/Q_0)}}=
1-3b\left(\frac52-\frac{\ell_1+\ell_2+\ell_3}{Y}\right)\gamma_0,\\
\label{eq:humpCG2}
{\cal C}_{G_{ij}}^{(2)}\!\!&\!\!=\!\!&\!\!1+\frac{1-\frac{b}{\sqrt{\beta_0}}
\left(5-3\frac{|\ln(x_ix_j)|}{\ln(Q/Q_0)}\right)\frac1{\sqrt{\ln(Q/Q_0)}}}
{3+9\left[\frac{\ln(x_i/x_j)}{\ln(Q/Q_0)}\right]^2
-2\sqrt{\frac{\beta_0}{\ln(Q/Q_0)}}-\frac{a}{\sqrt{\beta_0}}
\left(5-3\frac{|\ln(x_ix_j)|}{\ln(Q/Q_0)}\right)\frac1{\sqrt{\ln(Q/Q_0)}}},\\
\!\!&\!\!=\!\!&\!\!1+\frac{1-b\left(5-3\frac{\ell_i+\ell_j}{Y}\right)\gamma_0}
{3+9\left(\frac{\ell_i-\ell_j}{Y}\right)^2-2\beta_0\gamma_0-a\left(5-3\frac{\ell_i+\ell_j}{Y}\right)\gamma_0},\\
\label{eq:humpCQ2}
{\cal C}_{Q_{ij}}^{(2)}\!\!&\!\!=\!\!&\!\!1+\frac{N_c}{C_F}
\left[{\cal C}_{G_{ij}}^{(2)}-1+\frac14(b-a)\gamma_0\left(5-3\frac{|\ln x_ix_j|}{\sqrt{\ln(Q/Q_0)}}\right)\right]\\
\!\!&\!\!=\!\!&\!\!1+\frac{N_c}{C_F}\left[{\cal C}_{G_{ij}}^{(2)}-1+\frac14(b-a)
\gamma_0\left(5-3\frac{\ell_i+\ell_j}{Y}\right)\right],\cr
D_{Q}^{(3)}\!\!&\!\!=\!\!&\!\!9+9\left[\frac{\ln(x_2/x_1)}{\ln(Q/Q_0)}\right]^2
+9\left[\frac{\ln(x_3/x_1)}{\ln(Q/Q_0)}\right]^2+9\left[\frac{\ln(x_3/x_2)}{\ln(Q/Q_0)}\right]^2
-\frac{6\beta_0}{\sqrt{\beta_0\ln(Q/Q_0)}}\\
\!\!&\!\!-\!\!&\!\!\frac{9a}{\sqrt{\beta_0}}
\left(\frac52-\frac{|\ln(x_1x_2x_3)|}{\ln(Q/Q_0)}\right)\frac1{\sqrt{\ln(Q/Q_0)}},\cr
\!\!&\!\!=\!\!&\!\!9\!+\!9\!\left(\frac{\ell_1\!-\!\ell_2}{Y}\right)^2
\!\!+\!9\!\left(\frac{\ell_1\!-\!\ell_3}{Y}\right)^2\!\!+\!9\left(\frac{\ell_2\!-\!\ell_3}{Y}\right)^2\!
-\!6\beta_0\gamma_0\!-\!9a\!\left(\frac52-\frac{\ell_1\!+\!\ell_2\!+\!\ell_3}{Y}\right)\!\gamma_0,\\
\label{eq:humpNQ3}
N_{Q}^{(3)}\!\!&\!\!=\!\!&\!\!\frac{N_c^2}{C_F^2}{\cal C}_{G_{123}}^{(3)}
\left[1-\frac{3a}{\sqrt{\beta_0}}\left(\frac52-\frac{|\ln(x_1x_2x_3)|}{\ln(Q/Q_0)}\right)\frac1{\sqrt{\ln(Q/Q_0)}}\right]\\
\!\!&\!\!=\!\!&\!\!\frac{N_c^2}{C_F^2}{\cal C}_{G_{123}}^{(3)}
\left[1-\frac{3a}{\sqrt{\beta_0}}\left(\frac52-\frac{\ell_1+\ell_2+\ell_3}{\ln(Q/Q_0)}\right)\frac1{\sqrt{\ln(Q/Q_0)}}\right].
\notag
\end{eqnarray}
\end{subequations}
\section{DLA solution of the 4-particle correlations}
\label{sec:4partcorr}
Below, we display the expressions related to subsection \ref{subsec:beyond}. In the l.h.s. of
the evolution equation (\ref{eq:4partcorr}), we define
\begin{eqnarray}\label{eq:4partAhat}
\hat{A}^{(4)}_{1234}\!\!&\!\!=\!\!&\!\!A^{(4)}_{1234}-\left(A^{(3)}_{123}-A_1A_2A_3\right)\!A_4-
\left(A^{(3)}_{134}-A_1A_3A_4\right)\!A_2-\left(A^{(3)}_{234}-A_2A_3A_4\right)\!A_1\\
\!\!&\!\!-\!\!&\!\!\left(A^{(3)}_{124}-A_1A_2A_4\right)\!A_3
-\left(A_{12}^{(2)}-A_1A_2\right)\!\left(A_{34}^{(2)}-A_3A_4\right)
-\left(A_{13}^{(2)}-A_1A_3\right)\!\left(A_{24}^{(2)}-A_2A_4\right)\cr
\!\!&\!\!-\!\!&\!\!\left(A_{14}^{(2)}-A_1A_4\right)\!\left(A_{23}^{(2)}-A_2A_3\right)
+\left(A_{12}^{(2)}-A_1A_2\right)A_3A_4+\left(A_{13}^{(2)}-A_1A_3\right)A_2A_4\cr
\!\!&\!\!+\!\!&\!\!\left(A_{14}^{(2)}-A_1A_4\right)A_2A_3+\left(A_{23}^{(2)}-A_2A_3\right)A_1A_4+
\left(A_{24}^{(2)}-A_2A_4\right)A_1A_3\cr
\!\!&\!\!+\!\!&\!\!\left(A_{34}^{(2)}-A_3A_4\right)A_1A_2-A_1A_2A_3A_4.
\notag
\end{eqnarray}
In the DLA solution (\ref{eq:sol4corr}) of the equation (\ref{eq:4partcorr}), 
we have introduced the expressions:
\begin{eqnarray}\label{eq:H1}
H_1\!\!&\!\!=\!\!&\!\!\left(\dot{{\cal C}}^{(2)}_{12}-1\right)+\left(\dot{{\cal C}}^{(2)}_{13}-1\right)+
\left(\dot{{\cal C}}^{(2)}_{14}-1\right)+\left(\dot{{\cal C}}^{(2)}_{23}-1\right)+
\left(\dot{{\cal C}}^{(2)}_{24}-1\right)+\left(\dot{{\cal C}}^{(2)}_{34}-1\right),\\\cr
\label{eq:H2}
H_2\!\!&\!\!=\!\!&\!\!\left(\dot{{\cal C}}^{(3)}_{123}-1\right)+\left(\dot{{\cal C}}^{(3)}_{124}-1\right)+
\left(\dot{{\cal C}}^{(3)}_{134}-1\right)+\left(\dot{{\cal C}}^{(3)}_{234}-1\right)+
\left(\dot{{\cal C}}^{(2)}_{14}-1\right)\left(\dot{{\cal C}}^{(2)}_{23}-1\right)\\
\!\!&\!\!+\!\!&\!\!\left(\dot{{\cal C}}^{(2)}_{34}-1\right)\left(\dot{{\cal C}}^{(2)}_{12}-1\right)+
\left(\dot{{\cal C}}^{(2)}_{13}-1\right)\left(\dot{{\cal C}}^{(2)}_{24}-1\right)
-2\left(\dot{{\cal C}}^{(2)}_{12}-1\right)-2\left(\dot{{\cal C}}^{(2)}_{13}-1\right)\cr
\!\!&\!\!-\!\!&\!\!2\left(\dot{{\cal C}}^{(2)}_{14}-1\right)-2\left(\dot{{\cal C}}^{(2)}_{23}-1\right)
-2\left({\dot{\cal C}}^{(2)}_{24}-1\right)-2\left(\dot{{\cal C}}^{(2)}_{34}-1\right),\cr\cr
\label{eq:H3}
H_3\!\!&\!\!=\!\!&\!\!1+\left(\dot{{\cal C}}^{(3)}_{123}-1\right)+\left(\dot{{\cal C}}^{(3)}_{124}-1\right)+
\left(\dot{{\cal C}}^{(3)}_{134}-1\right)+\left(\dot{{\cal C}}^{(3)}_{234}-1\right)+
\left(\dot{{\cal C}}^{(2)}_{14}-1\right)\left(\dot{{\cal C}}^{(2)}_{23}-1\right)\\
\!\!&\!\!+\!\!&\!\!\left(\dot{{\cal C}}^{(2)}_{34}-1\right)\left(\dot{{\cal C}}^{(2)}_{12}-1\right)+
\left(\dot{{\cal C}}^{(2)}_{13}-1\right)\left(\dot{{\cal C}}^{(2)}_{24}-1\right)
-\left(\dot{{\cal C}}^{(2)}_{12}-1\right)-\left(\dot{{\cal C}}^{(2)}_{13}-1\right)\cr
\!\!&\!\!-\!\!&\!\!\left(\dot{{\cal C}}^{(2)}_{14}-1\right)-\left(\dot{{\cal C}}^{(2)}_{23}-1\right)
-\left(\dot{{\cal C}}^{(2)}_{24}-1\right)-\left(\dot{{\cal C}}^{(2)}_{34}-1\right).\notag
\end{eqnarray}
\bibliographystyle{plain}

\bibliography{mybib}

\begin{thebibliography}{10}
\bibitem{Ramos:2011tw}
Redamy~Perez Ramos, Vincent Mathieu, and Miguel-Angel Sanchis-Lozano.
\newblock {Three-particle correlations in QCD parton showers}.
\newblock arXiv:1104.1973 [hep-ph].

\bibitem{Fritzsch:1972jv}
Harald Fritzsch and Murray Gell-Mann.
\newblock {Current algebra: Quarks and what else?}
\newblock {\em eConf}, C720906V2:135--165, 1972.

\bibitem{Gross:1973id}
D.~J. Gross and Frank Wilczek.
\newblock {ULTRAVIOLET BEHAVIOR OF NON-ABELIAN GAUGE THEORIES}.
\newblock {\em Phys. Rev. Lett.}, 30:1343--1346, 1973.

\bibitem{Hanson:1975fe}
G.~Hanson et~al.
\newblock {Evidence for Jet Structure in Hadron Production by e+ e-
  Annihilation}.
\newblock {\em Phys. Rev. Lett.}, 35:1609--1612, 1975.

\bibitem{Berger:1978rr}
Christoph Berger et~al.
\newblock {JET ANALYSIS OF THE UPSILON (9.46) DECAY INTO CHARGED HADRONS}.
\newblock {\em Phys. Lett.}, B82:449, 1979.

\bibitem{Dokshitzer:1991wu}
Yuri~L. Dokshitzer, Valery~A. Khoze, Alfred~H. Mueller, and S.~I. Troian.
\newblock Gif-sur-Yvette, France: Ed. Frontieres (1991) 274 p. (Basics of).

\bibitem{Azimov:1984np}
Yakov~I. Azimov, Yuri~L. Dokshitzer, Valery~A. Khoze, and S.~I. Troyan.
\newblock {Similarity of Parton and Hadron Spectra in QCD Jets}.
\newblock {\em Z. Phys.}, C27:65--72, 1985.

\bibitem{Azimov:1985by}
Yakov~I. Azimov, Yuri~L. Dokshitzer, Valery~A. Khoze, and S.~I. Troyan.
\newblock {Humpbacked QCD Plateau in Hadron Spectra}.
\newblock {\em Zeit. Phys.}, C31:213, 1986.

\bibitem{Akrawy:1990ha}
M.~Z. Akrawy et~al.
\newblock {A Study of coherence of soft gluons in hadron jets}.
\newblock {\em Phys. Lett.}, B247:617--628, 1990.

\bibitem{Abbiendi:1999ki}
G.~Abbiendi et~al.
\newblock {Intermittency and correlations in hadronic Z0 decays}.
\newblock {\em Eur. Phys. J.}, C11:239--250, 1999.

\bibitem{:2008ec}
T.~Aaltonen et~al.
\newblock {Two-Particle Momentum Correlations in Jets Produced in $p\bar{p}$
  Collisions at $\sqrt{s}$ = 1.96-TeV}.
\newblock {\em Phys. Rev.}, D77:092001, 2008.

\bibitem{Aaltonen:2008yn}
T.~Aaltonen et~al.
\newblock {Measurement of the $k_T$ Distribution of Particles in Jets Produced
  in $p\bar{p}$ Collisions at $\sqrt{s}=1.96$-TeV}.
\newblock {\em Phys. Rev. Lett.}, 102:232002, 2009.

\bibitem{PerezRamos:2005nh}
Redamy Perez-Ramos and Bruno Machet.
\newblock {MLLA inclusive hadronic distributions inside one jet at high energy
  colliders}.
\newblock {\em JHEP}, 04:043, 2006.

\bibitem{PerezRamos:2007cr}
Redamy P{\'e}rez~Ramos, Francois Arleo, and Bruno Machet.
\newblock {Next-to-MLLA corrections to single inclusive kt- distributions and
  2-particle correlations in a jet}.
\newblock {\em Phys. Rev.}, D78:014019, 2008.

\bibitem{Arleo:2007wn}
Francois Arleo, Redamy P{\'e}rez~Ramos, and Bruno Machet.
\newblock {Hadronic single inclusive kt distributions inside one jet beyond
  MLLA}.
\newblock {\em Phys. Rev. Lett.}, 100:052002, 2008.

\bibitem{Dokshitzer:1982ia}
Yuri~L. Dokshitzer, Victor~S. Fadin, and Valery~A. Khoze.
\newblock {On the Sensitivity of the Inclusive Distributions in Parton Jets to
  the Coherence Effects in QCD Gluon Cascades}.
\newblock {\em Z. Phys.}, C18:37, 1983.

\bibitem{Fong:1990ph}
C.~P. Fong and B.~R. Webber.
\newblock {TWO PARTICLE CORRELATIONS AT SMALL x IN QCD JETS}.
\newblock {\em Phys. Lett.}, B241:255, 1990.

\bibitem{Fong:1990nt}
C.~P. Fong and B.~R. Webber.
\newblock {One and two particle distributions at small x in QCD jets}.
\newblock {\em Nucl. Phys.}, B355:54--81, 1991.

\bibitem{Ramos:2006dx}
Redamy~Perez Ramos.
\newblock {Two particle correlations inside one jet at 'modified leading
  logarithmic approximation' of quantum chromodynamics. I: Exact solution of
  the evolution equations at small x}.
\newblock {\em JHEP}, 06:019, 2006.

\bibitem{Acton:1992gd}
P.~D. Acton et~al.
\newblock {A Study of two particle momentum correlations in hadronic Z0
  decays}.
\newblock {\em Phys. Lett.}, B287:401--412, 1992.

\bibitem{DeWolf:1995pc}
E.~A. De~Wolf, I.~M. Dremin, and W.~Kittel.
\newblock {Scaling laws for density correlations and fluctuations in
  multiparticle dynamics}.
\newblock {\em Phys. Rept.}, 270:1--141, 1996.

\bibitem{Kittel:2005fu}
W.~Kittel and E.~A. De~Wolf.
\newblock {Soft multihadron dynamics}.
\newblock Hackensack, USA: World Scientific (2005) 652 p.

\bibitem{Konishi:1979cb}
K.~Konishi, A.~Ukawa, and G.~Veneziano.
\newblock {Jet Calculus: A Simple Algorithm for Resolving QCD Jets}.
\newblock {\em Nucl. Phys.}, B157:45--107, 1979.

\bibitem{Dokshitzer:1978hw}
Yuri~L. Dokshitzer, Dmitri Diakonov, and S.~I. Troian.
\newblock {Hard Processes in Quantum Chromodynamics}.
\newblock {\em Phys. Rept.}, 58:269--395, 1980.

\bibitem{Ramos:2006mk}
Redamy~Perez Ramos.
\newblock {Single inclusive distribution and two-particle correlations inside
  one jet at 'modified leading logarithmic approximation' of quantum
  chromodynamics. II: Steepest descent evaluation at small X}.
\newblock {\em JHEP}, 09:014, 2006.

\bibitem{Dokshitzer:1988bq}
Yuri~L. Dokshitzer, Valery~A. Khoze, and S.~I. Troian.
\newblock {COHERENCE AND PHYSICS OF QCD JETS}.
\newblock {\em Adv. Ser. Direct. High Energy Phys.}, 5:241--410, 1988.

\bibitem{Dokshitzer:1982xr}
Yuri~L. Dokshitzer, Victor~S. Fadin, and Valery~A. Khoze.
\newblock {Double Logs of Perturbative QCD for Parton Jets and Soft Hadron
  Spectra}.
\newblock {\em Zeit. Phys.}, C15:325, 1982.

\bibitem{Khoze:1996dn}
Valery~A. Khoze and Wolfgang Ochs.
\newblock {Perturbative QCD approach to multiparticle production}.
\newblock {\em Int. J. Mod. Phys.}, A12:2949--3120, 1997.

\bibitem{Fadin:1983dl}
V.~S. Fadin.
\newblock Double logarithmic asymptotics of the cross-sections of e+ e-
  annihilation into quarks and gluons.
\newblock {\em Sov. J. Nucl. Phys.}, 37:245, 1983.

\bibitem{Koba:1972ng}
Z.~Koba, Holger~Bech Nielsen, and P.~Olesen.
\newblock {Scaling of multiplicity distributions in high-energy hadron
  collisions}.
\newblock {\em Nucl. Phys.}, B40:317--334, 1972.

\bibitem{Dokshitzer:1993dc}
Yuri~L. Dokshitzer.
\newblock {Improved QCD treatment of the KNO phenomenon}.
\newblock {\em Phys. Lett.}, B305:295--301, 1993.

\bibitem{Malaza:1985jd}
E.~D. Malaza and B.~R. Webber.
\newblock {MULTIPLICITY DISTRIBUTIONS IN QUARK AND GLUON JETS}.
\newblock {\em Nucl. Phys.}, B267:702, 1986.

\bibitem{Lupia:1998nc}
Sergio Lupia, Wolfgang Ochs, and Jacek Wosiek.
\newblock {Poissonian limit of soft gluon multiplicity}.
\newblock {\em Nucl. Phys.}, B540:405--433, 1999.

\bibitem{Chekanov:2001sj}
S.~Chekanov et~al.
\newblock {Multiplicity moments in deep inelastic scattering at HERA}.
\newblock {\em Phys. Lett.}, B510:36--54, 2001.

\bibitem{Abbiendi:2006qr}
G.~Abbiendi et~al.
\newblock {QCD coherence and correlations of particles with restricted momenta
  in hadronic Z decays}.
\newblock {\em Phys. Lett.}, B638:30--38, 2006.

\end{thebibliography}

%%%%%%%%%%%%%%%%%%%%%%%%%%%%%%%%%%%%%%%%%%%%%%%%%%%%%%%%%%%%%%%%%%%%%%%%%%%
\end{document}